\documentclass[french]{article}
\def\ifFrench{\iftrue}

\usepackage{RR}
\usepackage{style-rapport-bref}
\usepackage{tabularx}

\RRdate{Mai 2006}

\RRauthor{Stéphane Huet, Pascale Sébillot et Guillaume Gravier
\thanks{\{shuet, sebillot, ggravier\}@irisa.fr}}

\authorhead{S. Huet, P. Sébillot \& G. Gravier} 
\RRtitle{Utilisation de la linguistique en reconnaissance de la parole
: un état de l'art}
\titlehead{Utilisation de la linguistique en reconnaissance de la parole} 

\RRetitle{\begin{otherlanguage*}{english} Using Linguistics in Speech
  Recognition: A State of the Art \end{otherlanguage*}} 

\RRabstract{To transcribe speech, automatic speech recognition systems
  use statistical methods, particularly hidden Markov model and N-gram
  models. Although these techniques perform well and lead to efficient
  systems, they approach their maximum possibilities. It seems thus
  necessary, in order to outperform current results, to use additional
  information, especially bound to language. However, introducing such
  knowledge must be realized taking into account specificities of
  spoken language (hesitations for example) and being robust to
  possible misrecognized words. This document presents a state of the
  art of these researches, evaluating the impact of the insertion of
  linguistic information on the quality of the transcription.}

\RRresume{Pour transcrire des documents sonores, les systèmes de
reconnaissance de la parole font appel à des méthodes statistiques,
notamment aux chaînes de Markov cachées et aux modèles N-grammes. Même
si ces techniques se sont révélées performantes, elles approchent du
maximum de leurs possibilités avec la mise à disposition de corpus de
taille suffisante et il semble nécessaire, pour tenter d'aller au-delà
des résultats actuels, d'utiliser des informations supplémentaires, en
particulier liées au langage. Intégrer de telles connaissances
linguistiques doit toutefois se faire en tenant compte des
spécificités de l'oral (présence d'hésitations par exemple) et en
étant robuste à d'éventuelles erreurs de reconnaissance de certains
mots. Ce document présente un état de l'art des recherches de ce type,
en évaluant l'impact de l'insertion des informations linguistiques sur
la qualité de la transcription.}

\RRmotcle{reconnaissance de la parole, langue parlée, corpus oral,
  traitement automatique des langues, modèle de langage, connaissances
  linguistiques, disfluences} 

\RRkeyword{speech recognition, spoken language, spoken corpus, natural
  language processing, language model, linguistics knowledge,
  disfluencies}

\RRprojets{TexMex et Metiss} 
\RRtheme{\THSym \THCog} 

\URRennes

\begin{document}

\makeRR 

\tableofcontents
\newpage

\section{Introduction}
\label{SecIntroduction}

De nombreux documents sonores contiennent de la parole. Rares sont les
émissions radiophoniques ou audiovisuelles sans discours, dialogues ou
encore commentaires, et il existe maintenant certaines bases de
données sonores très volumineuses. En France, l'Institut national de
l'audiovisuel collecte ainsi chaque année 80~000 heures de programmes
radiophoniques ou télévisuels. Une des plus grandes archives digitales
au monde, contenant les témoignages des survivants de la Shoah,
contient quant à elle plus de 115~000 heures de discours non
contraints, provenant de 52~000 locuteurs s'exprimant dans 32 langues
différentes \cite{Franz03}. Face à la taille des données à traiter, le
recours à des méthodes automatiques, notamment à celles de
reconnaissance de la parole, facilite la manipulation des documents et
apporte une aide pour certaines tâches, telles que l'indexation.

L'objectif d'un système de reconnaissance automatique de la parole
(RAP) est de transcrire automatiquement un signal sonore en texte. Un
tel système cherche dans un premier temps à reconnaître des mots, en
se basant uniquement sur des critères d'ordre acoustique, sans essayer
d'interpréter le \og contenu \fg\ transmis par l'ensemble de ces
mots. L'analyse du signal conduit alors à un ensemble d'hypothèses sur
la succession des mots prononcés, auquel sont adjoints des scores dits
acoustiques. Une seconde étape choisit la meilleure hypothèse en ne
considérant plus le signal comme une suite de sons mais plutôt comme
une succession de mots porteurs d'information. L'utilisation de la
linguistique, en tant que science du langage, s'inscrit naturellement
dans ce contexte puisque des connaissances sur la morphologie, la
syntaxe ou le sens semblent pouvoir guider la sélection de la
meilleure hypothèse. Néanmoins, bien souvent les systèmes de RAP ne
s'appuient dans leur choix que sur des critères principalement
statistiques, et ce, essentiellement pour des raisons historiques.

Pendant une longue période, les linguistes et les adeptes du
traitement automatique des langues (TAL) ont en effet délaissé les
méthodes empiriques permettant d'estimer la probabilité d'observation
d'un phénomène à partir d'une grande collection de documents ou
corpus. Noam Chomsky dit ainsi en 1969 : \textit{But it must be
recognized that the notion ``probability of a sentence'' is an
entirely useless one, under any known interpretation of this
term}\footnote{Mais il doit être reconnu que la notion de \og
probabilité de phrase\fg\ est absolument inutile, et ce, quelle que
soit l'interprétation de ce terme.}. Le point de vue des linguistes
s'intéressant à l'écrit s'est alors éloigné des préoccupations des
chercheurs en reconnaissance de la parole, à tel point que Frederick
Jelinek, en 1988, alors à IBM, déclara : \textit{Anytime a linguist
leaves the group the recognition rate goes up}\footnote{À chaque fois
qu'un linguiste quitte le groupe, le taux de reconnaissance
augmente.}. Les méthodes statistiques, à base de N-grammes, se sont en
effet révélées beaucoup plus efficaces que les solutions proposées par
les linguistes pour choisir la meilleure hypothèse de mots.

Depuis la fin des années 80, les méthodes statistiques ont toutefois
commencé à atteindre leurs limites dans l'amélioration des systèmes de
RAP, notamment avec la mise à disposition de corpus de tailles
satisfaisantes. Dans le même temps, le domaine du TAL faisait de plus
en plus appel aux modèles probabilistes. Une des pistes envisageables
pour améliorer les performances de la reconnaissance de la parole peut
donc consister à employer davantage de linguistique dans les systèmes
de RAP.

Ce document se propose de faire une synthèse de l'introduction de
connaissances linguistiques au cours de la reconnaissance de la
parole. Il expose des tentatives effectuées pour utiliser certaines
informations telles que la morphologie, la syntaxe ou la sémantique
dans les différentes étapes du processus de transcription. Une
première partie décrit le fonctionnement général d'un système de RAP,
en insistant sur ses limitations à traiter certains phénomènes. Elle
présente la succession des étapes nécessaires, à savoir l'extraction
d'informations numériques pertinentes à partir du son, la conversion
de ces valeurs en plusieurs hypothèses possibles de succession de mots
et enfin le choix de la meilleure hypothèse. La deuxième partie
examine les propriétés de la langue parlée. Les méthodes de TAL sont
en effet souvent appliquées à des documents qui restent dans le
domaine de l'écrit, comme des ouvrages ou des articles de journaux ;
or, les documents analysés par les systèmes de RAP sont d'une autre
nature. La dernière section expose à quel niveau du processus de
transcription les connaissances linguistiques peuvent être mobilisées.
Elle se focalise particulièrement sur l'introduction de connaissances
linguistiques au sein des modèles de langages, un des deux
constituants, avec le modèle acoustique, d'un système de RAP. Si le
modèle acoustique utilise des ressources purement acoustiques, se
limitant donc, sur le plan de la linguistique, à la phonétique et à la
phonologie, le modèle de langage peut au contraire prendre en compte
des connaissances linguistiques plus variées puisque son rôle est
justement d'examiner les informations véhiculées par les hypothèses de
mots. Ceci peut donc conduire à l'exploitation de morphologie, de
syntaxe, de sémantique ou encore de pragmatique.

\newpage

\section{Principes de la reconnaissance de la parole}
\label{SecPrincRP}

L'objectif d'un système de RAP est d'extraire les mots prononcés à
partir du signal acoustique. Le résultat produit représente la
finalité d'une application de dictée vocale mais peut également être
utilisé par d'autres dispositifs. La transcription constitue ainsi une
source d'informations pour indexer des documents audio ou
audiovisuels. Les interfaces vocales homme-machine intègrent quant à
elles un module de compréhension de la parole en sus d'un système de
RAP.

La reconnaissance de la parole est un problème complexe, notamment du
fait de la grande variabilité des signaux à traiter. Après avoir
présenté les principales difficultés de la transcription, nous
exposons la modélisation qui est employée pour décoder le signal
acoustique. Nous évoquons ensuite sous quelles formes les systèmes de
RAP produisent leurs résultats et nous terminons cette section par une
description des techniques utilisées pour évaluer ces systèmes.

\subsection{Difficultés de la transcription}
\label{subSecDiffTrans}

Les difficultés de la transcription de la parole sont dues pour une
grande part à la diversité des signaux à traiter. La parole produite
pour une même phrase prononcée peut ainsi varier d'un individu à un
autre. Outre le fait que chaque individu possède une voix qui lui est
propre, on rencontre d'importantes différences telles que les
variations homme/femme, le régionalisme ou encore les difficultés de
prononciation rencontrées par des locuteurs non natifs. Cette
variabilité est qualifiée d'\emph{inter-locuteurs}. Il existe
également une variabilité, dite \emph{intra-locuteur}, correspondant à
une modification de la parole produite par un même individu. Cette
variabilité peut concerner aussi bien les caractéristiques de la voix,
dans les cas d'un rhume ou d'un état émotionnel, ou bien la qualité
d'élocution, selon que la parole intervient lors d'un discours formel
ou d'un dialogue spontané. En sus des variabilités au niveau de la
parole prononcée par le locuteur, les conditions d'enregistrement
peuvent dégrader le signal qui sera traité par le système de RAP. Il
peut par exemple y avoir une modification de la qualité du signal,
notamment si celui-ci doit transiter par un canal de communication qui
a une bande passante limitée, comme une ligne téléphonique. De même,
l'environnement acoustique peut être disparate. Le bruit de fond peut
être plus ou moins important et de natures diverses (musique, paroles
d'autres locuteurs, parasites du micro, bruits de bouche...).

Par ailleurs, le lexique des documents à transcrire est un autre
facteur-clé influençant la qualité des résultats et dépendant de
chaque application. La taille du vocabulaire peut être très réduite
(moins de 100 mots) dans le cas d'un système de navigation dans un
menu, moyenne (quelques milliers de mots) pour des recherches
d'information dans une base de données dans un domaine précis, ou
large (plusieurs dizaines de milliers de mots) pour faire de la dictée
vocale \cite{Stolcke97}.

La parole humaine, on le voit donc, est très variable. La modélisation
d'une telle variation étant difficile à faire de manière compacte et
la compréhension des mécanismes cognitifs intervenant dans la
reconnaissance de la parole étant limitée, les systèmes de RAP
utilisent essentiellement des méthodes statistiques. Ces méthodes
extraient automatiquement les informations sur le langage et la
relation entre le son et les mots prononcés à partir de corpus dans
lesquels les textes sont alignés avec les signaux acoustiques. Elles
utilisent ainsi des modèles dont les paramètres sont appris sur des
corpus d'apprentissage \cite{Deshmukh99}.

Pour réduire les difficultés impliquées par la variabilité
inter-locuteurs, certains systèmes de RAP requièrent de la part de
chaque locuteur une prononciation préalable d'un certain nombre de
mots. Toutefois, ce procédé étant contraignant, les systèmes sont
généralement indépendants du locuteur. Afin d'augmenter la
\emph{robustesse} vis-à-vis du locuteur et de l'environnement, ils
combinent plusieurs modèles statistiques \cite{Jouvet96,
Gauvain05}. Ainsi, dans le cas de la variabilité inter-locuteurs, un
modèle spécifique peut être créé pour les hommes, un autre pour les
femmes. De même, des mécanismes d'adaptation permettent de modifier le
traitement acoustique en fonction des caractéristiques de la voix du
locuteur. Dans le cas de changements de conditions d'enregistrement,
les systèmes de RAP peuvent avoir un modèle particulier pour les
entretiens téléphoniques et avoir des détecteurs \og bruit/parole\fg\
ou \og musique/parole\fg\ pour ne chercher à transcrire que les
segments du signal contenant de la parole. En outre, un filtrage
adaptatif peut être mené pour éliminer le bruit du signal.

En ce qui concerne le lexique, afin d'avoir un espace de recherche
raisonnable lors du décodage du signal acoustique, les systèmes de RAP
ont un vocabulaire fermé. Ils utilisent actuellement un lexique
beaucoup plus important que ceux employés auparavant, qui pouvaient se
limiter à quelques dizaines de mots. Les systèmes de RAP les plus
perfectionnés, dits à très grand vocabulaire, ont ainsi un lexique
dont le volume avoisine les 65~000, voire 200~000~mots. L'ensemble des
mots reconnaissables est choisi en fonction de l'application. Dans le
cas de reconnaissance d'un dialogue spontané, il pourra ainsi être
utile d'inclure dans le lexique des mots qui n'en sont pas vraiment
mais qui ont pourtant une fréquence élevée, tels que le \og
\textit{euh}\fg\ marquant une hésitation (\textit{cf.}
section~\ref{subSecPhHesitation}). Si l'on prend l'exemple d'une
transcription d'émissions d'actualité diffusées à la radio, le
vocabulaire pourra inclure les mots rencontrés récemment. Ce domaine
d'application se heurte à des difficultés particulières dues à
l'apparition des noms propres (noms de personnes ou de lieu) au gré de
l'actualité. De manière à pouvoir modifier le vocabulaire pris en
compte par le système de RAP, la reconnaissance doit être
\emph{flexible}, ce qui signifie qu'elle doit autoriser l'introduction
de mots dans le dictionnaire qui n'ont pas été utilisés au cours de
l'apprentissage des paramètres des modèles.

Notons au passage l'ambiguïté du terme \emph{mot}
\cite{Polguere03}. Il peut par exemple désigner un sens précis ou bien
un signe linguistique. En reconnaissance de la parole, \emph{mot}
désigne un \emph{mot-forme} défini par son orthographe. Ainsi, deux
flexions ou dérivations d'un même lemme, \textit{e.g.}  \og
\textit{mange}\fg\ et \og \textit{manges}\fg, seront considérées comme
deux mots différents. De même, deux homographes appartenant à deux
catégories différentes (\textit{e.g.} \og \textit{mérite}
[{\footnotesize VERBE}]\fg\ et \og \textit{mérite} [{\footnotesize
NOM}]\fg) ou deux sens différents (\textit{e.g.}  \og \textit{avocat}
[{\footnotesize AUXILIAIRE DE JUSTICE}]\fg\ et \og \textit{avocat}
[{\footnotesize FRUIT COMESTIBLE}]\fg) ne représenteront pas le même
mot \cite{Jelinek97}. Par la suite, chaque emploi du terme \emph{mot}
désignera en réalité un \emph{mot-forme}.

Malgré les difficultés rencontrées, les systèmes de RAP parviennent à
décoder le signal acoustique avec d'assez bonnes performances. La
transcription de mots prononcés de manière isolée, par un locuteur
unique, est une technologie ancienne et bien maîtrisée. Son champ
d'application est néanmoins très restreint puisque le locuteur doit
marquer une pause brève entre chaque mot. Les chercheurs et ingénieurs
en parole ont par la suite développé des systèmes de transcription de
parole continue, ce qui a augmenté considérablement la complexité du
problème. Les systèmes actuels les plus performants reconnaissent
toutefois sans erreur plus de 90\,\% des mots d'une émission d'actualité
en anglais \cite{Pallett03} et plus de 88\,\% des mots pour une émission
du même type en français \cite{Gauvain05}. Dans des situations plus
complexes à analyser, comme des conversations téléphoniques, où chacun
des locuteurs n'a pas préparé son discours et est donc sujet à de
nombreuses hésitations, les systèmes de RAP peuvent reconnaître sans
erreur jusqu'à 80\,\% des mots \cite{Pallett03, Gauvain04}.

\subsection{Modélisation statistique de la reconnaissance de la parole}
\label{subSecModelRP}

Le signal sonore à étudier peut être interprété comme une version de
la phrase prononcée qui serait passée par un canal de
communication. Ce canal introduit du \og bruit\fg\ dans la version
originale. L'objectif d'un système de RAP est de modéliser le canal de
manière à retrouver la phrase prononcée après décodage
(Fig.~\ref{figCanalBruite}). Ceci revient à chercher parmi un très
grand nombre de phrases sources potentielles celle qui a la plus
grande probabilité de générer la phrase \og bruitée\fg\
\cite{Jurafsky00}. Autrement dit, dans cette métaphore du \emph{canal
bruité}, un système de RAP cherche à trouver la séquence de mots la
plus probable $W$ parmi toutes les séquences d'un langage
$\mathcal{L}$, étant donné le signal acoustique $A$.

\FigurePS{!htbp}{10cm}{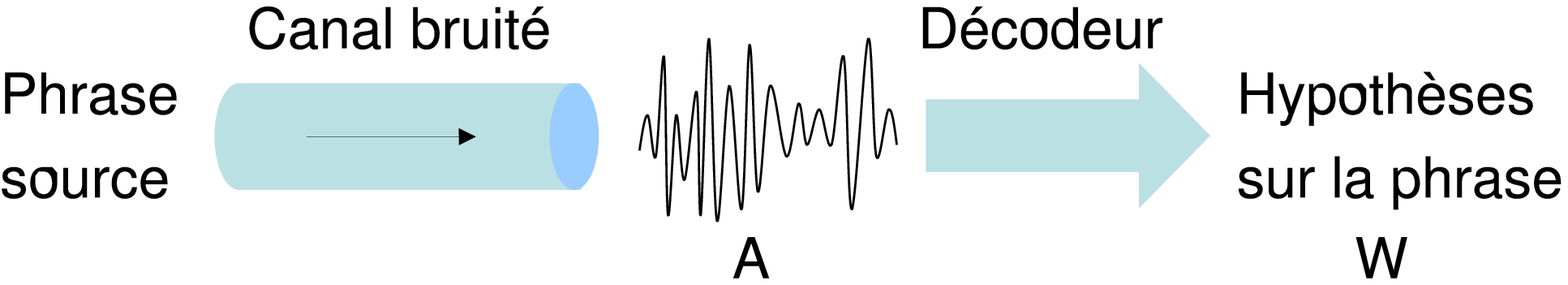}{Modèle du canal bruité}{figCanalBruite}

L'entrée acoustique $A$ représente une séquence d'ob\-ser\-va\-tions
$a_1 \ldots a_t$, ob\-te\-nue en dé\-cou\-pant l'entrée par exemple
toutes les 10\,millisecondes et en associant à chaque morceau les
coefficients représentant l'enveloppe spectrale du signal. La sortie
$W$ du système de RAP est une chaîne de mots $w_1 \ldots w_n$
appartenant à un vocabulaire fini.

En utilisant une formalisation statistique issue de la théorie de
l'information, le problème de la reconnaissance de la parole se ramène
alors à chercher :
\begin{equation} 
\hat{W}=\arg\max_{W \in \mathcal{L}} \; P(W|A)
\end{equation}
Cette équation peut être réécrite sous la forme suivante à l'aide de
la formule de Bayes \cite{Jelinek97} :
\begin{equation} \label{eqWbay}
\hat{W}=\arg\max_{W \in \mathcal{L}} \; \frac{P(W) P(A|W)}{P(A)}
\end{equation}
où $P(W)$ est la probabilité que $W$ soit prononcée, $P(A|W)$ est la
probabilité que le locuteur émette les sons $A$ en souhaitant prononcer
les mots $W$ et $P(A)$ est la probabilité moyenne que $A$ soit
produit. $\hat{W}$ étant estimé en fixant $A$, $P(A)$ n'intervient pas
et l'équation (\ref{eqWbay}) devient :
\begin{equation} \label{eqWbaySimpl}
\hat{W}=\arg\max_{W \in \mathcal{L}} \; P(W)P(A|W)
\end{equation}

Le problème de la reconnaissance de la parole se ramène ainsi à
l'extraction des indices acoustiques $A$, au calcul de la
vraisemblance d'observation $P(A|W)$ ainsi que de la probabilité
\textit{a priori} $P(W)$, et à la recherche de la séquence de mots $W$
la plus probable. Pour ce faire, la transcription se décompose en
plusieurs modules (Fig.~\ref{figRAP}) :
\begin{itemize}
\item l'extraction de caractéristiques produisant $A$,
\item l'utilisation du \emph{modèle acoustique} (MA) calculant
  $P(A|W)$ et cherchant les hypothèses $W$ qui sont vraisemblablement
  associées à $A$,
\item l'utilisation du \emph{modèle de langage} (ML) calculant $P(W)$
  pour choisir une ou plusieurs hypothèses sur $W$ en fonction de
  connaissances sur la langue.
\end{itemize}
\noindent Pour évaluer $P(W)$, le ML doit disposer au préalable des
hypothèses $W$ établies par le MA sur les mots prononcés. Néanmoins,
les systèmes de RAP actuels ne se limitent pas à une juxtaposition
séquentielle des deux modules ; de manière à utiliser le plus tôt
possible les informations sur la langue, ils font appel au ML dès
qu'une hypothèse de mot est proposée par le MA et non à la fin du
traitement de la totalité du signal.

\FigurePS{!htbp}{10cm}{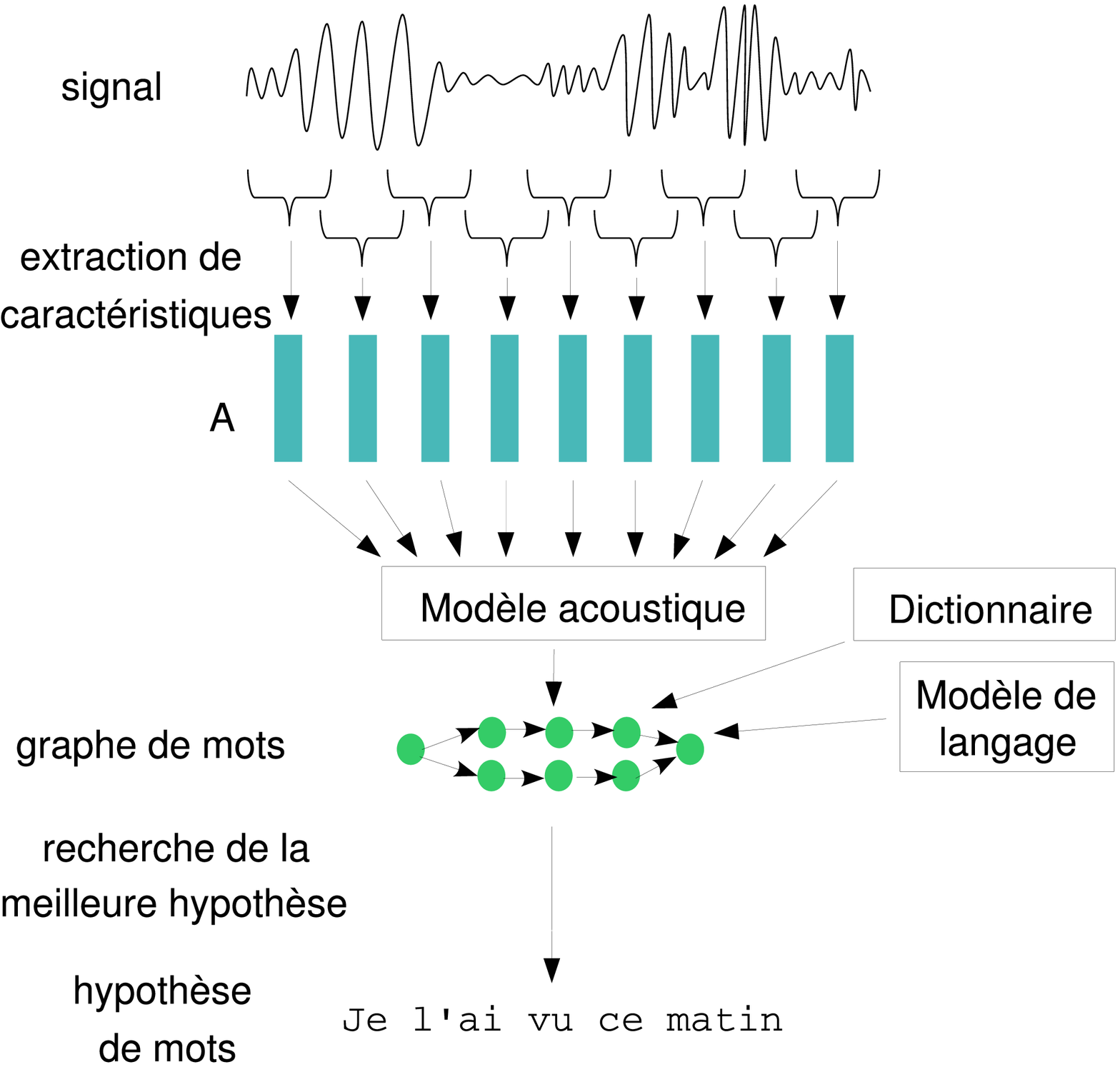}{Constituants d'un système de
transcription}{figRAP}

Les sections suivantes décrivent les principes de fonctionnement de
chacun de ces modules.

\subsubsection{Extraction de caractéristiques}
\label{subSecCaract}

Le signal sonore à analyser se présente sous la forme d'une onde dont
l'intensité varie au cours du temps. La première étape du processus de
transcription consiste à extraire une succession de valeurs
numériques suffisamment informatives sur le plan acoustique pour
décoder le signal par la suite.

Le signal est susceptible de contenir des zones de silence, de bruit
ou de musique. Ces zones sont tout d'abord éliminées afin de n'avoir
que des portions du signal utiles à la transcription, \textit{i.e.},
celles qui correspondent à de la parole. Le signal sonore est ensuite
segmenté en ce que l'on qualifie de \emph{groupes de souffle}, en
utilisant comme délimiteurs des pauses silencieuses suffisamment
longues (de l'ordre de 0,3~s). L'intérêt de cette segmentation est
d'avoir un signal sonore continu de taille raisonnable par rapport aux
capacités de calculs des modèles du système de RAP ; dans la suite du
processus de transcription, l'analyse se fera séparément pour chaque
groupe de souffle.

Pour repérer les fluctuations du signal sonore, qui varie généralement
rapidement au cours du temps, le groupe de souffle est lui-même
découpé en fenêtres d'étude de quelques millisecondes (habituellement
de 20 ou 30\,ms). De manière à ne pas perdre d'informations
importantes se trouvant en début ou fin de fenêtres, on fait en sorte
que celles-ci se chevauchent, ce qui conduit à extraire des
caractéristiques toutes les 10\,ms environ.

À partir du signal contenu dans chaque fenêtre d'analyse sont
calculées des valeurs numériques caractérisant la voix humaine. À
l'issue de cette étape, le signal devient alors une succession de
vecteurs dits acoustiques, de dimension souvent supérieure ou égale à
39.

\subsubsection{Modèle acoustique}
\label{subSecMA}

Une étape suivante consiste à associer aux vecteurs acoustiques, qui
sont, comme nous venons de le voir, des vecteurs numériques, un
ensemble d'hypothèses de mots, \textit{i.e.}, des symboles. En se
référant à l'équation~(\ref{eqWbaySimpl}) de la modélisation
statistique, cela revient à estimer $P(A|W)$. Les techniques qui
permettent de calculer cette valeur forment ce qu'on appelle le modèle
acoustique.

L'outil le plus utilisé pour la modélisation du MA est la chaîne de
Markov cachée (désignée aussi sous le terme de HMM pour \textit{Hidden
Markov Model}). Les HMM ont en effet montré dans la pratique leur
efficacité pour reconnaître la parole. Même s'ils présentent quelques
limitations pour modéliser certaines caractéristiques du signal, comme
la durée ou la dépendance des observations acoustiques successives,
les HMM offrent un cadre ma\-thé\-ma\-tique bien défini pour calculer
les probabilités $P(A|W)$ \cite{Rabiner89}. Les MA font intervenir
trois niveaux de HMM (Fig.~\ref{figMA}).

\FigurePS{!htbp}{10cm}{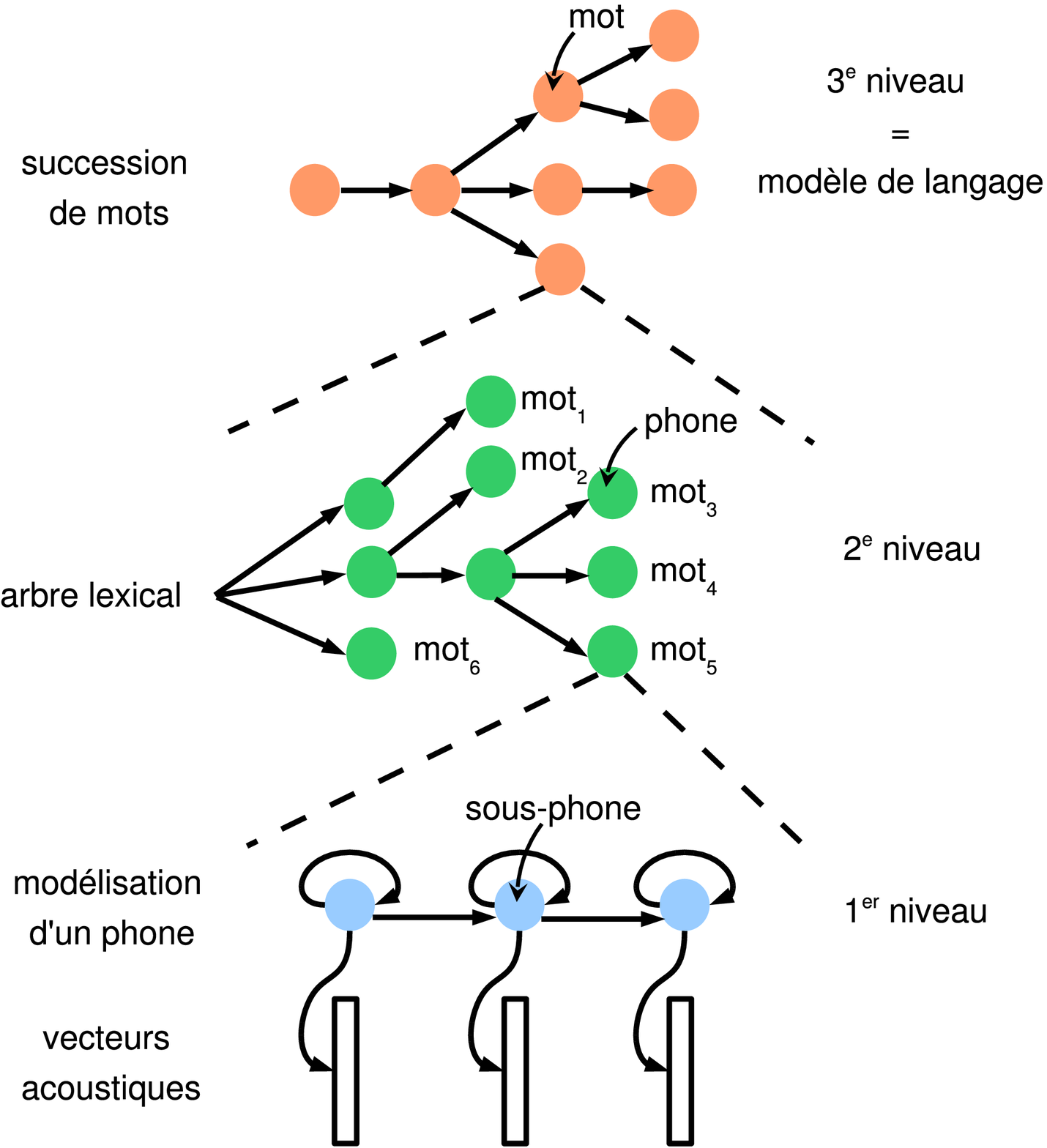}{Niveaux de modélisation du modèle
  acoustique}{figMA}

Ils cherchent dans un premier temps à reconnaître les types de son,
autrement dit à identifier des \emph{phones}\footnote{Sons prononcés
par un locuteur et définis par des caractéristiques précises.}. Pour
ce faire, ils modélisent un phone par un HMM, généralement à trois
états représentant ses début, milieu et fin. La variable cachée est
alors un \emph{sous-phone} et les observations sont des vecteurs
acoustiques, \textit{i.e.}, des vecteurs continus. Pour calculer les
probabilités d'observation dans chaque état, deux approches sont
souvent envisagées, l'une basée sur la représentation des densités de
probabilité par des gaussiennes et l'autre reposant sur des réseaux de
neurones. Ces différentes méthodes établissent des hypothèses sur la
probabilité des phones prononcés. Or, l'objectif des MA est de
déterminer une succession de mots. Les MA utilisent à cette fin un
dictionnaire de prononciations, qui effectue la correspondance entre
un mot et ses prononciations. Comme un mot est susceptible d'être
prononcé de différentes manières, selon son prédécesseur et son
successeur, ou tout simplement selon les habitudes du locuteur, il
peut y avoir plusieurs entrées dans ce lexique pour un même mot. Les
indications sont données au moyen des \emph{phonèmes}\footnote{Unité
linguistique associée à un type de prononciation d'une langue
donnée. Le phonème final /p/ pourra par exemple être prononcé en
français par le phone [b] dans l'expression \og \textit{grippe
aviaire}\fg\ et par le phone [p] dans \og \textit{grippe du
poulet}\fg.} caractéristiques de la prononciation. Sur la
figure~\ref{figExtraitDicoPrononciations}, les phonèmes sont
transcrits dans le système de représentation SAMPA.

\begin{figure}
\begin{center}
\begin{tabular}{l l}
\hline
adorateurs & a d O R a t 9 R z \\
adorateurs & a d O R a t 9 R \\
adoration & a d O R a s j o\string~ \\
adore &  a d O R @ \\
adore & a d O R \\
\hline
\end{tabular}
\end{center}
\caption{Extrait d'un dictionnaire de prononciations}
\label{figExtraitDicoPrononciations}
\end{figure}

Le deuxième niveau de HMM modélise les mots à partir des HMM
représentant des phones et du lexique de prononciations. Il se
présente sous la forme d'un arbre lexical contenant initialement tous
les mots du vocabulaire, progressivement élagué au fur et à mesure que
sont reconnus des phones. Puisque les HMM de premier niveau modélisent
des phones et non des phonèmes, les phonèmes disponibles dans le
dictionnaire de prononciations sont convertis en phones afin de
reconnaître des mots. Des règles de transformation dépendant du
contexte d'apparition du phonème sont alors utilisées.

Le troisième niveau modélise enfin la succession des mots $W$ au sein
d'un groupe de souffle et peut alors incorporer les connaissances
apportées par le ML sur $W$. Pour établir ce HMM équivalent à un
graphe de mots, le HMM correspondant à l'arbre lexical est dupliqué à
chaque fois que le MA effectue l'hypothèse qu'un nouveau mot a été
reconnu \cite{Ortmanns97}.

Le fonctionnement du MA que nous venons de décrire se heurte à un
problème majeur : l'espace de recherche du HMM de plus haut niveau
devient fréquemment considérable, surtout si le vocabulaire est
important et si le groupe de souffle à analyser contient plusieurs
mots. Des algorithmes issus de la programmation dynamique permettent
de calculer efficacement les probabilités ; il s'agit principalement
de l'algorithme de Viterbi et le décodage par pile, appelé aussi
décodage A*. De plus, il est fait recours très régulièrement à
l'élagage pour ne conserver que les hypothèses susceptibles d'être les
plus intéressantes \cite{Deshmukh99}.

Le rôle du MA consiste ainsi à aligner le signal sonore avec des
hypothèses de mots en utilisant uniquement des indices d'ordre
acoustique. Il inclut dans son dernier niveau de modélisation les
informations sur les mots apportées par le ML.

\subsubsection{Modèle de langage}
\label{subSecML}

Le ML a pour objectif de trouver les séquences de mots les plus
probables, autrement dit celles qui maximisent la valeur $P(W)$ de
l'équation~(\ref{eqWbaySimpl}). Si l'on se réfère au HMM de plus haut
niveau du MA (Fig.~\ref{figMA}), les valeurs $P(W)$ correspondent aux
probabilités de succession de mots.

\paragraph{Fonctionnement d'un modèle de langage\\} 
\label{subSubSecFoncML} 

En posant $W = w_1^n = w_1 \ldots w_n$, où $w_i$ est le mot de rang $i$
de la séquence $W$, la probabilité $P(W)$ se décompose de la manière
suivante :
\begin{equation} 
P(w_1^n) = P(w_1) \prod_{i=2}^n P(w_i|w_1 \ldots w_{i-1})
\end{equation}

L'évaluation de $P(W)$ se ramène alors au calcul des valeurs $P(w_i)$
et $P(w_i|w_1^{i-1})$ qui s'obtiennent respectivement à l'aide des
égalités :
\begin{equation}
P(w_i) = \frac{C(w_i)}{\sum_{w \in \mathcal{V}} C(w)}
\end{equation}
\begin{equation}
\label{eqCoomptageNGrammes}
P(w_i|w_1^{i-1}) = \frac{C(w_1^i)}{\sum_{w_i} C(w_1^i)}
\end{equation}
où $\mathcal{V}$ est le vocabulaire utilisé par le système de RAP, et
$C(w_i)$ et $C(w_1^i)$ représentent les nombres d'occurrences
respectifs du mot $w_i$ et de la séquence de mots $w_1^i$ dans le
corpus d'apprentissage.

Malheureusement, pour prédire la suite de mots $w_1^n$, le nombre des
paramètres $P(w_i)$ et $P(w_i|w_1^{i-1})$ du ML à estimer augmente de
manière exponentielle avec $n$. Dans le but de réduire ce nombre,
$P(w_i|w_1^{i-1})$ est modélisé par un \emph{modèle N-gramme},
\textit{i.e.}, une chaîne de Markov d'ordre $N-1$ (avec $N~>~1$), à
l'aide de l'équation suivante :
\begin{equation} 
P(w_i|w_1^{i-1}) \approx  P(w_i|w_{i-N+1}^{i-1})
\end{equation}

\noindent Cette équation indique que chaque mot $w_i$ peut être prédit
à partir des $N-1$ mots précédents. Pour $N = 2, 3$ ou $4$, on parle
respectivement de modèle \emph{bigramme}, \emph{trigramme} ou
\emph{quadrigramme}. Pour $N = 1$, le modèle est dit \emph{unigramme}
et revient à estimer $P(w_{i})$. Généralement, ce sont les modèles
bigrammes, trigrammes et quadrigrammes qui sont utilisés dans les ML
des systèmes de RAP.

Cette approche rencontre des limites du fait de l'absence de
nombreuses séquences de mots de taille $N$, appelées des
\emph{N-grammes}, dans les corpus d'apprentissage, bien que ceux-ci
puissent être de taille conséquente. En effet, même en ayant une
valeur de $N$ réduite, de nombreux mots seront rares, voire absents du
corpus. On dit à ce sujet que les mots suivent une loi de Zipf,
stipulant que la fréquence d'apparition d'un mot décroît rapidement
avec son rang d'apparition. Pour pallier cette difficulté, il est fait
appel à des méthodes statistiques de \emph{lissage}.

Un premier procédé de lissage, connu sous le nom de
\emph{discounting}, consiste à retrancher au comptage des N-grammes
une certaine valeur qui sera en suite redistribuée vers le comptage
des N-grammes absents du corpus d'apprentissage. Il existe de
nombreuses méthodes de \textit{discounting}.  Un procédé très simple
consiste par exemple à ajouter un à l'ensemble des comptages, y
compris ceux associés aux N-grammes absents.

Un autre moyen d'effectuer le lissage est d'utiliser les fréquences
d'apparition des N-grammes d'ordres inférieurs si le N-gramme étudié
est peu présent dans le dictionnaire. On distingue alors la technique
de l'\emph{interpolation linéaire} de celle du \emph{repli} (appelée
aussi \emph{backoff}).

L'interpolation linéaire consiste à évaluer la probabilité
$\hat{P}(w_i|w_{i-N+1}^{i-1})$ en faisant une combinaison linéaire des
probabilités calculées pour les N-grammes d'ordre inférieur. Dans le
cas d'un modèle trigramme par exemple, le calcul s'effectue à partir
des probabilités unigramme, bigramme et trigramme de la manière
suivante~:
\begin{equation}
\label{eqInterpolLineaire}
\hat{P}(w_i|w_{i-2}w_{i-1}) = \lambda_1 P(w_i|w_{i-2}w_{i-1}) +
\lambda_2 P(w_i|w_{i-1}) + \lambda_3 P(w_i)
\end{equation}
avec :
\begin{equation}
\sum_{k=1}^3 \lambda_k = 1
\end{equation}
de manière à ce que $\hat{P}$ demeure une probabilité. $P$ est
calculée au moyen de l'équation~(\ref{eqCoomptageNGrammes}) utilisant
les comptages. Les valeurs $\lambda_k$ sont estimées de façon à
maximiser la vraisemblance de $\hat{P}$ sur un corpus de
test, différent du corpus d'apprentissage.

Le repli effectue lui aussi une combinaison linéaire avec les
probabilités de N-grammes d'ordre inférieur mais, à la différence de
l'interpolation linéaire, le recours aux N-grammes de tailles plus
réduites n'est pas systématique \cite{Katz87}. Le calcul des
probabilités s'exprime de la manière suivante~:
\begin{equation} 
\hat{P}(w_i|w_{i-N+1}^{i-1}) = \left\{ \begin{array}{ll}
\tilde{P}(w_i|w_{i-N+1}^{i-1}) & \textrm{si } C(w_{i-N+1}^i) > 0\\
\alpha(w_{i-N+1}^{i-1}) \times \hat{P}(w_i|w_{i-N+2}^{i-1}) &
\textrm{sinon}
\end{array} \right.
\end{equation}

De même que pour les $\lambda_i$ de
l'équation~(\ref{eqInterpolLineaire}), les coefficients $\alpha$ sont
calculés pour que $\hat{P}$ soit une probabilité. Le symbole
\string~ sur le $P$ sert à indiquer que $\tilde{P}$ est souvent obtenu
à partir d'un procédé de \textit{discounting}.

Au-delà de ces techniques de lissage, il existe de nombreuses
variantes basées sur leurs principes, parmi lesquelles figure la
version modifiée du lissage de Kneser-Ney qui, d'après des études
empiriques, donne de bons résultats \cite{Chen98}.

Les modèles N-grammes, qu'ils utilisent ou non des techniques de
lissage, sont des méthodes statistiques dont le nombre de paramètres à
estimer est très grand, ce qui nécessite de disposer d'une grande
quantité de données d'apprentissage. Depuis les débuts d'utilisation
des N-grammes, de nombreux textes de natures différentes ont été
collectés, ce qui a largement profité à l'amélioration de ces
techniques. Toutefois, cette évolution suit depuis quelque temps une
asymptote. Selon une estimation informelle d'IBM, les performances des
modèles bigrammes n'enregistrent plus de gain important au-delà de
quelques centaines de millions de mots, tandis que les modèles
trigrammes semblent saturer à partir de quelques milliards de
mots. Or, dans plusieurs domaines d'application de systèmes de RAP, de
tels volumes de données ont déjà été collectés \cite{Rosenfeld00}.

Pour tenter d'améliorer les performances des ML, des évolutions du
mode de calcul des probabilités par les modèles N-grammes ont été
envisagées. Le principal reproche qui est fait aux modèles N-grammes
est l'hypothèse qui a permis l'élaboration des premiers ML performants
et utilisables, à savoir la prise en compte d'un historique $h_i$ de
taille limitée. Le calcul des probabilités est en effet réalisé au
moyen d'une égalité du type :
\begin{equation} 
P(w_1^n) = P(w_1) \prod_{i=2}^n P(w_i|h_i)
\end{equation}
Plusieurs études ont été menées pour examiner un historique plus
étendu que $w_{i-N+1}^{i-1}$ \cite{Goodman01}.

\paragraph{Modifications de l'historique dans le calcul
  des probabilités \\} 
\label{subSubSecModifHisto}

La prise en compte d'un historique de très grande taille est
susceptible d'améliorer les performances du ML. Toutefois, les
modifications envisagées doivent tenir compte du fait que parmi
l'ensemble des historiques possibles, beaucoup deviennent rares voire
inexistants dans le corpus d'apprentissage quand on augmente le nombre
de mots pris en compte.

Un moyen très simple pour étendre les N-grammes est d'utiliser
l'interpolation linéaire (\textit{cf.}
équation~(\ref{eqInterpolLineaire})) pour combiner des modèles
N-grammes d'ordres différents. Ceci permet d'avoir un historique
étendu mais aussi de faire face à la rareté des données dans le corpus
d'apprentissage. Ces modèles sont nommés modèles \emph{polygrammes}
\cite{Kuhn94, Gallwitz96}. Cette méthode présente deux
inconvénients. Elle nécessite tout d'abord l'évaluation de nombreux
paramètres puisque les calculs de probabilité sont effectués à partir
de plusieurs tailles d'historique. De plus, il n'est pas possible
d'augmenter indéfiniment la taille de l'historique, même en utilisant
des méthodes de lissage perfectionnées. On considère souvent à ce
sujet que les modèles N-grammes d'ordre supérieur à 5 n'apportent pas
de gain par rapport aux modèles d'ordres inférieurs \cite{Goodman01}.

Une autre variation des modèles N-grammes consiste à ignorer certaines
positions dans l'historique ; il s'agit alors de modèles
\emph{skipping}. L'interpolation linéaire de modèles 5-grammes
\textit{skipping} peut ainsi avoir la forme :
\begin{equation}
\lambda_1 P(w_i|w_{i-4}w_{i-3}w_{i-2}w_{i-1}) + \lambda_2
P(w_i|w_{i-4}w_{i-3}w_{i-1}) + \lambda_3
P(w_i|w_{i-4}w_{i-2}w_{i-1})
\end{equation}
Ces types de modèles permettent de prendre en compte des historiques
qui sont proches mais non strictement identiques au contexte
courant. Cette propriété est particulièrement importante quand on
augmente la taille des N-grammes car il devient alors de plus en
plus rare de trouver deux historiques identiques \cite{Goodman01}. 

Les modèles dits \emph{permugrammes} adoptent une approche similaire
en ce qui concerne les historiques, en effectuant une permutation des
mots pris dans le contexte. Ils redéfinissent le calcul des
probabilités par l'équation :
\begin{equation}
P(W) = P(w_{\pi(1)}) \prod_{i=2}^{n} P(w_{\pi(i)}|w_{\pi(i-N+1)}
\ldots w_{\pi(i-1)})
\end{equation}
où $\pi$ est une permutation qui réordonne la succession des
mots. Cette équation conduit ainsi à des historiques de la forme $h_i
= w_{i-3},w_{i+1}$ \cite{Schukat95}. L'utilisation de ces modèles peut
parfois apporter une légère amélioration des performances des
N-grammes. Ils augmentent toutefois de manière importante l'espace
de recherche ainsi que le nombre de paramètres à calculer.

D'autres études ont considéré l'historique non pas comme une
succession de mots mais comme une suite de groupes de mots. Cette
approche s'appuie sur une segmentation en séquences de mots, en fixant
une taille maximum $M$ pour ces séquences. Le calcul de la probabilité
$P(w_1w_2w_3)$ devient par exemple, avec $M = 3$ et en prenant un
historique de longueur maximale 2 :
\begin{equation}
P(w_1w_2w_3) = max \left\{ \begin{array}{l} 
P([w_1w_2w_3]) \\
P([w_1]) P([w_2w_3]|[w_1]) \\
P([w_1w_2]) P([w_3]|[w_1w_2]) \\
P([w_1]) P([w_2]|[w_1]) P([w_3]|[w_1][w_2])
\end{array} \right\}
\end{equation}
Il existe une variante de ce calcul, remplaçant $max$ par la somme de
toutes les segmentations possibles. Ces modèles, qualifiés de
\emph{multigrammes}, n'ont pas permis pour l'instant d'améliorer
véritablement les performances par rapport aux ML classiques
\cite{Bimbot95, Deligne95, Alain05}.

Un autre type d'extension des N-grammes repose sur un historique à
longueur variable, ce qui permet de faire des distinctions
supplémentaires dans certains cas ambigus, tout en conservant un
nombre raisonnable de paramètres à estimer. Les N-grammes pris en
compte dans le calcul des probabilités sont alors présentés sous la
forme d'un arbre (Fig.~\ref{figVarigrammes}). Dans le cas des modèles
N-grammes classiques, où la taille de l'historique est fixe, cet arbre
a une profondeur fixe égale à $N - 1$. Dans le cas des modèles
\emph{varigrammes} au contraire, la profondeur varie selon les
branches puisque des n\oe uds proches dans l'arbre et associés à des
probabilités conditionnelles similaires sont fusionnés \cite{Siu00,
Kneser96, Niesler96b}.
\FigurePS{!htbp}{10cm}{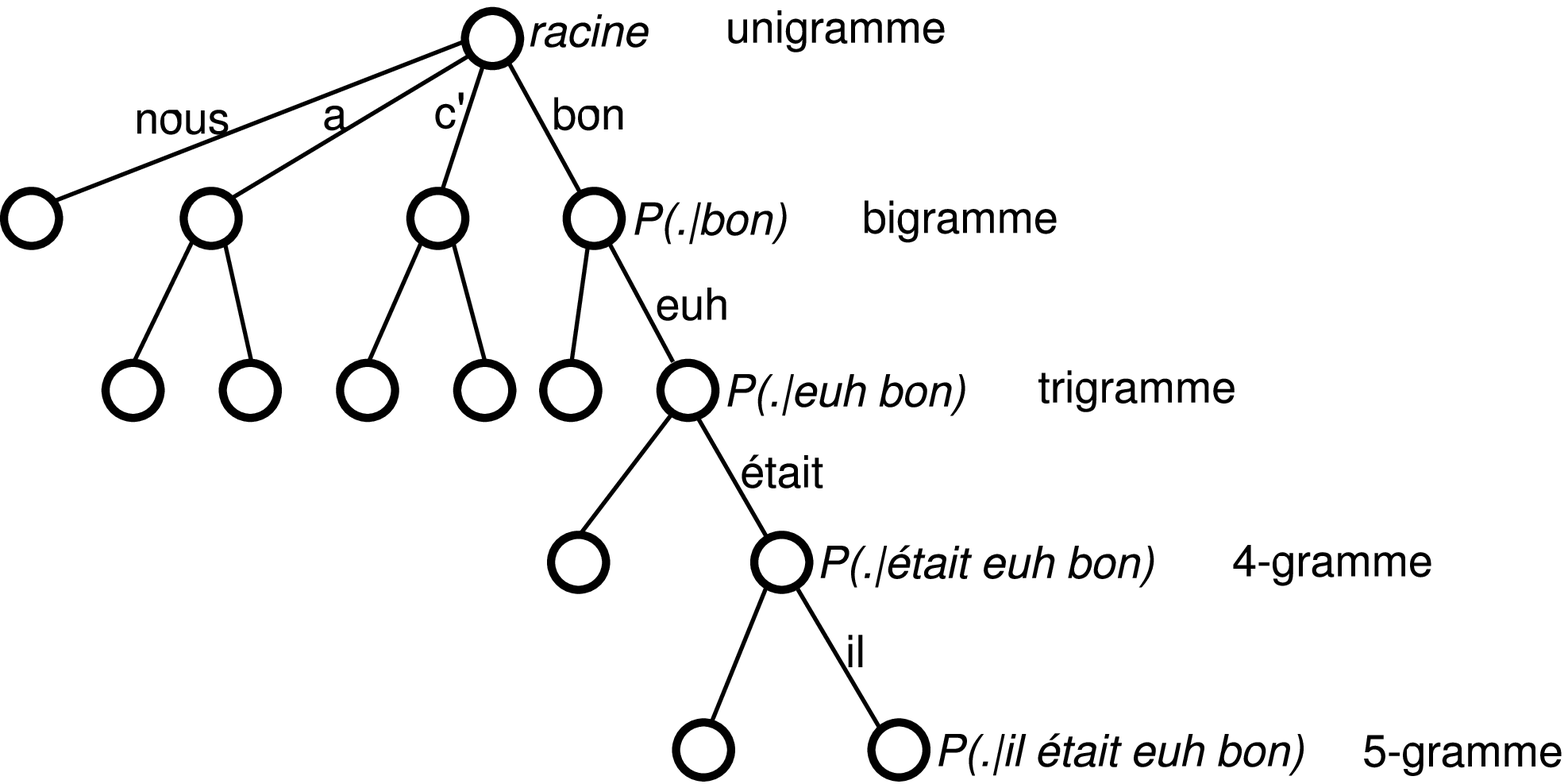}{Représentation sous forme
d'arbre d'un modèle varigramme}{figVarigrammes} La conception des
modèles varigrammes conduit à une diminution de la taille des ML, sans
modifier leurs performances. Les modèles dits \emph{x-grammes}
poursuivent également le même objectif, \textit{i.e.}, la fusion des
historiques très proches. Leur particularité est de modéliser les ML
par des automates à états finis.

Enfin, on peut citer l'utilisation des modèles à base de
\emph{cache}. Ces modèles conduisent à l'utilisation d'historiques de
l'ordre de plusieurs centaines de mots, \textit{i.e.}, de taille
beaucoup plus importante que les modèles N-grammes. Ils sont basés
sur le principe que si un locuteur utilise un mot, la probabilité
qu'il l'utilisera à nouveau dans un futur très proche augmente
considérablement. Ils redéfinissent alors le calcul des probabilités
$P(W)$ par l'équation :
\begin{equation}
P(W) = \tilde{P}(w_1) \prod_{i=2}^{n}
\tilde{P}(w_i|w_{i-N+1} \ldots w_{i-1})
\end{equation}
où les probabilités $\tilde{P}$ sont évaluées à partir des comptages
des mots dans le cache et non plus dans un corpus d'apprentissage
comme cela est le cas pour les modèles N-grammes. Les modèles à base
de cache sont systématiquement combinés avec des modèles
N-grammes. Ils peuvent contribuer à améliorer les performances des ML,
mais ce, au détriment de la vitesse de calcul des probabilités lors de
la transcription \cite{Goodman01, Kuhn90, Clarkson97}.

On le voit donc, de nombreuses tentatives ont été faites pour essayer
de saisir les dé\-pen\-dances à longue distance, \textit{i.e.}, les
relations entre les mots qui sont séparés par plus de $N-1$ mots. Les
techniques qui viennent d'être présentées permettent souvent
d'améliorer légèrement les performances par rapport à des modèles
trigrammes. Toutefois, les progrès sont bien souvent peu
significatifs, alors que leur utilisation engendre un accroissement de
la complexité des calculs lors de la reconnaissance ou une
augmentation importante du nombre de paramètres à estimer
\cite{Goodman01}.


Cette section a présenté les principes des ML, la dernière étape
intervenant dans l'association du signal acoustique à une succession
de mots. Il nous reste à voir sous quelles formes se présentent les
sorties du processus de transcription de la parole.

\subsection{Sorties des systèmes de transcription}
\label{subSecSortiesSTP}

Comme dit précédemment, le rôle d'un système de RAP est de produire
une transcription d'un signal sonore. Le résultat pourra donc être
naturellement un texte. Toutefois, selon le cadre d'utilisation d'un
tel système, il existe d'autres types de sortie envisageables.

\paragraph{Texte brut} 
Lorsqu'un système de RAP produit un texte, celui-ci correspond à la
succession de mots qui a obtenu la plus haute probabilité de la part
du système. Ce texte est organisé sous la forme de groupes de souffle,
le décodage de la parole se basant sur la détection de pauses
silencieuses. Cette forme purement textuelle laisse envisager la
possibilité d'appliquer des techniques de TAL pour améliorer les
résultats produits.

Dans le cas où l'on dispose d'une transcription de référence, obtenue
généralement manuellement, celle-ci peut être alignée avec le texte
produit par le système de RAP (Fig.~\ref{figAlignementTrans}) pour
effectuer des calculs de performance (\textit{cf.}
section~\ref{subSecEval}). L'alignement est obtenu en faisant
correspondre un maximum de mots des deux transcriptions au moyen d'un
algorithme de type Viterbi. Trois types d'erreur sont alors distingués
:
\begin{itemize}
\item les insertions, repérées par des I, correspondant à un ajout
  d'un mot de la transcription automatique par rapport à la
  transcription de référence,
\item les suppressions, marquées par des D (pour \textit{deletion}),
  associées à un mot manquant dans la transcription automatique,
\item les substitutions, représentées par des S, indiquant un
  remplacement d'un mot de la transcription de référence par un autre
  mot présent dans la transcription automatique.
\end{itemize}

\begin{figure}
\begin{center}
\begin{tabular}{l}
\hline
\verb+REF: il AURA ALORS  face à lui une fronde syndicale *** UNIE+\\
\verb+HYP: il **** VALEUR face à lui une fronde syndicale EST PUNI+\\
\\
\verb+         D    S                                      I   S  +\\
\hline
\end{tabular}
\end{center}
\caption{Alignement de la transcription automatique (HYP) et de la
  transcription de référence (REF)}
\label{figAlignementTrans}
\end{figure}

Lorsque des traitements doivent être opérés après le processus de
transcription, on choisit parfois de conserver non pas la meilleure
hypothèse, mais plutôt la liste des $N$ meilleures hypothèses. Ceci
laisse la possibilité de réordonner les hypothèses, si on dispose de
connaissances supplémentaires par la suite.

\paragraph{Graphes de mots}
La liste des $N$ meilleures hypothèses présente de nombreuses
redondances, les éléments de la liste différant bien souvent par un
seul mot. Une sortie plus compacte qui lui est souvent préférée est le
graphe de mots (Fig.~\ref{figGrapheMots}). Ce graphe peut être une
variante du HMM de plus haut niveau du MA (\textit{cf.}
section~\ref{subSecMA}). Les arcs sont valués par les probabilités
établies par le MA et le ML, et représentent des hypothèses ou sur les
mots prononcés ou sur la présence de pauses silencieuses (notées par
\og \textit{sil} \fg\ sur la figure~~\ref{figGrapheMots}). Les n\oe
uds sont quant à eux associés aux instants possibles où un mot se
termine et un autre débute. Puisque les informations concernant
l'instant de prononciation des mots sont peu employées, le graphe est
souvent compacté en supprimant les arcs et les n\oe uds qui
représentent les mêmes hypothèses de succession de mots
\cite{Deshmukh99}. De surcroît, la taille du graphe de mots construit
pouvant être considérable, il peut être utile de l'élaguer en fixant
par exemple un nombre maximum de n\oe uds. Un autre critère possible
d'élagage est le nombre $k$ d'hypothèses retenues sur les successions
de mots.

\FigurePS{!htbp}{10cm}{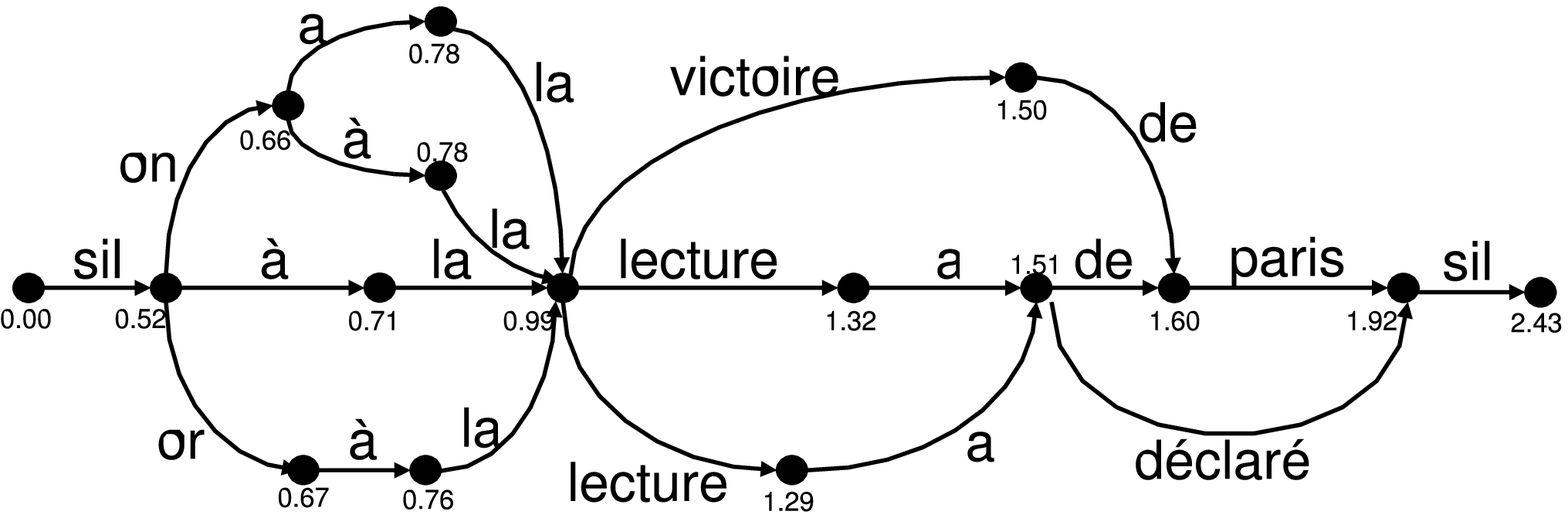}{Exemple de graphe de mots
  (non valué)}{figGrapheMots}

\paragraph{Réseaux de confusion}
Cet autre type de sortie correspond à un compactage des graphes de
mots, en conservant davantage d'informations que les $k$ meilleures
hypothèses. La construction des réseaux de confusion, également
appelés \og saussisses\fg\ (Fig.~\ref{figReseauConfusion}), consiste à
aligner les hypothèses de succession de mots, un peu comme on le
ferait si l'on souhaitait aligner la transcription automatique avec la
transcription de référence
(Fig.~\ref{figAlignementTrans}). L'extraction des meilleures
hypothèses se fait ici en tentant de minimiser les erreurs
d'insertion, de suppression ou de substitution, et non pas en
déterminant la succession de mots associée aux plus faibles
probabilités \textit{a posteriori}, comme dans le cas des graphes de
mots. Ceci est une propriété intéressante des réseaux de confusion
dans la mesure où les systèmes de RAP cherchent plutôt à obtenir une
transcription ayant le minimum d'erreurs qu'un texte ayant un bon \og
score\fg\ de probabilité \cite{Mangu00}.

\FigurePS{!htbp}{9.5cm}{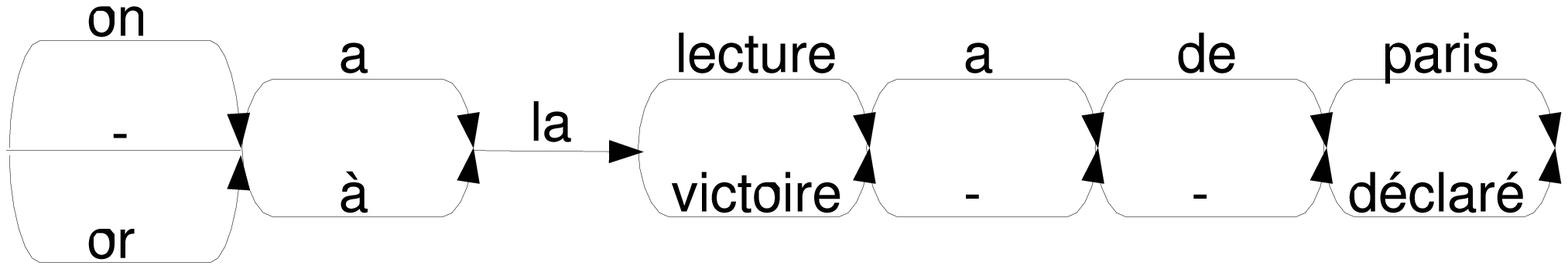}{Exemple de réseau de
confusion}{figReseauConfusion}

À l'issue de la présentation des différents types de sorties de la
transcription, la section suivante s'intéresse à la description de
méthodes de mesure de la qualité des résultats produits.

\subsection{Méthodes d'évaluation}
\label{subSecEval}

La mesure qui est communément employée pour mesurer la qualité d'une
transcription est le \emph{taux d'erreur sur les mots} défini par :
\begin{equation}
\textrm{taux d'erreur }= \frac{\textrm{nb d'insertions} + \textrm{nb
    de suppressions} + \textrm{nb de substitutions}}{\textrm{nb de
    mots dans la transcription de référence}}
\end{equation}

Dans ce calcul revenant à déterminer la distance d'édition entre la
transcription automatique et la transcription de référence, les coûts
d'insertion, de suppression ou de substitution ont une valeur fixe qui
ne dépend pas du nombre de caractères erronés dans la transcription
automatique. La transcription étant envisagée comme une succession de
mots-formes (\textit{cf.} section~\ref{subSecDiffTrans}), ceci
implique qu'une simple erreur d'accord en genre et en nombre aura le
même coût qu'une substitution par un mot qui n'a aucun rapport sur le
plan acoustique ou sémantique avec le mot correct. On voit ainsi que
le taux d'erreur, même s'il est très utilisé, n'indique pas
directement si le résultat produit est correct du point de vue des
informations transmises. Bien entendu, s'il devient faible, la
probabilité qu'il s'est produit une erreur sur les mots informatifs
diminue mais ce n'est pas systématique.

Le taux d'erreur sur les mots permet d'évaluer globalement le système
de RAP. Il existe d'autres mesures, telles que l'\emph{entropie
croisée} ou la \emph{perplexité}, pour mesurer la qualité du ML seul
\cite{Jurafsky00}. Pour définir la première, précisons pour débuter la
notion d'\emph{entropie}.

L'entropie, notée $H$, est une mesure d'information qui se calcule
pour une séquence de mots $w_1^n$ de la manière suivante :
\begin{equation}
H(w_1^n) = - \sum_{w_1^n \in \mathcal{L}} P(w_1^n) \log_2P(w_1^n)
\end{equation}

Pour un langage $\mathcal{L}$, elle s'obtient par :
\begin{eqnarray}
H(\mathcal{L}) & = & \lim_{n \to \infty} \frac{1}{n} H(w_1^n)
\\
& = & \lim_{n \to \infty} - \frac{1}{n} \sum_{w_1^n \in \mathcal{L}}
P(w_1^n) \log_2 P(w_1^n)
\end{eqnarray}

La mesure d'\emph{entropie croisée}, directement dérivée de l'entropie
et souvent abusivement appelée de manière identique, permet d'évaluer
la performance d'un ML. Si l'estimation de la probabilité par un ML
est notée $\hat{P}$ et si $P$ est la distribution réelle de
probabilité, l'entropie croisée de ce modèle se calcule par :
\begin{equation}
H(P,\hat{P}) = \lim_{n \to \infty} - \frac{1}{n} \sum_{w_1^n \in
\mathcal{L}} P(w_1^n) \log_2 \hat{P}(w_1^n)
\end{equation}

En supposant que le langage $\mathcal{L}$ possède de bonnes propriétés,
le théorème de Shannon-McMillan-Breiman permet d'écrire :
\begin{equation}
H(P,\hat{P}) = \lim_{n \to \infty} - \frac{1}{n} \log_2 \hat{P}(w_1^n)
\end{equation}

En pratique, pour comparer la performance de deux ML, on compare les
entropies croisées calculées pour ces deux modèles sur un corpus de
test $T$ aussi grand que possible, de manière à approximer au mieux la
limite vers l'infini :
\begin{equation}
H_T(P,\hat{P}) = - \frac{1}{t} \log_2 \hat{P}(T)
\end{equation}
où $t$ représente le nombre de mots du corpus de test. Cette équation
est très proche de l'objectif de la modélisation statistique du ML
(\textit{cf.} équation~(\ref{eqWbaySimpl})), \textit{i.e.}, trouver
$\hat{P}$ maximisant $\hat{P}(\mathcal{L})$. Plus l'entropie croisée
est faible, meilleur est ainsi le modèle.

Dans la plupart des études évaluant les ML, la \emph{perplexité} $PP$
est souvent préférée à l'entropie croisée. Elle se calcule sur un
ensemble de test $T$ à partir de :
\begin{equation}
PP_T(\hat{P}) = 2^{H_T(P,\hat{P})}
\end{equation}

Bien que très utilisé pour la comparaison des ML, ce critère possède
des limites. Une baisse importante de la perplexité ne correspond pas
forcément à une baisse de même ordre du taux d'erreur sur les mots, si
on intègre le ML dans un système de RAP ; seules des réductions de la
perplexité d'au moins 10-20\,\% semblent être vraiment significatives
\cite{Rosenfeld00}. Son principal inconvénient est qu'il favorise les
modèles accordant une plus grande probabilité aux mots présents dans
le corpus de test, et ignore la manière avec laquelle sont distribuées
les probabilités aux autres mots. Or, ces mots peuvent conduire, lors
du décodage par le système de RAP, à des probabilités plus élevées que
celles obtenues pour les mots corrects \cite{Clarkson99}. Dans
certaines situations, on peut ainsi observer une augmentation du taux
d'erreur en même temps qu'une diminution de la perplexité. Certains
auteurs préconisent donc d'utiliser plutôt l'entropie croisée, qui
semble plus corrélée avec le taux d'erreur \cite{Goodman01}.

\paragraph{}
Cette section~\ref{SecPrincRP} dédiée aux principes de fonctionnement
des systèmes de RAP nous a donné l'occasion d'évoquer les limitations
des ML les plus populaires, \textit{i.e.}, les modèles
N-grammes. Avant de voir en détail comment il est possible d'intégrer
davantage de connaissances linguistiques pour améliorer les
performances de la transcription, la section suivante se focalise sur
les spécificités de la langue parlée.

\newpage

\section{Caractéristiques de la langue parlée}
\label{secCaractLangParle}

Les documents traités par les systèmes de RAP sont par définition
prononcés par des locuteurs et relèvent donc du domaine de la langue
parlée. Ces types de documents ont malheureusement été peu étudiés par
les linguistes et les adeptes du TAL, comparativement aux textes
écrits. Cette constatation trouve sa source dans le mode de pensée de
la culture occidentale, qui établit très souvent que l'écrit,
considéré comme prestigieux, est seul digne d'intérêt. En outre, peu
de corpus oraux existent et ceux qui sont disponibles sont souvent de
taille peu importante. Les plus volumineux d'entre eux pour le
français ne possèdent ainsi que quelques millions d'occurrences. Ceci
a pour résultat que les systèmes de RAP actuels effectuent
l'apprentissage des ML sur un corpus oral de taille réduite,
accompagné d'un corpus écrit beaucoup plus volumineux, au risque de
paramétrer les modèles avec la langue utilisée dans le journal
\textit{Le Monde} pour le français ou celui du \textit{Wall Street
Journal} pour l'anglais \cite{Veronis04}.

Deux questions se posent alors. Peut-on modéliser la langue parlée à
l'aide de corpus de la langue écrite, disponibles en plus grands
volumes que les corpus oraux ?  Quel est le comportement du TAL et du
processus de transcription par rapport aux phénomènes spécifiques de
la langue parlée ? Pour répondre à ces questions, nous précisons tout
d'abord ce qui définit la langue parlée par rapport à la langue
écrite. Nous présentons ensuite les principales caractéristiques de
cette langue parlée en ce qui concerne son vocabulaire et sa syntaxe,
avant de décrire les perturbations provoquées par les phénomènes
d'hésitation très fréquents qu'elle contient. Nous abordons ensuite la
constitution des corpus oraux, en montrant brièvement sous quelles
formes ils se présentent et comment ils peuvent être annotés. Nous
terminons cette section en étudiant le comportement des systèmes de
RAP vis-à-vis des dialogues spontanés, qui sont la forme de l'oral se
distinguant le plus des formes conventionnelles de l'écrit.

\subsection{Langue parlée et langue écrite}
\label{subSecLPetLE}

La \emph{langue parlée} est définie comme étant ce qui est prononcé
par des locuteurs à l'oral et fait donc appel à la voix et à
l'oreille. La \emph{langue écrite} représente quant à elle ce qui
s'écrit et implique donc l'usage de la main et des yeux. La langue
parlée et la langue écrite exploitent ainsi des canaux différents qui
ont des contraintes importantes sur leur mode de production
\cite{Melis00}.

Le canal oral impose une certaine linéarité lors son émission et de
son écoute, même si on peut parfois avoir recours à des dispositifs
d'enregistrement permettant de faire des retours en arrière. L'oral,
produit en continu, exclut en général toute forme de préparation
préalable, de planification. La configuration typique de l'oral est
caractérisée par une réception immédiate par un ou plusieurs
interlocuteurs, qui ont la possibilité de réagir au cours même de la
production. Ceci peut obliger le locuteur à adapter son discours en
fonction de ces réactions.

Le canal écrit au contraire exploite la page et fait intervenir deux
dimensions : la largeur et la hauteur, facilitant ainsi les retours à
des éléments antérieurs. Le texte est généralement produit en différé
et peut donc être formulé et reformulé avant d'être livré aux
destinataires. Les récepteurs ne sont pas au contact du scripteur et
celui-ci se trouve en général contraint de proposer un texte
désambiguïsé, informatif et structuré.

Il existe en outre une opposition entre l'oral et l'écrit au niveau de
la segmentation des productions. La \emph{prosodie}, qui recouvre des
informations sur l'intonation, l'accent, les pauses et même le débit,
joue un rôle important dans la segmentation de la langue orale, sans
toutefois permettre une structuration aussi importante que ce qu'on
peut attendre pour les textes écrits. Il n'existe en effet pas à
l'oral de démarcation aussi nette que celle permise par les signes de
ponctuation ; on ne peut vraiment d'ailleurs y parler de phrase
(\textit{cf.}  section~\ref{subSubSecFormesCorpus}). Les procédés de
mise en page de l'écrit, à savoir les alinéas, les paragraphes ou
encore les sections, autorisent au contraire une structuration de
documents de taille importante.

Les modes de production des langues parlée et écrite, que nous venons
d'évoquer, font que l'écrit est stable et la parole instable. Cette
caractéristique a ainsi conduit la culture occidentale à valoriser
l'écrit sur l'oral, même si finalement l'écrit peut être vu uniquement
comme un simple système de codage de la langue parlée au moyen de
signes visibles. Alors que les lois, les contrats, les textes
religieux fondamentaux sont des documents écrits, l'oral ne nécessite
pas un apprentissage à l'école aussi formel que l'écrit. Avec
l'invention de l'imprimerie, qui a grandement favorisé la diffusion
des textes, le code écrit a pris de plus en plus
d'importance. L'apparition des ordinateurs et d'Internet a fait
croître de manière exponentielle la production et la diffusion de
documents écrits, mais favorise également l'apparition de nouvelles
formes de documents, comme les courriels ou les blogs. Dans le même
temps, les médias ont permis la diffusion massive de discours
télévisés ou bien encore d'émissions radiophoniques. L'opposition
entre langue parlée et langue orale est ainsi loin d'être aussi claire
que celle présentée au début de cette section et on assiste à une
hybridation des codes \cite{Polguere03}. Si l'on peut facilement
opposer les dialogues aux romans littéraires, on ne trouve pas
toujours de contrastes aussi forts entre la langue parlée et la langue
écrite, notamment en ce qui concerne l'interactivité et la préparation
des documents. Le discours télévisé du chef de l'état à la nation sera
par exemple préparé et n'autorise pas l'intervention des
récepteurs. Le courrier électronique autorise quant à lui davantage
d'interactions avec les destinataires et se rapproche plus du dialogue
oral.

La langue qui est employée dans les documents écrits ou oraux peut
subir d'importantes variations selon les régionalismes, les périodes
considérées, les groupes sociaux et culturels ou encore les registres
\cite{Melis00}. Dans le domaine de l'oral, la linguiste
Blanche-Benveniste définit six genres majeurs : les conversations face
à face, les conversations par téléphone, les débats et entrevues en
public, les émissions de radio ou de télévision, les discours non
préparés et enfin les discours préparés \cite{Blanche-Benveniste97}.

Les langues parlée et écrite correspondent donc à deux codes
différents puisqu'elles n'utilisent pas le même canal de
communication. On ne peut ainsi considérer l'oral comme une simple
forme dégradée de l'écrit. De même, bien que certaines écritures
soient phonographiques et codent en partie les sons, comme cela est le
cas du français, l'écrit est loin d'être une seule transcription de
l'oral, ne serait-ce que parce que le scripteur choisit la manière de
présenter les informations \cite{Gardes04}. Les différences entre
l'oral et l'écrit ne sont toutefois pas suffisantes pour qu'il faille
les étudier de manière totalement indépendante
\cite{Blanche-Benveniste90}.

Les sections suivantes se proposent de dégager des phénomènes
revenant plus fréquem\-ment à l'oral qu'à l'écrit.

\subsection{Vocabulaire et syntaxe}
\label{subSecVocaSynt}

Même s'il est incorrect de dire que l'oral est une forme relâchée de
l'écrit, le registre de la langue parlée est généralement moins
soutenu dans la mesure où elle est souvent employée dans des cadres
moins formels. Du point de vue lexical, la langue parlée se fonde sur
un ensemble familier de mots, plus restreint que celui de la langue
écrite. Une étude comparant deux corpus de 20 millions d'occurrences
de mots chacun, le premier correspondant à des journaux écrits et le
second à des émissions journalistiques radiophoniques et télévisées
transcrites manuellement, a ainsi obtenu 127~000 mots distincts pour
l'oral et 215~000 pour l'écrit, et ce, en considérant une même année
de production. Toujours selon cette même étude, l'utilisation des
catégories grammaticales suit des répartitions différentes dans les
deux corpus. Un taux plus important de pronoms et un taux moins
important de noms sont ainsi observés à l'oral \cite{Gendner02}. On y
observe aussi l'introduction de \og petits\fg\ mots, notamment appelés
ligateurs, marqueurs de discours ou encore inserts, comme par exemple
\og \textit{quoi}\fg, \og \textit{bon}\fg, \og \textit{donc}\fg, \og
\textit{enfin}\fg\ ou \og \textit{genre}\fg.

Au niveau de la syntaxe, il existe des tournures propres à l'oral. Si
en français le \og \textit{pas}\fg\ est facultatif à l'écrit dans la
négation \og \textit{ne ... pas}\fg, c'est le \og \textit{ne}\fg\ qui
le devient à l'oral. Le \og \textit{il}\fg\ prend également à l'oral
un caractère facultatif dans les formules \og \textit{il faut} \fg\ et
\og \textit{il y a}\fg. On peut aussi citer l'invariabilité du \og
\textit{c'est}\fg\ dans des expressions telles que \og \textit{c'est
les voisins qui sonnaient}\fg. Ces expressions sont dues au contexte
d'utilisation de la langue orale qui favorise l'interactivité et
impose donc un temps limité de formulation des idées. Pour la même
raison, et même si cela est loin d'être systématique, les accords du
participe passé seront moins respectés à l'oral qu'à l'écrit
\cite{Melis00}.

Il semble également que l'ordre des mots soit un peu plus souple dans
la langue parlée que dans l'écrite, bien que les langues rigides sur
l'ordre des mots comme le français ou l'anglais le soient encore à
l'oral \cite{Antoine01}. Les éléments régis par le verbe peuvent, à
l'oral, se placer avant le groupe sujet-verbe, comme dans l'exemple
\og \textit{les haricots j'aime pas}\fg. On observe aussi davantage de
clivées (\og \textit{c'est le coiffeur qui est content}\fg), de
pseudos-clivées (\og \textit{ce qui l'intéresse c'est le pognon}\fg),
de doubles marquages (\og \textit{moi j'en ai jamais vu en Suisse des
immeubles}\fg) ou encore de dislocations (\og \textit{j'ai choisi la
bleue de robe}\fg) \cite{Blanche-Benveniste90}.

Ces phénomènes ne justifient pas toutefois de proposer une grammaire
spéciale pour la langue parlée. Pour le français, Blanche-Benveniste
indique ainsi que la syntaxe de l'oral ne diffère en rien de celle de
l'écrit, sauf en termes de proportions \cite{Blanche-Benveniste90,
Benzitoun04b}. Bien que cela soit plus rare qu'à l'oral, on observe
également à l'écrit des clivées ou même des erreurs d'accord de
participe passé.

Cette section a montré les principales différences que l'on
pouvait observer entre l'oral et l'écrit du point de vue du lexique et
de la syntaxe. Toutefois, elle a occulté les perturbations de la
syntaxe par les phénomènes d'hésitation, plus nombreux dans la langue
parlée. La section suivante étudie l'influence de ces phénomènes.

\subsection{Phénomènes d'hésitation}
\label{subSecPhHesitation}

Les marques d'hésitation ne sont pas propres à l'oral puisqu'elles se
manifestent également dans les nouvelles formes de communication
écrite comme les courriels, les forums, les chats ou encore les SMS
\cite{Benzitoun04}. Elles sont également présentes dans les brouillons
des textes écrits sous la forme de ratures. Néanmoins, elles sont
avant tout caractéristiques de l'oral spontané. Il est ainsi estimé
qu'elles représentent environ 5\,\% des mots dans les corpus spontanés
et moins de 0,5\,\% des mots dans les corpus de parole lue ou préparée
(\textit{cf.} section~\ref{subSecCorpOraux}). Elles ne peuvent
donc être ignorées lors de la transcription.

Les phénomènes d'hésitation correspondent généralement à un travail de
formulation de la part du locuteur, dans la mesure où le discours est
composé au fur et à mesure de sa production. Ils peuvent cependant
avoir d'autres rôles tels que la manifestation d'un doute, l'envoi
d'un signal au récepteur pour garder la parole ou au contraire la lui
céder, le marquage de frontières dans le discours, ou bien encore
l'expression d'un stress ou d'une décontraction.

Comme les marques d'hésitation introduisent une rupture dans la
continuité du discours, ou plus exactement dans le déroulement
syntagmatique, elles sont également appelées \emph{disfluences}. Il
existe bien d'autres termes utilisés dans la littérature pour les
nommer, certains ne désignant pas exactement les mêmes phénomènes :
\emph{turbulences}, \emph{faux départs}, \emph{lapsus},
\emph{inattendus structurels}, \emph{spontanéités}, \emph{modes de
production de la langue parlée}, \emph{marques de réparation},
\emph{marques du travail de formulation}, \emph{extragrammaticalités},
\textit{etc}. Une telle profusion de termes témoigne de la discordance
entre certains linguistes sur ce qui relève ou non de l'hésitation à
l'oral. Toutefois, depuis quelques années, plusieurs études se sont
intéressées aux phénomènes d'hésitations et une nomenclature commence
à se dégager.

Ces phénomènes peuvent être de natures différentes. Ils incluent par
exemple les \emph{pauses silencieuses}, mais pas toutes, certaines
jouant d'autres rôles que l'hésitation comme la hiérarchisation et la
structuration des constituants, ou encore la mise en valeur
stylistique de certains syntagmes. Les \emph{pauses remplies},
correspondant au \og \textit{um}\fg\ ou au \og \textit{uh}\fg\ en
anglais et au \og \textit{euh}\fg\ en français, ou encore les
\emph{allongements vocaliques} en fin de mots sont des marques
d'hésitation très employées \cite{Henry02, Candea00}. On peut aussi
citer \cite{Shriberg01} :
\begin{itemize}
\item les répétitions : \og \textit{tous les - les jours}\footnote{Le
  \og \textit{-}\fg\ représentant une pause silencieuse.}\fg,
\item les suppressions : \og \textit{c'est - il est arrivé lundi}\fg,
\item les substitutions : \og \textit{tous les jours - toutes les
  semaines}\fg,
\item les insertions : \og \textit{je suis convaincu - je suis
  intimement convaincu}\fg,
\item les erreurs d'articulation : \og \textit{en jouin - en juin}\fg.
\end{itemize}
Ces quatre derniers types de disfluence sont parfois désignés sous
le terme d'autocorrections \cite{Henry02}.
 
Il est possible de faire une autre classification des phénomènes
d'hésitation, en distinguant \cite{Pallaud04} :
\begin{itemize}
\item les bribes, correspondant à une reprise à partir de syntagmes
  inachevés : \og \textit{il a quand même un - une fibre pédagogique}
 \fg,
\item les amorces, associées aux mots inachevés : \og \textit{c'est pas
  malho - c'est pas malhonnête}\fg.
\end{itemize}

Les différents phénomènes d'hésitation sont rarement produits seuls
mais plutôt en combinaison. Il peut ainsi y avoir des répétitions de
fragments de mots (\og \textit{on le re- re- revendique encore une
fois}\fg) \cite{Henry03}. De même, pour assurer un rôle d'hésitation,
les pauses silencieuses sont généralement associées à des allongements
vocaliques ou à des pauses remplies \cite{Campione04}.

La fréquence d'apparition des marques d'hésitation est très variable
selon le locuteur et le contexte d'élocution. Dans les dialogues
homme-machine, on constate généralement moins de disfluences que dans
les dialogues entre deux humains. Il semble également que les hommes
produisent en moyenne davantage de disfluences que les femmes. La
fréquence des marques d'hésitation est de plus dépendante de la
position dans la phrase. Les disfluences ont ainsi tendance à se
produire plutôt en début de phrase, lorsque l'effort de planification
est le plus important. Pour la même raison, leur taux augmente souvent
avec la longueur de la phrase \cite{Shriberg01}. Les phénomènes
d'hésitation affectent enfin différemment les mots. En distinguant
deux types de mots, à savoir les \emph{lexicaux}, représentant ceux
qui ont une charge lexicale pleine\footnote{En général les noms, les
adjectifs, les verbes et les adverbes.} et les \emph{grammaticaux},
participant à la structuration de la langue\footnote{Les pronoms, les
déterminants, les prépositions, les conjonctions et les verbes
auxiliaires.}, on constate que les répétitions concernent plus souvent
les mots grammaticaux et que les amorces affectent plus fréquemment
les lexicaux \cite{Pallaud04, Henry02a}.

En analysant les disfluences, des psycholinguistes ont mis en évidence
plusieurs régions (Fig.~\ref{figDisfluences})\cite{Shriberg94,
Shriberg01} :
\begin{itemize}
\item le \emph{reperandum}, désignant une partie ou la totalité de la
  séquence qui sera abandonnée au profit de la réparation,
\item le \emph{point d'interruption}, marquant une rupture dans la
  fluidité du discours,
\item l'\emph{interregnum}, pouvant être des pauses remplies, mais
  aussi des termes d'édition, comme \og \textit{ben}\fg,
  \og \textit{hein}\fg, \og \textit{tu vois}\fg, \og \textit{disons pour
  simplifier}\fg, \og \textit{je sais pas moi}\fg\ ou encore
  \og \textit{je me rappelle plus du nom}\fg,
\item la \emph{réparation}, représentant la partie corrigée du
  \textit{reparandum}.
\end{itemize}

\FigurePS{!htbp}{7cm}{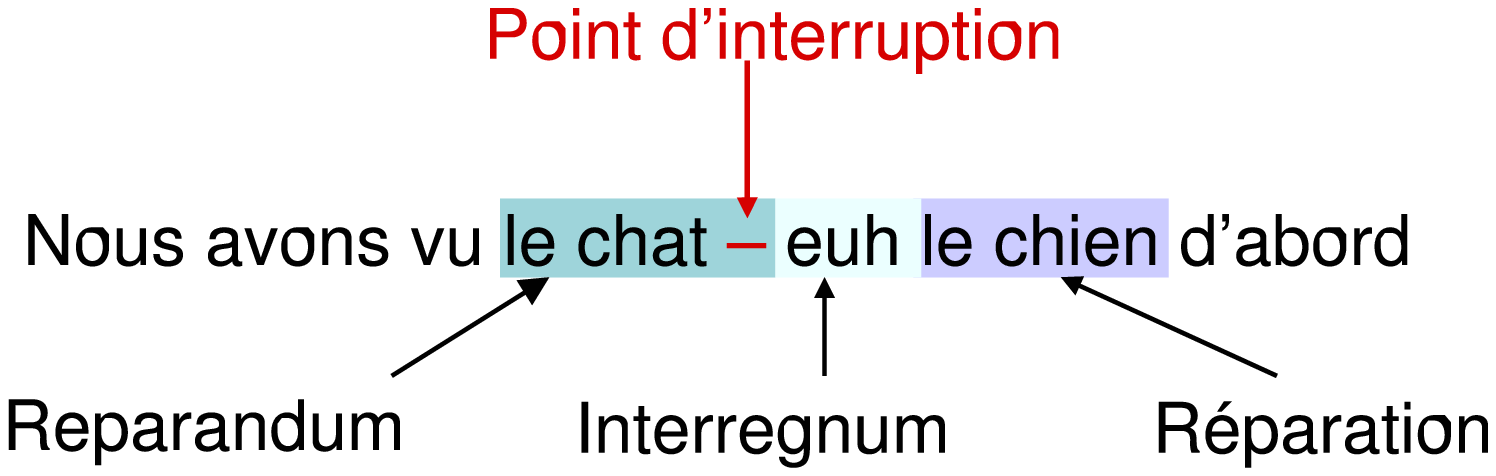}{Régions dans une disfluence
(d'après \cite{Shriberg01})}{figDisfluences}

\noindent La segmentation des disfluences en régions est liée à des
modifications des propriétés acoustiques et phonétiques. On observe
ainsi généralement un allongement des syllabes avant le point
d'interruption, ou au contraire, dans le cas d'une détection d'erreur
de la part du locuteur, un raccourcissement des syllabes. Cette
structuration permettrait de réduire les énoncés à un oral \og propre\fg,
proche de l'écrit, en éliminant le \textit{reparandum} et
l'\textit{interregnum}. Malheureusement, cette approche ne rencontre
pas l'unanimité des linguistes étudiant l'oral.

\subsection{Corpus oraux}
\label{subSecCorpOraux}

On dispose à l'heure actuelle de corpus écrits de taille
volumineuse. Le \textit{British National Corpus} (BNC) contient ainsi
100 millions d'occurrences pour l'anglais, tandis que la base
\textit{Frantext} comporte 210 millions d'occurrences pour le
français.  Les pages disponibles sur Internet, bien que souvent
bruitées, peuvent être également considérées comme une source
gigantesque d'informations \cite{Veronis04}. Il n'existe
malheureusement pas pour l'oral de corpus de taille comparable. Le
plus grand, à savoir la partie orale du BNC, contient 10 millions
d'occurrences en anglais. On peut citer pour le français de Belgique
le corpus Valibel comptant près de 4 millions d'occurrences et pour le
français de l'hexagone, le corpus Corpaix comptant environ 2,5
millions d'occurrences \cite{Campione05}, et le \textit{Corpus de
référence du français parlé} comptant 440~000 occurrences
\cite{delic04}.

Les corpus oraux, constitués au moyen de transcriptions manuelles,
concernent aussi bien de la parole \emph{lue}, \emph{préparée} ou
encore \emph{spontanée}. La parole lue est associée à la lecture à
haute de voix de textes, qui bien souvent appartiennent au domaine de
la langue écrite. La parole spontanée correspond à des situations de
la vie quotidienne où le locuteur parle sans avoir préparé au
préalable ce qu'il allait dire et en s'adaptant constamment à ses
interlocuteurs. La parole préparée est une forme intermédiaire par
rapport aux deux précédentes et est plus délicate à définir. Elle
désigne des prises de parole où le locuteur a réféchi auparavant aux
idées qu'il souhaite transmettre ; il peut même disposer d'un texte
contenant une partie de son discours, qu'il n'ira pas toutefois
jusqu'à lire mot à mot pour pouvoir interagir avec les
interlocuteurs. Les questions posées par un journaliste et les
entretiens donnés par un homme politique sont deux exemples de
production de parole préparée.

Selon les conventions adoptées, les corpus peuvent avoir plusieurs
formats ; ils peuvent contenir des informations précises sur la
prosodie, être ponctués ou bien encore avoir des annotations sur des
événements autres que la prononciation des mots. Accessoirement, ils
sont souvent étiquetés pour faciliter leur exploitation. Ils sont
alors segmentés en mots ou groupes de mots, auxquels sont associées
des informations (\textit{e.g.} syntaxiques ou sémantiques) sur le
rôle de ces constituants dans le groupe de souffle courant. La forme
choisie du corpus dépend de l'utilisation prévue mais aussi des
ressources humaines disponibles, compte tenu de la difficulté et du
temps nécessaire pour effectuer une transcription
manuelle. L'indication de la prosodie dans les corpus oraux, bien que
souhaitable, demande notamment un temps considérable et se limite
souvent à quelques informations. Même si des logiciels tels que
\textit{Transcriber} simplifient l'alignement du son avec la
transcription \cite{Barras01}, il a été estimé qu'environ 40 heures de
travail sont nécessaires pour faire une transcription orthographique
précise d'une heure de parole, et ce, avec de bonnes conditions
d'enregistrement et sans alternance de locuteurs \cite{Campione01b,
Veronis04}. La suite de cette section décrit sous quelles formes
peuvent se présenter les corpus oraux, avant de présenter la manière
dont ils peuvent être étiquetés.

\subsubsection{Formes des corpus oraux}
\label{subSubSecFormesCorpus}

La transposition du signal sonore sous la forme d'un autre canal de
communication, à savoir le texte, nécessite certaines
conventions. Elle se heurte à des difficultés qui sont principalement
: la présence d'événements non linguistiques dans le flux audio,
l'existence de phénomènes propres à la langue parlée et la
segmentation du discours.

La perception des mots à partir du signal sonore peut tout d'abord
être perturbée par des événements non linguistiques qu'il est
intéressant d'annoter dans la mesure où ils peuvent rendre la parole
inaudible. Ces phénomènes peuvent être liés aux conditions
d'enregistrement, tels les parasites ou les bruits de fond, ou bien
encore être produits par les locuteurs, comme les inspirations, les
rires ou les toux.

Des événements propres à la langue parlée nécessitent également des
conventions. Les chevauchements entre les discours prononcés en même
temps par plusieurs locuteurs sont ainsi problématiques, dans la
mesure où il faut indiquer la simultanéité dans le texte, qui a un
format séquentiel. Les phénomènes d'hésitation nécessitent quant à eux
une attention particulière de la part du transcripteur car ils sont
naturellement ignorés lors de la compréhension des dialogues. Ils
présentent pourtant un intérêt pour les linguistes, qui en les
observant peuvent voir la production du langage en train de se
réaliser. Il faut donc adopter des règles pour indiquer les pauses
silencieuses ou encore les amorces. Il existe aussi plusieurs types de
prononciation pour certains mots, ce qu'il est utile d'indiquer. Ceci
peut conduire à faire une transcription \emph{phonétique} plutôt
qu'une transcription plus classique, dite \emph{orthographique}, sous
forme de mots. On choisit généralement d'avoir une transcription
orthographique, avec une convention pour distinguer les prononciations
de mots problématiques, tels que les sigles.

Une des difficultés majeures de la constitution des corpus oraux
réside dans leur segmentation. Pour faciliter leur lisibilité, ils ne
peuvent en effet se présenter sous la forme d'une suite ininterrompue
de mots. Un premier niveau de segmentation consiste à distinguer les
\emph{tours de parole}, \textit{i.e.}, les suites de mots prononcées
par un locuteur donné avant qu'il ne cède la parole. Il n'existe pas
d'unité qui soit aussi nettement définie que la phrase pour la langue
écrite. L'unité de la langue parlée, que l'on qualifie
d'\emph{énoncé}, peut ainsi représenter l'ensemble du tour de
parole. Certains linguistes et psycholinguistes caractérisent aussi un
énoncé comme étant délimité par la prosodie. D'autres le définissent
comme une succession de mots représentant une idée cohérente ; dans
les directives d'annotation \textit{metadata} adoptées dans les
évaluations \textit{Rich Transcription} conduites par le NIST, on
parle ainsi de \textit{SU}, associée à plusieurs significations :
\textit{Sentential Unit}, \textit{Syntactic Unit}, \textit{Semantic
Unit} ou \textit{Slash Unit} \cite{Strassel03}.

Malgré la subjectivité que cela entraîne de la part du transcripteur,
certaines recommandations de transcription demandent d'indiquer les
signes de ponctuation dans les corpus oraux, ce qui les rapproche de
textes plus conventionnels.  La prosodie et les pauses silencieuses
apportent alors des informations précieuses quant à la ponctuation. La
figure~\ref{figCorpTransc} montre ainsi un extrait de corpus transcrit
avec des ponctuations.
\begin{figure}
\begin{center}
\begin{tabularx}{\textwidth}{X}
\hline 
\og \textit{Le gouvernement américain va présenter, [i] euh, au conseil
de sécurité de l' !ONU, [i] un projet d(e) résolution, [mic] un texte
qui permettrait, [i] la levée d(e) l'embargo Noëlle
\string^\string^Véli.}\fg \\
\\
où \og [\textit{i}]\fg\ marque une inspiration, \og \textit{!ONU}\fg\ 
représente un acronyme, \og \textit{d(e)}\fg\ indique une absence de
prononciation du \og \textit{e}\fg, \og [\textit{mic}]\fg\ témoigne de la
présence d'un bruit de micro et \og \textit{\string^\string^Véli}\fg\ 
précise que le mot est inconnu par le transcripteur. \\
\hline
\end{tabularx}
\end{center}
\caption{Extrait d'un corpus transcrit selon les conventions de
Transcriber}
\label{figCorpTransc}
\end{figure}
D'autres conventions de transcription recommandent au contraire de ne
pas indiquer de signes de ponctuation car le découpage effectué par le
transcripteur en syntagmes ou en phrases préjuge de l'analyse à faire
\cite{Blanche-Benveniste90}. Les pauses silencieuses se révèlent de
plus imprécises pour détecter ces ponctuations, y compris quand le
corpus transcrit correspond à un texte lu \cite{Campione02}. Pour
améliorer la lisibilité des transcriptions, Blanche-Benveniste propose
une segmentation plus fine sous forme de grille. Lors de la production
de la parole, il existe en effet un travail de formulation important
qui vient perturber la succession des mots et le discours s'apparente
souvent à une énumération. Ce type de présentation conduit à
positionner l'un en dessous de l'autre des éléments prononcés
successivement et qui occupent une même place syntaxique
(Fig.~\ref{figMiseGrille}) \cite{Blanche-Benveniste90}.

\begin{figure}
\begin{center}
\begin{tabular}{l l l}
\hline
\og \textit{ils ont des ouvriers euh} & \textit{payés}  \\
& \textit{spécialisés} & \textit{sup-} \\
& & \textit{sur les chantiers de fouille}\fg\ \\
\hline
\end{tabular}
\end{center}
\caption{Mise en grille d'un énoncé (d'après \cite{Guenot05})}
\label{figMiseGrille}
\end{figure}

\subsubsection{Étiquetage des corpus oraux}
\label{subSubSecEtiqCorpusOraux}

Pour faciliter l'utilisation des corpus, qu'ils soient oraux ou
écrits, un étiquetage morpho-syntaxique est souvent réalisé. Cet
étiquetage consiste à attribuer à chaque mot, voire à des groupes de
mots dans le cas de locutions, une étiquette que l'on appelle
\emph{partie du discours} ou \emph{PoS} (pour \textit{Part of
Speech}). Une PoS correspond à une propriété grammaticale dans une
phrase, telle que nom, verbe, adjectif, préposition, \textit{etc.},
que l'on peut préciser par le genre, le nombre, la personne,
\textit{etc.} Le choix des PoS à considérer n'est pas fixe pour une
langue donnée. Les étiqueteurs peuvent ainsi avoir des jeux
d'étiquettes divers pour des mots très employés \cite{Campione05}.

Devant la taille des corpus à analyser, l'étiquetage automatique se
révèle indispensable. Les premiers programmes d'étiquetage automatique
sont apparus pour l'écrit dès les années 50. Ils étaient basés sur la
production manuelle de règles. Depuis les années 80, certains
étiqueteurs utilisent des méthodes statistiques, telles que les HMM
\cite{Brants00} ou les arbres de décision \cite{Schmid94, Schmid95},
pour prédire la probabilité d'attribution des étiquettes. Mais ce
n'est que récemment que des études ont été menées sur l'étiquetage de
corpus oraux.

Il n'existe pas à l'heure actuelle, tout du moins à notre
connaissance, d'étiqueteurs conçus spécifiquement pour l'oral. Des
systèmes prévus initialement pour étiqueter des documents écrits ont
été utilisés sur des corpus oraux, que ce soit pour l'anglais
\cite{Garside95}, le français \cite{Valli99, Campione05}, le suédois
\cite{Nivre01}, le néerlandais \cite{VanEynde00}, l'espagnol
\cite{Moreno03}, le portugais \cite{Mendes03}, l'italien
\cite{Panunzi04} ou encore le japonais \cite{Uchimoto02}. Ces
étiqueteurs sont souvent adaptés à l'oral en modifiant légèrement leur
comportement pour certains mots et en appliquant au corpus un
traitement préalable pour éliminer certaines caractéristiques de
l'oral, telles que les amorces. Dans le cas où le corpus ne contient
pas de signe de ponctuation, les marques de pauses peuvent être
remplacées par des points de suspension, \textit{i.e.}, par la
ponctuation la plus neutre possible par rapport au fonctionnement de
l'étiqueteur. Les amorces de mots sont en général ignorées par
l'étiqueteur car il est parfois difficile de deviner le mot
prononcé. Il peut être également utile d'éliminer des événements non
lexicaux, tels que \og \textit{hein}\fg\ et \og \textit{euh}\fg, ou
encore des événements non linguistiques comme le signalement
d'applaudissements. De plus, certains mots apparaissent beaucoup plus
fréquemment à l'oral qu'à l'écrit, notamment les contractions de mots
et les interjections ; les mots caractéristiques de l'oral sont ainsi
insérés dans le lexique de l'étiqueteur. De même, on ajoute
manuellement des règles ou on modifie manuellement les probabilités
d'assignation des étiquettes pour les mots qui ont un comportement
différent à l'écrit et à l'oral.

Pour mesurer les performances de l'étiquetage automatique, un extrait
de corpus étiqueté automatiquement est comparé avec le même extrait
étiqueté manuellement. L'évaluation est une opération délicate à
effectuer dans la mesure où l'étiquetage de référence, \textit{i.e.},
l'étiquetage manuel, peut différer selon les annotateurs pour certains
mots problématiques. Il est le plus souvent choisi de considérer comme
\emph{acceptable} une étiquette dès qu'elle relève d'un point de
discussion entre linguistes. Le critère de performance est alors le
pourcentage d'étiquettes acceptables. La comparaison des performances
entre étiqueteurs n'est pas non plus chose aisée car le jeu
d'étiquettes ou la segmentation en unités à étiqueter peuvent différer
\cite{Adda99}. En ce qui concerne l'étiquetage de l'écrit, les
performances sont supérieures à 95\,\% d'étiquettes correctes. En
utilisant des étiqueteurs conçus initialement pour l'écrit et adaptés
pour l'oral, des expériences ont permis d'atteindre 98,75\,\% pour le
français \cite{Valli99}, entre 95\,\% et 97\,\% pour le suédois
\cite{Nivre01}, 94,3\,\% pour le néerlandais \cite{VanEynde00} ou
encore 98,3\,\% pour l'espagnol \cite{Moreno03}.

Les performances sont donc très proches de ce qui est observé pour
l'écrit, contrairement à ce que l'on aurait pu penser. L'explication
qui est souvent donnée pour justifier ces résultats est que les
étiqueteurs se basent sur des phénomènes locaux et se révèlent donc
peu sensibles aux phénomènes propres à la langue parlée. Une analyse
qualitative des erreurs montre toutefois que l'étiquetage est mis en
erreur par certaines particularités des corpus oraux comme l'absence
de ponctuation, les répétitions ou les chevauchements entre des mots
prononcés simultanément \cite{Valli99}.

Outre l'étiquetage par des PoS, il existe d'autres niveaux
d'annotation envisageables pour les corpus oraux. On peut ainsi citer
:
\begin{itemize}
\item l'annotation pragmatique caractérisant les actes de discours,
  tels que la question, le conseil, la confirmation, les
  remerciements... \cite{Leech97},
\item l'annotation stylistique caractérisant la présentation du
  discours et de la pensée, \textit{e.g.} par narration, discours
  direct, discours indirect, discours indirect libre... \cite{Leech97}
\item l'annotation syntaxique indiquant les dépendances entre mots ou
groupes de mots \cite{Benzitoun04b, Benzitoun05}.
\end{itemize}
Toutes ces annotations sont beaucoup moins courantes que l'étiquetage
morphosyntaxique et ne peuvent être obtenues convenablement par des
méthodes automatiques. L'analyse syntaxique de l'oral par des
grammaires pose ainsi de nombreux problèmes. Aux difficultés
rencontrées pour la langue écrite et qui sont également présentes dans
la langue parlée, comme l'ambiguïté des analyses possibles, s'ajoutent
des problèmes bien spécifiques à l'oral, tels que la présence de
disfluences, le respect plus lâche des règles de la langue et
l'absence de segmentation bien claire en phrases. Les distorsions de
la langue parlée requièrent une grande robustesse de la part des
analyseurs syntaxiques. Pour concevoir ces systèmes, une approche
similaire à la démarche adoptée pour l'étiquetage morphosyntaxique
consiste à utiliser un analyseur développé pour l'écrit et à lui
adjoindre des procédures traitant les extragrammaticalités de l'oral
\cite{Boufaden98}.

\subsection{Transcription automatique des dialogues spontanés}
\label{subSecTransAutoDiaSpont}

La communauté de la reconnaissance de la parole s'intéresse de plus en
plus aux dialogues spontanés, pour lesquels les locuteurs s'expriment
sans préparer leur discours, ce qui entraîne une production accrue de
disfluences. Les campagnes d'évaluation conduites par le NIST depuis
1987 témoignent de cette évolution \cite{Pallett03}. Les premières
évaluations ont ainsi porté sur des applications au vocabulaire très
réduit et avec une grammaire spécifiée, puis sur la lecture de
journaux. C'est avec la transcription d'émissions d'actualité que les
systèmes de RAP se sont heurtés à des documents plus variés. Ces
émissions comportent en effet des discours préparés mais aussi des
interviews ou des reportages, dans lesquels on retrouve fréquemment
des dialogues spontanés, dans des conditions d'enregistrement
bruitées. D'autres évaluations portent désormais sur la transcription
de la base de données \textit{Switchboard}, constituée de dialogues
téléphoniques spontanés sur des sujets particuliers, ou plus récemment
sur les échanges enregistrés entre les participants d'une réunion.

Les phénomènes caractéristiques de la langue parlée gênent
considérablement la transcription automatique. L'ordre plus flexible
des mots perturbe par exemple les modèles de langages qui basent leur
calcul sur les séquences de mots (\textit{cf.}
section~\ref{subSecML}). Les disfluences compliquent quant à elles
l'analyse du signal et viennent perturber le calcul des probabilités
par les modèles N-grammes. Les amorces et les contractions de mots
sont également difficiles à modéliser, le vocabulaire ayant une taille
limitée. Enfin, les marqueurs de discours tels que \og \textit{en
fait}\fg, \og \textit{enfin}\fg, \og \textit{voilà}\fg\ sont
particulièrement problématiques car ils sont beaucoup plus fréquents à
l'oral et sont souvent très mal articulés. Une étude menée sur
10~heures d'interviews télévisées a montré que les disfluences,
représentant 8\,\% du corpus, expliquent 12\,\% du taux d'erreur
(\textit{cf.}  section~\ref{subSecEval}) global sur la
transcription. Ce taux suggère que les disfluences accroissent la
difficulté de la transcription mais n'exercent pas un effet majeur sur
les segments voisins \cite{AddaDecker04}.

La segmentation du signal acoustique en groupes de souffle n'est pas
non plus sans poser de problème. Le groupe de souffle étant en effet
justifié par le mode de fonctionnement des systèmes de RAP et
uniquement défini d'après des indices acoustiques, à savoir la
détection de pauses, les unités sur lesquelles ont à opérer les ML ne
sont pas aussi cohérentes que les phrases de la langue écrite. Une
segmentation plus linguistique peut ainsi être envisagée avant
d'appliquer des ML. Les meilleures perplexités obtenues sur un corpus
de test non segmenté, avec un ML utilisant une segmentation
linguistique plutôt qu'une segmentation acoustique, illustrent
l'utilité de cette adaptation \cite{Meteer96}. Toutefois, la détection
des segments linguistiques, effectuée notamment à l'aide d'indices sur
les classes grammaticales et la prosodie, restent une tâche difficile
puisque les meilleurs systèmes détectent les fins de segments avec des
taux d'erreurs variant entre 30\,\% et 50\,\%, selon le type de
discours à analyser \cite{Liu04}.

Pour traiter les dialogues spontanés, il est nécessaire d'apporter des
modifications importantes au processus de transcription. Un système de
RAP initialement conçu pour être appliqué sur des émissions
d'information transcrivait ainsi des conversations téléphoniques avec
un taux d'erreur de l'ordre de 50\,\%. Après plusieurs raffinements,
ce taux a finalement pu être ramené à 21\,\% \cite{Gauvain04}. Parmi
les adaptations à effectuer pour prendre en compte la parole
spontanée, on peut citer :
\begin{itemize}
\item au niveau du modèle acoustique, l'ajout de trois phones associés
  aux pauses, aux hésitations et aux respirations ;
\item au niveau du dictionnaire de prononciations, l'ajout de répétitions
  et de contractions pour certains mots et l'intégration des
  interjections ;
\item au niveau du modèle de langage, l'inclusion des inspirations et
  des pauses remplies dans l'historique des modèles N-grammes.
\end{itemize}

De manière à limiter les perturbations engendrées par les disfluences,
certains systèmes de RAP cherchent à les détecter et à les
éliminer. Les disfluences qui sont alors principalement corrigées sont
celles où le \textit{reparandum} et la réparation
(Fig.~\ref{figDisfluences}, page~\pageref{figDisfluences}) sont très
proches au niveau lexical \cite{Dowding93, Johnson04b}. Certaines
sont toutefois plus problématiques à détecter et peuvent
de surcroît apporter des informations pour la prédiction des mots. Il
a été ainsi constaté qu'une suppression des pauses remplies dans les
corpus d'apprentissage et de test entraînait une augmentation de la
perplexité \cite{Stolcke96b}. Une alternative est donc d'utiliser les
disfluences comme source d'informations pour la transcription
\cite{Siu96}.

\paragraph{}
La section~\ref{secCaractLangParle} a exposé les caractéristiques de
la langue parlée, en montrant brièvement les formes sous lesquelles se
présentent les corpus oraux. La section suivante décrit comment des
connaissances linguistiques peuvent être introduites dans des systèmes
de RAP. Les techniques utilisées à cette fin doivent pouvoir
s'intégrer dans le schéma de fonctionnement que nous avons décrit en
section~\ref{SecPrincRP} et être suffisamment flexibles pour ne pas
être perturbées par les phénomènes de l'oral.

\newpage

\section{La linguistique pour la reconnaissance de la parole}
\label{secIntroConnaissancesLing}

Comme nous l'avons vu en section~\ref{subSecModelRP}, les systèmes de
RAP adoptent une modélisation hiérarchique pour décoder la parole. Le
MA identifie à partir des caractéristiques extraites du signal des
phones puis des mots ; le ML est quant à lui chargé de reconnaître les
successions de mots les plus probables pour un groupe de souffle. Ce
mode de fonctionnement présente des similarités avec celui de l'être
humain qui, pour reconnaître le sens véhiculé par la parole, identifie
successivement des phones, des syllabes, des mots et des \og
phrases\fg\ \cite{Allen94}. Mais à la différence de l'être humain qui
utilise différents niveaux de connaissances linguistiques, les
informations sur le langage dans les systèmes de RAP se limitent
essentiellement à la connaissance des phones et des phonèmes d'une
langue, à la réalisation d'un dictionnaire de prononciations et à
l'apprentissage de modèles de langages sur des corpus oraux. Il est
notamment souvent reproché aux ML les plus employés, \textit{i.e.},
aux modèles N-grammes, de ne considérer de manière arbitraire qu'un
historique de $N-1$ mots. Il existe en effet bien des configurations
où leurs hypothèses de conception sont mises en défaut. Ainsi, dans
l'exemple suivant : \og \textit{les oiseaux sur l'arbre
chantaient}\fg, un modèle quadrigramme n'aura pas les éléments utiles,
dans son historique de trois mots, pour prédire l'accord du verbe \og
\textit{chantaient}\fg, du fait de l'insertion du complément de lieu
entre le sujet et le verbe.

Une expérience menée par Brill \textit{et al.}  corrobore l'hypothèse
d'un apport de la linguistique à l'amélioration de la reconnaissance
de la parole. Elle a consisté à analyser les ressources qu'un être
humain utiliserait pour corriger des trans\-crip\-tions
au\-to\-ma\-tiques \cite{Brill98}. Trois corpus de parole ont été à
cette fin traités par un système de RAP pour produire les listes des
dix meilleures hypothèses (\textit{cf.}
section~\ref{subSecSortiesSTP}) associées à chaque groupe de
souffle. Il a alors été demandé à des sujets humains de sélectionner
parmi chacune de ces listes l'hypothèse qui leur semblait la plus
juste. Les choix effectués par les sujets se sont souvent révélés
judicieux puisqu'ils ont permis une nette diminution du taux d'erreur
par rapport à la meilleure hypothèse désignée par le système de RAP
pour chacun des groupes de souffle. Un questionnaire sur les
connaissances utilisées pour faire leur sélection a montré que les
humains se basaient principalement sur des informations linguistiques
telles que l'emploi correct des prépositions et des déterminants, les
accords en genre et en nombre, l'examen du temps pour les verbes, la
connaissance de syntagmes idiomatiques ou encore l'analyse de la
structure des hypothèses proposées.

Si l'introduction de connaissances linguistiques supplémentaires est
donc souhaitable, elle peut s'effectuer de diverses manières. D'une
part, elle implique des choix qui influent sur le processus de
transcription. Si l'on souhaite utiliser au plus tôt des informations
sur le langage, certains modes de couplage MA-ML sont ainsi à
privilégier par rapport à d'autres. Il existe également plusieurs
modes possibles de calcul de probabilités de successions de
mots. D'autre part, les techniques diffèrent suivant le type de
connaissances envisagé. Les méthodes introduisant des informations
morphologiques ne sont ainsi pas les mêmes que celles prenant en
compte de la sémantique. Cette section présente dans un premier temps
les techniques adaptant le processus de transcription à l'introduction
de la linguistique, avant de rentrer plus en détail dans les méthodes
spécifiques à chaque type de connaissances.

\subsection{Adaptation du processus de transcription}
\label{subSecMobLingRAP}

Les ML classiques ne manipulent que des mots et se limitent à un
historique restreint dans leurs calculs de probabilités. L'intégration
de connaissances linguistiques conduit à traiter des informations
supplémentaires au cours de la transcription, telles que la
dé\-com\-po\-si\-tion morphologique des mots à reconnaître, ou bien
encore à effectuer des calculs de probabilités prenant en compte
l'intégralité du groupe du souffle courant, comme cela est le cas de
certaines méthodes d'analyse syntaxique. Le mode de fonctionnement du
système de RAP doit alors être modifié afin d'introduire ces
connaissances sans trop dégrader la rapidité de la transcription. Les
adaptations nécessaires sont effectuées au niveau de l'interface MA-ML
et du calcul des probabilités par le ML, ce qui fait l'objet des deux
sections suivantes.

\subsubsection{Intégration du modèle acoustique avec le modèle de langage}
\label{subSubSecIntMAML}

L'intervention du MA et du ML dans le processus de transcription se
révèle souvent plus complexe que la description qui en a été faite en
section~\ref{subSecModelRP}. Pour des raisons de rapidité, le décodage
de la parole est généralement réalisé en plusieurs passes, chaque
passe correspondant à une application d'un MA et d'un ML et conduisant
à la création d'un graphe de mots. Au fur et à mesure des passes, les
graphes de mots produits sont de plus en plus réduits, tandis que les
MA et les ML sont de plus en plus informatifs et, par conséquent,
lents. On utilisera par exemple plutôt un ML trigramme dans une
première passe et un ML quadrigramme dans une seconde passe. L'intérêt
de ces systèmes multipasses consiste ainsi à pouvoir utiliser des
modèles complexes au niveau des dernières passes, sans engendrer une
augmentation trop importante du temps de calcul. En sus du mode
d'organisation général du processus de transcription, le système de
RAP peut être adapté au niveau de l'interface entre un MA et un ML
donnés. Il existe ainsi trois types d'intégration des modules : le
couplage étroit, le couplage lâche et le couplage modéré
\cite{Harper94}.

Un système à \emph{couplage étroit} intègre toutes les connaissances
dans un ensemble de processus interdépendants et non séparables. Dans
un tel système, le ML est directement intégré dans le MA, ce qui est
permis par certaines structures de ML. Celui-ci doit ainsi être
particulièrement rapide puisque le nombre d'hypothèses examinées par
le MA est très grand et n'a pas encore été réduit. Les ML qui prennent
uniquement en compte des contraintes locales sont ainsi bien adaptés à
ce type de couplage car ils facilitent la mise en \oe uvre
d'algorithmes dynamiques. Un modèle N-grammes, à condition que $N$ ne
soit pas trop grand, ou encore une grammaire à états finis
\cite{Moore99}, peuvent par exemple convenir à ce type de couplage.

Les systèmes à couplage étroit présentent l'avantage d'utiliser au
plus tôt les connaissances linguistiques du ML pour réduire l'espace
de recherche du MA. Ils sont néanmoins difficiles à faire évoluer
puisque ce couplage impose des contraintes fortes sur le
fonctionnement du ML. En outre, la forte imbrication des composants
complique grandement l'évaluation de l'impact de chaque connaissance
sur les performances du système.

Un système à \emph{couplage lâche} isole le MA et le ML en modules
relativement indépendants et communiquant entre eux. Le rôle du ML est
alors de filtrer ou réordonner les hypothèses fournies par le
MA. Trois possibilités sont généralement envisagées pour faire
l'interface entre le MA et le ML : utiliser un graphe de mots
\cite{Chow89b, Harper99}, une liste des N meilleures hypothèses
\cite{Chow89}, ou bien encore un \emph{graphe d'homophones}
\cite{Gauvain05, Bechet99}. Dans ce dernier type d'interface, chaque
mot de la meilleure hypothèse est remplacé par tous ses homophones
possibles, \textit{i.e.}, par tous les mots du langage qui ne sont pas
discriminables au moyen d'indices acoustiques. Cette approche convient
parfaitement à la détection des erreurs dues à des problèmes d'accord
en genre et en nombre.

Généralement, les systèmes à couplage lâche sont plus faciles à faire
évoluer que ceux à couplage étroit et on ne constate pas d'explosion
combinatoire quand on augmente le nombre de connaissances prises en
compte. De plus, ils n'imposent pas de contraintes importantes sur la
structure du ML et permettent d'évaluer facilement les performances
des MA et des ML, qui se présentent sous la forme de modules bien
différenciés. Ils ont toutefois le désavantage de ne pas réduire
l'espace de recherche lors du décodage par le MA.

Un système à \emph{couplage modéré} a un comportement intermédiaire
entre les deux pré\-cé\-dents. Il utilise le ML pour guider le MA,
sans y être intégré. Il se différencie du couplage lâche par le fait
que le ML ne peut pas être supprimé sans modifier le processus de
recherche conduit par le MA. Deux approches ont été envisagées pour le
couplage modéré : l'approche descendante et l'approche ascendante
\cite{Hauenstein94}. Dans l'\emph{approche descendante}, le ML est
invoqué à des points de décision où il prédit des hypothèses. Le MA
est ensuite chargé de sélectionner la meilleure hypothèse. Un
analyseur syntaxique LR peut par exemple être utilisé pour prédire des
phones qui sont ensuite vérifiés par un HMM de phones
\cite{Kita89}. Les phones qui constituent un mot sont alors spécifiés
par des règles de grammaire. Dans l'\emph{approche ascendante}, les
scores acoustiques sont calculés en premier et le ML est appliqué pour
vérifier les hypothèses, en réduisant éventuellement le nombre de
candidats acoustiques. Cette organisation est très proche de celle des
systèmes à couplage lâche mais il est fait appel au ML à chaque point
de décision et non pas à la fin de l'analyse du groupe de
souffle. L'approche ascendante a été par exemple mise en \oe uvre par
un analyseur syntaxique tabulaire par îlots \cite{Beutler03}. Les
hypothèses de mots sont alors fournies les unes après les autres par
le MA au ML. Quand celui-ci échoue dans son analyse syntaxique, il
demande au MA de produire des hypothèses supplémentaires à des
endroits précis du groupe de souffle. Une autre possibilité envisagée
a consisté à construire dynamiquement le ML par un réseau représentant
une grammaire, de manière à n'inclure que les transitions nécessaires
à l'analyse du groupe de souffle courant \cite{Moore89}.

Les systèmes à couplage modéré constituent donc un compromis original
avec les deux autres types. Malheureusement, ils posent de nombreux
problèmes d'ingénierie pour faire interagir les modules et imposent
des contraintes sur la structure des ML telles que la nécessité d'une
analyse gauche-droite.

Au terme de la présentation des trois modes de couplage du MA et du
ML, les deux alternatives qui paraissent les plus intéressantes sont
donc le couplage étroit et le couplage lâche, le premier autorisant
l'intégration de connaissances linguistiques au plus tôt, le second
imposant moins de contraintes sur la structure et la rapidité du
ML. Les systèmes à couplage modéré semblent quant à eux être une
approche difficile à mettre en \oe uvre. Il existe d'ailleurs peu de
systèmes de RAP à ce jour qui soient basés sur cette
organisation. Cette section a présenté les manières de réaliser
l'interface MA-ML en vue d'introduire des informations linguistiques ;
la suivante décrit comment des ML prenant en compte des connaissances
diverses peuvent être agencés pour calculer des probabilités de
succession de mots.

\subsubsection{Linguistique et modèle de langage}
\label{subSubSecLingML}

Dans les systèmes de RAP actuels, le ML contient généralement un
modèle N-grammes. Il existe plusieurs schémas d'intégration de
connaissances linguistiques supplémentaires dans ce modèle
\cite{Chappelier99}.

Une première solution consiste à construire un nouveau ML, plus
linguistique, que l'on utilise en remplacement d'un modèle N-grammes.
Ceci implique qu'il possède des propriétés communes avec le modèle
N-grammes ; il doit être notamment capable de fournir un score et de
faire une analyse gauche-droite en un temps raisonnable. Ce type de
solution est généralement envisagé quand on dispose de connaissances
\textit{a priori} sur un domaine particulier, mais pas d'un corpus
d'apprentissage suffisant, ou quand le nouveau ML intègre
simultanément plusieurs types de connaissances, y compris des
connaissances lexicales qui seraient redondantes avec celles apportées
par un modèle N-grammes \cite{Wang02b}.

Une deuxième possibilité est d'utiliser les connaissances
linguistiques pour lisser le calcul des probabilités des modèles
N-grammes \cite{Jurafsky95}. Ceci se fait par exemple en utilisant une
grammaire pour générer des phrases qui viennent ensuite enrichir le
corpus d'apprentissage du modèle N-grammes \cite{Wang00}. Les
probabilités d'un modèle N-grammes peuvent également être évaluées
directement à partir du nouveau ML \cite{Stolcke94}.

Dans une troisième solution, les connaissances sont utilisées
séquentiellement. L'intérêt est que si chaque source d'informations
est susceptible d'améliorer les performances globales du système de
RAP, chacune a une influence et une complexité de calcul
potentiellement très différentes. Par exemple, un modèle bigramme
pourra réduire la perplexité d'un facteur de dix, avec peu de calculs,
tandis qu'un ML apportant des informations plus complètes sur la
syntaxe et la sémantique pourra demander beaucoup de calculs pour une
réduction de perplexité moindre \cite{Schwartz90}. Dans cette approche
séquentielle, les modèles les plus rapides sont utilisés en premier
lieu pour produire l'ensemble des hypothèses les plus probables. Cet
ensemble, qui se présente sous la forme d'une liste des meilleures
hypothèses \cite{Chow89} ou d'un graphe de mots \cite{Su92}, est
ensuite filtré et réordonné au moyen des sources de connaissance
restantes.

Une dernière solution consiste à combiner plusieurs modèles, apportant
chacun des connaissances, pour constituer un seul ML. Cette
combinaison peut se faire à l'aide de l'interpolation linéaire ou du
repli, dont les principes ont déjà été exposés (\textit{cf.}
section~\ref{subSubSecFoncML}). Dans le cas de l'interpolation
linéaire, technique la plus utilisée du fait de sa simplicité, la
combinaison de $M$ modèles différents, associés aux distributions de
probabilité $P_k$ avec $k = 1 \ldots M$, s'effectue de la manière
suivante :
\begin{equation}
P(w_i|w_1^{i-1}) = \sum_{k=1}^M \lambda_k P_k(w_i|w_1^{i-1})
\end{equation}
où $\sum_{k=1}^M = 1$. En ce qui concerne le repli, le calcul pour
combiner M modèles, avec M fixé ici à 2 pour simplifier l'équation,
s'écrit \cite{Niesler96} :
\begin{equation} 
P(w_i|w_1^{i-1}) = \left\{ \begin{array}{ll}
P_1(w_i|w_1^{i-1}) & \textrm{si } w_i \in \Phi_1(w_1^{i-1}) \\
\alpha(w_1^{i-1}) \times P_2(w_i|w_1^{i-1}) & \textrm{sinon}
\end{array} \right.
\end{equation}
où $\Phi_k(w_1^{i-1})$ représente l'ensemble des mots dans le contexte
$w_1^{i-1}$ pour lequel le $k^{i\grave{e}me}$ modèle est à utiliser en
priorité, et $\alpha$ est un coefficient de normalisation. L'intérêt
de cette méthode est d'utiliser d'abord les modèles les plus
informatifs quand on dispose de suffisamment d'informations dans le
contexte courant. Une autre possibilité pour combiner des modèles est
d'utiliser des \emph{modèles exponentiels}, appelés encore \emph{à
entropie maximale} \cite{Rosenfeld96, Goodman01}. Ceux-ci évaluent les
probabilités sous la forme :
\begin{equation}
P(w_i|w_1^{i-1}) = \frac{\exp (\sum_k \lambda_k f_k(w_1^i))}{z(w_1^{i-1})}
\end{equation}
où $z$ est une fonction de normalisation telle que :
\begin{equation}
\sum_{w_i} P(w_i|w_1^{i-1}) = 1
\end{equation}
Les $\lambda_k$ sont des coefficients obtenus grâce à un algorithme
d'apprentissage, tandis que les $f_k$ sont des fonctions de
contraintes retournant typiquement 0 ou 1. Le principal intérêt de ce
type de modèle est que les $f_k$ permettent de représenter des modèles
N-grammes, des modèles à base de cache (\textit{cf.}
section~\ref{subSubSecModifHisto}), des modèles N-classes
(\textit{cf.} section~\ref{subSecMorphologie}) ou encore des modèles
\textit{triggers} (\textit{cf.}  section~\ref{ssSecASyntax}). Dans le
cas d'une fonction pour le trigramme $w_a w_b w_c$, on aura ainsi :
\begin{equation} 
f_{w_a w_b w_c}(w_1^i) = \left\{ \begin{array}{ll}
1 & \textrm{si } w_{i-2}=w_a, w_{i-1}=w_b \textrm{ et } w_{i}=w_c\\
0 & \textrm{sinon }
\end{array} \right.
\end{equation}
Les modèles exponentiels permettent ainsi d'intégrer plusieurs sources
d'informations de manière élégante. Leur temps d'apprentissage est
cependant extrêmement long et ils se révèlent assez lents lors de leur
utilisation. En outre, mis à part avec les modèles \textit{triggers},
il semble qu'ils n'aient pas encore permis de réduire la perplexité de
manière significative \cite{Goodman01}.

Après avoir décrit les adaptations possibles du processus de
transcription pour intégrer de nouvelles informations, nous nous
tournons désormais vers la présentation des sortes de connaissances
linguistiques pouvant compléter celles généralement apportées par les
MA et les ML, des méthodes qui ont déjà été envisagées pour les
utiliser et de leur influence sur la qualité de la transcription
produite.

\subsection{Quelles connaissances linguistiques ?}
\label{subSecQuellesConnaissancesLing}

Cette section fait un tour d'horizon des types d'informations
linguistiques qui ont été pris en compte en RAP. Les MA ne pouvant
inclure que des connaissances concernant l'acoustique, nous nous
consacrons plus particulièrement ici aux expériences menées dans le
cadre de l'intégration de ces connaissances au sein de ML. La
présentation est structurée selon une typologie à cinq niveaux
généralement reconnue en linguistique :
\begin{itemize}
\item la \emph{phonologie} et la \emph{phonétique} qui étudient les
  sons,
\item la \emph{morphologie} qui étudie la structure des mots,
\item la \emph{syntaxe} qui étudie la structure des syntagmes et des
phrases,
\item la \emph{sémantique} qui étudie les sens des mots, des
  locutions, des phrases ou des textes,
\item la \emph{pragmatique} qui étudie la relation entre le langage et
  son contexte d'utilisation.
\end{itemize}

Les techniques d'insertion se heurtent à plusieurs difficultés ; elles
doivent être robustes aux distorsions de l'oral, suffisamment rapides
selon le type de couplage MA-ML envisagé et doivent pouvoir s'intégrer
dans un ML statistique pour sélectionner les hypothèses les plus
probables. Ce dernier point complique notamment la prise en compte de
connaissances symboliques au cours de la transcription
\cite{Antoine99}.


\subsubsection{Phonologie et phonétique}
\label{subSecPhonologiePhonetique}

Les ensembles de phonèmes et de phones d'une langue sont finis et il
est possible d'établir des règles quant à la succession de leurs
éléments. En français, l'emploi consécutif des deux phones [d] et [s]
est par exemple illicite. Les connaissances issues de la phonologie et
de la phonétique présentent la particularité de pouvoir être prises en
compte à la fois dans le MA, le dictionnaire de prononciations et le
ML.

Les MA les plus performants utilisent par exemple davantage
d'informations sur le contexte de prononciation des phones que ce qui
a été présenté en section~\ref{subSecMA} ; les états des HMM peuvent
être des \emph{triphones}, prenant en considération les influences des
phones précédant et suivant le phone courant, plutôt que des
phones. La prédiction des phénomènes de liaison entre les mots, dans
des séquences telles que \og \textit{les enfants}\fg\ \cite{Boula03},
peut également être envisagée pour affiner l'utilisation du
dictionnaire de prononciations par le MA. Cette section se
consacrant plutôt à l'amélioration du ML, nous ne décrivons toutefois
pas davantage les modifications possibles du MA et du dictionnaire de
prononciations.

Peu d'informations sur la phonologie et la phonétique ont été
intégrées dans le ML. La seule étude que nous ayons rencontrée porte
sur la détermination des événements impossibles \cite{Langlois03}. Ce
sont des successions de mots qui ne peuvent se produire du fait de
contraintes sur la langue. En français, l'expression \og \textit{je
aime}\fg\ est par exemple interdite. La détermination de ces
événements permet d'affiner le calcul des probabilités des modèles
N-grammes, qui s'effectue au moyen de
l'équation~(\ref{eqCoomptageNGrammes}) (\textit{cf.}
page~\pageref{eqCoomptageNGrammes}), en dénombrant l'ensemble des
événements $S$ (pour \textit{Seen}) présents dans le corpus
d'apprentissage. Les techniques de lissage prennent en compte les
séquences de mots absentes de ce corpus mais elles ne distinguent pas
les événements $U$ (pour \textit{Unseen}) non rencontrés mais pourtant
possibles des événements $I$ impossibles. Il paraît intéressant
d'intégrer uniquement $S$ et $U$ dans l'évaluation des
probabilités. Une des contraintes examinées dans l'étude, d'ordre
phonologique, repose sur des règles d'élision, imposant que certains
mots terminés par une voyelle ne peuvent être suivis par une voyelle
(\textit{e.g.}  \og \textit{le arbre}\fg). Cette contrainte a permis
de considérer comme impossible une fraction très modeste (0,1\,\%) de
l'ensemble des bigrammes et n'apporte donc pas de gain significatif au
niveau de la qualité de la transcription.

\subsubsection{Morphologie}
\label{subSecMorphologie}

La structuration des mots constitue une source d'informations
intéressante pour des langues morphologiquement très riches comme le
turc, le russe ou l'arabe ou même pour des langues à haut taux de
flexion comme le français. Dans le cas des langues agglutinantes
notamment, pour avoir une couverture lexicale similaire à celle
obtenue pour l'anglais, il est nécessaire d'envisager un nombre
considérable de mots ; le fait de décomposer les mots en plusieurs
constituants élémentaires permet de réduire le nombre d'événements à
envisager lors du calcul des probabilités. Dans le cas des langues
flexionnelles, l'analyse morphologique permet de rassembler dans une
même classe des mots qui jouent le même rôle dans la phrase, comme \og
\textit{mangeait} \fg\ et \og \textit{mangerons}\fg\ en français.

Les informations morphologiques peuvent être introduites dans un ML en
utilisant des modèles N-grammes basés sur des classes (on parle alors
de \emph{modèles N-classes}), et non plus des modèles basés sur des
mots. Les classes contiennent alors l'ensemble des mots possédant le
même lemme\footnote{Forme \og simple \fg\ d'un mot obtenue par un
processus de \emph{lemmatisation}. Le lemme associé à un verbe
conjugué sera par exemple sa forme à l'infinitif ; pour un adjectif ou
un nom, ce sera sa forme au masculin singulier. Cette notion permet
d'associer à un même lemme l'ensemble de mots qui ne se distinguent
que par la flexion.}. Généralement, un mot est associé à un seul lemme
mais il peut arriver en cas d'ambiguïté qu'il corresponde à
plusieurs. Il en existe ainsi deux pour le participe passé \og
\textit{plu} \fg\ : \og \textit{pleuvoir}\fg\ et \og
\textit{plaire}\fg. 

Dans un modèle N-classes, si $\mathcal{C}_i$ représente l'ensemble des
classes $c_i$ auxquelles peut appartenir un mot $w_i$, le calcul des
probabilités se fait de la manière suivante :
\begin{eqnarray}
P(w_1^n) & = & \sum_{c_1 \in \mathcal{C}_1 \ldots c_n \in
  \mathcal{C}_n} P(w_1^n c_1^n) \\
& = & \sum_{c_1 \in \mathcal{C}_1 \ldots c_n \in
  \mathcal{C}_n} \prod_{i=1}^n P(w_i|w_1^{i-1}c_1^i) P(c_i|w_1^{i-1}c_1^{i-1})
\end{eqnarray}
où $P(w_i|w_1^{i-1}c_1^i)$ est appelée la \emph{probabilité lexicale}
et $P(c_i|w_1^{i-1}c_1^{i-1})$ la \emph{probabilité contextuelle}. En
supposant que la probabilité de $w_i$ dépend uniquement des classes
$c_i$ auxquelles il peut appartenir dans le groupe de souffle, on
obtient :
\begin{equation} 
\label{eqApproxProbaLex}
P(w_1^n) \approx \sum_{c_1 \in \mathcal{C}_1 \ldots c_n \in
  \mathcal{C}_n} \prod_{i=1}^n P(w_i|c_i) P(c_i|w_1^{i-1}c_1^{i-1})
\end{equation} 
De manière similaire aux modèles N-grammes de mots qui ne prennent en
compte que les $N-1$ mots précédents dans le calcul,
$P(c_i|w_1^{i-1}c_1^{i-1})$ est approximé en ne considérant que les
$N-1$ classes attribuées précédemment :
\begin{equation} \label{eqApproxMLclasses}
P(w_1^n) \approx \sum_{c_1 \in \mathcal{C}_1 \ldots c_n \in
  \mathcal{C}_n} \prod_{i=1}^n P(w_i|c_i) P(c_i|c_{i-N+1}^{i-1})
\end{equation} 

L'intérêt principal des modèles N-classes est de réduire
considérablement le nombre d'événements possibles par rapport aux
modèles N-grammes puisque le nombre total de classes est généralement
beaucoup plus petit que la taille du vocabulaire du système de
RAP. Ceci permet de limiter le recours aux techniques de lissage, qui
introduisent des approximations lors du calcul des probabilités. 

Les modèles N-classes utilisant les lemmes sont généralement combinés,
par interpolation linéaire, avec des modèles N-classes à base de PoS
(\textit{cf.} section~\ref{ssSecPoS}) \cite{Elbeze90, Maltese92}. Ceci
permet d'améliorer légèrement la perplexité par rapport à un simple
modèle N-classes de PoS. Le comportement n'a toutefois pas été testé
dans des systèmes de RAP.

Une alternative aux modèles N-classes est le \emph{ML factorisé}. Ce
ML décompose chaque mot $w_i$ en $k$ caractéristiques $f_i^{1:k}$,
aussi appelées facteurs. Elles représentent des informations
morphologiques, syntaxiques ou sémantiques sur le mot, en plus du mot
lui-même. Les ML factorisés probabilistes utilisant une approximation
trigramme calculent les probabilités grâce à l'expression :
\begin{equation}
P(f_1^{1:k}, f_2^{1:k} \ldots f_n^{i:k}) = \prod_{i=3}^n
P(f_i^{1:k}|f_{i-2}^{1:k}, f_{i-1}^{1:k})
\end{equation}
Ces ML factorisés ont notamment été appliqués à l'arabe, avec comme
facteurs le radical\footnote{Support morphologique d'un mot. C'est la
partie qui contient le sens d'un mot, après avoir supprimé tout ce qui
relevait de la flexion. Par exemple, les radicaux respectifs de \og
\textit{déstabiliser} \fg\ et \og \textit{nationaliser}\fg\ sont \og
\textit{déstabilis-}\fg\ et \og \textit{nationalis-}\fg.}, la
racine\footnote{Constituant d'un mot qui porte la partie principale de
son sens. À la différence du radical, la racine ne peut pas être
décomposée en d'autres éléments porteurs de sens ou morphologiquement
simples. Ainsi, les racines respectives de \og \textit{déstabiliser}
\fg\ et \og \textit{nationaliser}\fg\ sont \og \textit{stabil-}\fg\ et
\og \textit{nation}\fg.} et la classe morphologique. Leur utilisation
au sein d'un système de RAP pour transcrire des émissions d'actualité
a permis de réduire le taux d'erreur sur les mots de 57,6\,\% à
56,1\,\% \cite{Vergyri04}. Un inconvénient principal est qu'ils
nécessitent des adaptations importantes pour être intégrés dans le
processus de transcription, celui-ci étant généralement conçu pour
opérer sur des mots, plutôt que sur des facteurs.

\subsubsection{Syntaxe}
\label{subSecSyntaxe}

L'information syntaxique peut être utilisée sous plusieurs formes :
les ML peuvent ainsi tenir compte des classes grammaticales attribuées
aux mots, d'une nouvelle segmentation du groupe de souffle en
réunissant plusieurs mots au sein d'un même constituant, ou encore
d'une analyse syntaxique du groupe de souffle. Nous présentons
successivement un état de l'art des tentatives faites pour intégrer
ces divers types de connaissances.

\paragraph{Parties du discours et classes statistiques\\}
\label{ssSecPoS}

Il existe des successions de parties de discours (PoS) (\textit{cf.}
section~\ref{subSubSecEtiqCorpusOraux}) qui se produisent de manière
rarissime dans une langue donnée. Il est ainsi très peu fréquent que
deux noms communs se suivent en français. Des ML basés sur ce principe
ont été conçus en considérant soit des classes syntaxiques déterminées
\textit{a priori} (en général des PoS souvent accompagnées
d'informations morphologiques sur le genre, le nombre, le temps ou
encore le mode), soit des classes produites automatiquement par des
méthodes statistiques.

Les PoS sont généralement intégrées au système de RAP au moyen de
modèles N-classes, où chaque PoS correspond à une classe
\cite{Maltese92, Goodman01}. Les modèles N-classes sont toutefois
moins performants que les modèles N-grammes, en considérant des
historiques de même longueur. On observe en revanche dans certains cas
une amélioration de la perplexité par rapport aux modèles N-grammes
quand on combine des modèles N-classes avec des modèles
N-grammes. Cette baisse de la perplexité reste cependant peu
importante même quand on dispose de suffisamment de données pour
apprendre les paramètres du ML. Diverses améliorations des modèles
N-classes ont donc été envisagées.

Dans une première solution, le mode d'intégration dans le système de
RAP des modèles N-classes utilisant les PoS est révisé. Au lieu d'être
employé sur le graphe de mots produit aux cours des passes
précédentes, le ML est appliqué de manière plus sélective sur le
graphe d'homophones établi à partir de la meilleure hypothèse trouvée
(\textit{cf.}  section~\ref{subSubSecIntMAML}). Le rôle du modèle
N-classes est alors de sélectionner un homophone possible parmi ceux
générés pour chaque hypothèse de mots \cite{Bechet99}. Ce mode est
particulièrement adapté pour corriger des fautes d'accord en genre et
en nombre, notamment en français où les formes d'un même mot au
singulier et au pluriel sont souvent homophones. L'application d'un
modèle N-classes adoptant cette méthode a permis de réduire le taux
d'erreur de 10,7\,\% à 10,5\,\% sur la transcription d'émissions
d'actualité en français \cite{Gauvain05}.

Une deuxième solution consiste à reconsidérer le mode de calcul des
probabilités. En modifiant l'approximation faite sur la probabilité
lexicale dans l'équation~(\ref{eqApproxProbaLex}) par :
\begin{equation}
P(w_i|w_1^{i-1}c_1^i) \approx P(w_i|c_{i-N+1}^i)
\end{equation}
une légère amélioration de l'entropie croisée a été observée
\cite{Goodman01}. On peut même aller jusqu'à supprimer les
approximations faites à la fois sur les probabilités lexicale et
contextuelle. Une étude \cite{Heeman99} propose ainsi de redéfinir
l'objectif d'un système de RAP (\textit{cf.}
section~\ref{subSecModelRP}), de manière à ce que les PoS $C$
associées aux mots $W$ à reconnaître soient considérées comme partie
intégrante de la sortie de la transcription et non plus comme des
objets intermédiaires. La finalité de la RAP revient alors à estimer :
\begin{equation} 
\hat{W}, \hat{C}=\arg\max_{W, C} P(W, C|A)
\end{equation}
En éliminant les approximations, le nombre d'événements à examiner
pour évaluer les probabilités augmente considérablement ; une méthode
basée sur des arbres de décision a donc été élaborée. Cette technique
a donné des résultats satisfaisants pour transcrire des dialogues
portant sur des sujets précis puisque le ML triclasse modifié a
conduit à une réduction du taux d'erreur de 26,0\,\% à 24,9\,\% par
rapport à des ML trigrammes, tandis que le ML triclasse conventionnel
faisait quant à lui augmenter le taux d'erreur.

Une troisième solution pour améliorer les modèles N-classes est de
modifier la construction des classes. Au lieu d'avoir une classe par
PoS, une possibilité est de regrouper au sein d'une même classe
l'ensemble des mots ayant les mêmes PoS possibles avec le même ordre
de vraisemblance. Ceci supprime l'ambiguïté des classes que l'on peut
associer aux mots. Bien que l'attribution d'une PoS à un mot soit une
technique relativement maîtrisée (\textit{cf.}
section~\ref{subSubSecEtiqCorpusOraux}), il demeure toujours des
erreurs qui peuvent venir perturber le calcul des probabilités. Un ML
triclasse utilisant ce type de classes, combiné avec un ML trigramme,
a permis de réduire le taux d'erreur sur les mots pour traiter de la
parole lue \cite{Samuelsson99}.

On peut également envisager un système à nombre très restreint de
classes, en poussant à l'extrême la propriété principale des modèles
N-classes, qui est de réduire le total des événements à envisager pour
le calcul des probabilités. Normalement, il est possible de considérer
d'une vingtaine à une centaine de PoS différentes ; dans un nouveau
système, on ne discerne que deux types de PoS : les \emph{classes
ouvertes}, correspondant aux mots lexicaux, et les \emph{classes
fermées}, correspondant aux mots grammaticaux (\textit{cf.}
section~\ref{subSecPhHesitation}). Cette distinction est faite avec
l'idée que la séquence des mots grammaticaux reflète les contraintes
syntaxiques du groupe de souffle, alors que la séquence de mots
lexicaux est contrôlée par des relations sémantiques entre les
mots. Les classes ouvertes et fermées sont généralement utilisées en
adaptant les modèles N-classes \cite{Isotani94, Geutner96}. Dans le
cas où on considère un historique de longueur deux, seuls les derniers
mots lexical et grammatical rencontrés sont considérés. Si $o_{i-1}$
est le dernier mot de classe ouverte et $f_{i-1}$ le dernier mot de
classe fermée dans $w_1^{i-1}$, le calcul des probabilités se fait
comme suit :
\begin{equation} 
P(w_i|w_1^{i-1}) \approx  \left\{ \begin{array}{ll}
P(w_i|w_{i-1}, o_{i-1}) & \textrm{si $w_{i-1}$ est un mot de classe fermée}\\
P(w_i|w_{i-1}, f_{i-1}) & \textrm{si $w_{i-1}$ est un mot de classe ouverte}\\
\end{array} \right.
\end{equation}
La combinaison d'un tel modèle avec un ML trigramme a permis de
réduire le taux d'erreur de 29,4\,\% à 29,0\,\% pour transcrire un corpus
de parole spontané en allemand \cite{Geutner96}. La prise en compte de
ces deux types de classes peut être faite différemment en utilisant un
modèle N-grammes classique pour les classes ouvertes et un modèle
spécialement conçu pour les mots de classes fermées \cite{Peng01}. Le
modèle spécifique prédit alors les mots grammaticaux à partir des
$M-1$ mots grammaticaux précédents. La conception de ce modèle est
justifiée par le fait qu'en anglais par exemple, les mots des classes
fermées représentent 30\,\% du langage écrit et sont en moyenne distants
de 1,9 mots. L'utilisation de ce modèle a permis une légère
amélioration de la perplexité par rapport à un ML trigramme.

Outre les modèles N-classes, la connaissance sur les PoS peut être
introduite grâce aux modèles de cache (\textit{cf.}
section~\ref{subSubSecModifHisto}). Un cache, limité à 200 mots, peut
être construit pour chaque PoS \cite{Kuhn90}. Le cache d'une PoS
donnée contiendra alors les derniers mots rencontrés, étiquetés par
cette PoS. Ces modèles sont conçus avec l'idée que chaque PoS a une
répartition particulière d'occurrences. Les mots lexicaux ont
ainsi tendance à apparaître par vagues, au gré des sujets traités,
tandis que les mots grammaticaux sont répartis plus uniformément. Ce
type de modèle possède des propriétés intéressantes puisque la
combinaison d'un modèle de cache avec un ML trigramme a conduit à une
réduction de perplexité d'un facteur supérieur à trois par rapport à
un modèle trigramme \cite{Kuhn90}. Un modèle assez similaire a été
également conçu pour discriminer les formes singulier et pluriel
homophones \cite{Bechet99}.

\paragraph{}
Une alternative aux PoS déterminées \textit{a priori} consiste à
construire des classes statistiques \cite{Brown92, Kneser93b,
Tamoto95, Farhat96, Jardino96}. L'objectif de ces classes est de
regrouper les mots qui ont le même \og comportement\fg. Elles peuvent
ainsi réunir ceux qui possèdent les mêmes PoS et qui sont liés sur le
plan sémantique, comme par exemple \og \textit{gens}\fg, \og
\textit{hommes}\fg, \og \textit{femmes} \fg\ et \og
\textit{enfants}\fg. Elles sont obtenues par des méthodes de
classification automatique, en cherchant notamment à maximiser la
perplexité sur un ensemble de test ou à maximiser l'information
mutuelle moyenne entre les classes. Lors du processus de construction
des classes statistiques, la position du mot peut être prise en
compte. Des classes dites \emph{prédictives} peuvent ainsi être créées
quand le mot est situé en position $w_i$ dans le calcul de
$P(w_i|w_1^{i-N+1})$, tandis que d'autres dites \emph{conditionnelles}
correspondent aux mots positionnés dans l'historique $w_1^{i-N+1}$
\cite{Yamamoto99}. Si on considère par exemple en anglais \og
\textit{a}\fg\ et \og \textit{an}\fg, ils peuvent suivre les mêmes
mots puisque ce sont deux articles indéfinis ; ils appartiennent par
conséquent à la même classe prédictive. En revanche, du fait qu'il
existe très peu de mots qui peuvent à la fois se positionner après \og
\textit{a}\fg\ et \og \textit{an}\fg, ils sont associés à deux classes
conditionnelles différentes. L'utilisation de deux types de classes
permet ainsi d'introduire des connaissances de granularité plus fine.

Généralement, on attribue une seule classe statistique à chaque mot,
contrairement aux ML à base de PoS où plusieurs PoS sont associables à
un mot. Cette propriété permet de simplifier le calcul des
probabilités de l'équation~(\ref{eqApproxMLclasses}) :
\begin{equation}
P(w_1^n) \approx \prod_{i=1}^n P(w_i|c_i) P(c_i|c_{i-N+1}^{i-1})
\end{equation} 
Des modèles triclasses utilisant des classes statistiques ont permis
de réduire la perplexité par rapport à des modèles trigrammes, que ce
soit pour l'anglais, le français ou l'allemand \cite{Jardino96}. Les
gains en perplexité sont plus importants pour des langues à haut taux
de flexion telles que le russe \cite{Whittaker01}. L'intégration des
classes statistiques dans des modèles varigrammes (\textit{cf.}
section~\ref{subSubSecModifHisto}) a conduit quant à elle à une baisse
du taux d'erreur relative de 7\,\% par rapport à des modèles
varigrammes sans classe pour transcrire de la parole lue en anglais
\cite{Blasig99}. Il semble que ces classes statistiques conduisent à
une réduction plus grande de perplexité que les PoS
\cite{Niesler98}. Toutefois, le fait que la perplexité diminue
davantage quand on combine les modèles N-classes à base de classes
statistiques avec ceux à base de PoS \cite{Kneser93b, Perraud2003}
laisse penser que les deux types de classes portent des informations
complémentaires. Les bons résultats des classes statistiques
pourraient être expliqués par le fait que celles-ci dépassent
généralement une taille critique, ce qui permet aux modèles N-classes
de ne pas faire intervenir des nombres d'occurrences faibles dans le
calcul des probabilités, contrairement aux modèles N-grammes. La
baisse du taux d'erreur, constatée en classifiant uniquement les mots
peu fréquents pour transcrire de la parole spontanée dans le domaine
du trafic aérien, va dans ce sens \cite{Farhat96}. Il apparaît
toutefois que l'association des mots rares à leur classe par des
méthodes automatiques est peu fiable, ce qui diminue l'intérêt des
modèles N-classes par rapport aux techniques de lissage pour évaluer
la probabilité des événements absents de l'ensemble d'apprentissage
\cite{Rosenfeld00b}.

En sus des modèles N-classes, les réseaux de neurones constituent un
autre mode de construction des ML à base de classes statistiques
\cite{bengio03}. Leur particularité est de représenter chaque mot non
plus par une classe discrète, mais par un vecteur de caractéristiques
à $m$ dimensions (\textit{e.g.}  $m = 30$), où $m$ est beaucoup plus
petit que la taille du vocabulaire. L'objectif du nouvel espace est
que deux mots qui jouent des rôles syntaxiques et sémantiques
similaires, comme \og \textit{chat} \fg\ et \og \textit{chien} \fg\
par exemple, aient des vecteurs de caractéristiques proches. La
projection de chaque mot dans le nouvel espace et les valeurs
$P(w_i|w_{i-N+1}^{i-1})$ sont déterminées simultanément lors de la
phase d'apprentissage d'un réseau de neurones, où la première couche
cachée représente les vecteurs de caractéristiques des mots
$w_{i-N+1}^{i-1}$ et les sorties les probabilités
$P(w_i|w_{i-N+1}^{i-1})$ pour chaque mot $w_i$ du vocabulaire. Les ML
à base de neurones se révèlent performants puisqu'ils ont permis de
réduire le taux d'erreur de 22,6\,\% à 21,8\,\% pour transcrire de la
parole conversationnelle en anglais, par rapport à un ML quadrigramme
\cite{Schwenk04}.

Enfin, une autre application envisageable de la connaissance des
classes statistiques est de réduire le nombre d'événements impossibles
(\textit{cf.}  section~\ref{subSecPhonologiePhonetique}). Les mots
sont regroupés en classes selon leurs ressemblances syntaxiques et
sémantiques et reçoivent l'étiquette de leur classe. L'ensemble des
suites possibles de deux classes consécutives étant connu, les
séquences impossibles sont celles qui sont absentes d'un corpus
représentatif de la langue, voire présentes de manière peu
significative si on tient compte des erreurs de typographie de ce
corpus. Cette technique a également été étendue aux PoS, en remarquant
que des règles de certaines langues prohibent des successions telles
que \og [{\footnotesize ARTICLE DÉFINI PLURIEL}] [{\footnotesize NOM
MASCULIN SINGULIER}]\fg. Dans l'expérience décrite dans
\cite{Langlois03}, la prise en compte de 200 classes statistiques et
de 233 PoS a ainsi permis de réduire de près de 15\,\% le nombre de
bigrammes possibles.

\paragraph{Multimots\\}
\label{ssSecGroupesMots}

Une autre utilisation de connaissances syntaxiques consiste à
regrouper plusieurs mots au sein d'unités d'ordre supérieur, comme des
locutions ou des syntagmes, et à les ajouter au vocabulaire du système
de RAP. Ces groupes de mots, que nous désignons par la suite par le
terme \emph{multimots}, peuvent être de natures très diverses. Il peut
s'agir de mots qui cooccurrent fréquemment dans un corpus, tels que
\og \textit{demain matin}\fg\ ou encore \og \textit{millions de
dollars}\fg, de mots composés comme \og \textit{New York}\fg\ ou \og
\textit{vice président}\fg, ou d'entités nommées concernant des dates
ou des noms de personne. Dans le cas d'applications ciblées, ce peut
être des expressions propres à un domaine telles que \og \textit{vous
êtes la bienvenue}\fg\ ou \og \textit{pouvez-vous s'il vous plaît me
mettre en contact avec}\fg.

Les multimots sont généralement sélectionnés par des méthodes
automatiques parmi l'ensemble des combinaisons possibles de mots du
vocabulaire du système de RAP. Il existe deux approches principales
pour les obtenir : les méthodes purement statistiques et les systèmes
à base de règles, même si cette distinction est parfois un peu
arbitraire, certaines techniques combinant les deux approches
\cite{Bechet99}. Parmi les méthodes purement statistiques figurent les
modèles multigrammes, déjà évoqués en
section~\ref{subSubSecModifHisto}, qui choisissent pour chaque groupe
de souffle la meilleure segmentation parmi plusieurs possibles, en
maximisant la probabilité d'observation. Ce type de modèle détermine
les multimots en ligne pour chaque nouveau groupe de souffle, en
fixant le nombre maximal de mots qu'ils peuvent contenir. La plupart
des méthodes statistiques prennent cependant en compte les multimots
en les sélectionnant hors ligne suivant un critère donné, tel que les
fréquences de cooccurrences, ou encore l'évaluation de la perplexité
par validation croisée, puis en les intégrant au vocabulaire du
système de RAP au même titre que les mots \cite{Suhm94, RIes96,
Kuo99}. Parmi les méthodes à base de règles, on peut citer celles
utilisant les automates à états finis \cite{Nasr99, Bechet99} ou les
grammaires non contextuelles probabilistes \cite{Gillett98, Wang00,
Mou02, Seneff03}.

L'intérêt principal des multimots par rapport aux mots est d'autoriser
la prise en compte de phénomènes entre mots distants, tels que les
accords en genre et en nombre, sans avoir à augmenter la taille de
l'historique des ML \cite{Bechet99}. Ces unités peuvent également être
utilisées avec profit pour améliorer la modélisation acoustique des
liaisons entre les mots. Elles permettent enfin d'introduire des
connaissances spécifiques à un domaine en indiquant des expressions à
reconnaître \cite{Kuo99} ou encore des règles pour identifier
certaines entités nommées \cite{Mou02}.

La qualité des multimots détectés diffère selon le domaine étudié
\cite{Kuo99}. Dans des applications ciblées, telles que le routage
d'appels téléphoniques, les constructions stéréotypées telles que \og
\textit{can you please get me}\fg\ sont très naturellement
modélisées. Le choix des multimots se révèle plus délicat dans le
cadre de la parole lue, où les constructions et le vocabulaire sont
riches et variés. L'apport des multimots dans des modèles trigrammes
est d'ailleurs plus important en domaines spécialisés.

L'ajout de multimots au vocabulaire ayant l'inconvénient d'augmenter
le nombre d'évé\-ne\-ments possibles, des modèles N-classes sont
souvent envisagés \cite{Deligne98, RIes96, Nasr99}. Les classes
peuvent indifféremment contenir des mots et des multimots
\cite{Deligne98, RIes96}, ce qui peut notamment permettre de regrouper
des expressions différant uniquement par la présence de disfluences
(\textit{cf.}  section~\ref{subSecPhHesitation}), comme \og
\textit{euh je veux}\fg\ et \og \textit{je veux}\fg. Ce type de modèle
a conduit à une réduction du taux d'erreur de 29,5\,\% à 27,9\,\% par
rapport à un ML trigramme pour transcrire de la parole spontanée en
allemand \cite{RIes96}. Les classes peuvent également ne concerner que
les multimots, ce qui est surtout le cas quand on utilise des
automates d'états finis ou des grammaires \cite{Nasr99, Gillett98,
Wang00, Mou02, Seneff03}. Les multimots reconnus par une même règle et
correspondant à un concept précis, tels que le type de précipitation
ou le nom d'une ville, se retrouvent ainsi au sein d'une même
classe. Cette méthode a permis de faire baisser le taux d'erreur par
rapport à des modèles N-classes standard pour des requêtes sur la
météo (de 18,3\,\% à 18,0\,\%) ainsi que dans le domaine du trafic
aérien (de 15,6\,\% à 15,0\,\%) \cite{Seneff03}.

\paragraph{Analyse syntaxique\\}
\label{ssSecASyntax}

Nous venons de voir, au travers de l'extraction de multimots, un
emploi un peu particulier de l'analyse syntaxique. Les techniques
présentées ici l'utilisent dans un cadre plus formel, en prenant en
compte la structure du groupe de souffle. Les \emph{grammaires non
contextuelles}, notamment leurs versions probabilistes, ont été le
premier type d'analyse envisagé dans les systèmes de RAP
\cite{Stolcke94, Jurafsky95, Seneff92, Seneff95, Linares04}. Le mode
d'attribution des probabilités pour ces grammaires est souvent jugé
limité puisqu'il ne dépend que du non-terminal de la partie gauche de
chacune des règles ; d'autres formes d'analyse sont souvent préférées.
Les \emph{grammaires lexicalisées} \cite{Jurafsky00} étendent ainsi
les grammaires non contextuelles en choisissant pour chaque
constituant détecté un mot jouant le rôle de tête. Le calcul des
probabilités est alors conditionné par la tête \cite{Chelba00} et
parfois par d'autres non-terminaux rencontrés auparavant dans
l'analyse \cite{Roark01, Charniak01}. Les \emph{grammaires de
dépendance} offrent elles aussi un plus haut degré de lexicalisation
que les formes non contextuelles, en déterminant les liens qui
s'établissent entre les mots (par exemple à l'aide de \emph{grammaires
de liens} \cite{Lafferty92, Berger98b}) ou en exprimant les
dépendances par des contraintes syntaxiques et sémantiques (on parle
alors de \emph{grammaires de dépendance par contraintes}
\cite{Harper95, Wang02b}). Si les règles de toutes ces grammaires sont
généralement déterminées \textit{a priori}, l'évaluation des
probabilités se fait automatiquement soit à partir de corpus annotés
syntaxiquement tels que le Penn Treebank, soit à partir de corpus
analysés par des méthodes automatiques.

Le principal avantage de ce type de connaissance est de prendre en
compte les dépendances syntaxiques entre les constituants d'un même
groupe de souffle, et ce, même si ces constituants se trouvent à des
positions assez éloignées. Le calcul des probabilités intègre alors
des informations plus précises que l'influence des $N-1$ mots
précédents sur le mot courant, comme cela est le cas des modèles
N-grammes.

La difficulté la plus importante à laquelle se trouvent confrontées
les méthodes d'analyse syntaxique concerne la conception de grammaires
suffisamment robustes. Il est déjà difficile d'élaborer pour l'écrit
des analyseurs syntaxiques ayant une large couverture, même si
certaines grammaires lexicalisées parviennent à une
précision\footnote{Définie par $\frac{\textrm{nb de constituants
communs à GOLD et à TEST}} {\textrm{nb de constituants dans TEST}}$,
où GOLD et TEST sont les arbres d'analyse obtenus respectivement
manuellement et automatiquement.}  et un rappel\footnote{Défini par
$\frac{\textrm{nb de constituants communs à GOLD et à TEST}}
{\textrm{nb de constituants dans GOLD}}$.} proches de 90\,\% pour
analyser une partie du corpus du \textit{Wall Street Journal}
\cite{Charniak00}. Les difficultés intrinsèques de l'oral
(\textit{cf.}  section~\ref{secCaractLangParle}), notamment le manque
de ponctuation, la présence de disfluences et l'absence éventuelle de
majuscules dans la transcription, ajoutées aux erreurs de
reconnaissance des systèmes de RAP, compliquent encore la réalisation
d'analyseurs. L'utilisation de l'analyse syntaxique par les systèmes
de RAP a, pour ces raisons, longtemps été confinée à des applications
homme-machine où les tournures de phrase et le vocabulaire étaient
très spécifiques \cite{Seneff92}.  Les méthodes d'analyse partielle,
qui ne nécessitent pas de construire un arbre syntaxique décrivant la
structure détaillée du groupe de souffle entier, sont particulièrement
adaptées pour concevoir des solutions robustes. Un ML a ainsi été
défini en segmentant les suites de mots à analyser en constituants non
récursifs, que l'on appelle \emph{chunks} \cite{Zechner98}.

La prise en compte de l'analyse syntaxique au niveau du ML s'envisage
différemment selon que l'on associe ou non des probabilités aux règles
de la grammaire. Dans le cas des versions non probabilistes, on
distingue trois types d'intégration pour guider le choix des
hypothèses de transcription. Une solution simple consiste à utiliser
l'analyse pour filtrer les N meilleures hypothèses fournies les unes
après les autres par le système de RAP, en arrêtant dès qu'une
hypothèse a pu être analysée entièrement. Une deuxième solution repose
sur le calcul du nombre de mots du groupe de souffle couverts par
l'analyse. Plus ce nombre est grand, meilleure est considérée
l'hypothèse \cite{Zechner98}. Dans une dernière solution, un modèle
N-grammes est construit en examinant les séquences de constituants de
l'analyse et non plus celles de mots. Dans le cas où on utilise des
\textit{chunks}, l'examen du groupe de souffle \og [{\footnotesize NP}
\textit{you}] [{\footnotesize VC} \textit{weren't born}]
[{\footnotesize ADVP} \textit{just}] [{\footnotesize NP} \textit{to
soak up}] [{\footnotesize NP} \textit{sun}]\footnote{Où {\footnotesize
NP} = groupe nominal, {\footnotesize VC} = verbe complexe et
{\footnotesize ADVP} = groupe adverbial.}\fg\ conduira ainsi à l'étude
de la séquence \og {\footnotesize NP} {\footnotesize VC}
{\footnotesize ADVP} {\footnotesize VC} {\footnotesize NP} \fg\
\cite{Zechner98}.

L'intégration des versions probabilistes dans les systèmes de RAP
dépend quant à elle du type de grammaire envisagé. En ce qui concerne
les grammaires non contextuelles probabilistes, il existe des
algorithmes rapides \cite{Jelinek91, Stolcke95} permettant d'estimer
avec exactitude les probabilités $P(w_1^i) = P(S
\stackrel{*}\rightarrow w_1 w_2 \ldots w_i \ldots)$ des chaînes
préfixes $w_1^i$. Ces techniques peuvent être utilisées pour définir
des probabilités de modèles bigrammes à partir de grammaires
\cite{Stolcke94, Jurafsky95}, en considérant que :
\begin{equation}
P(w_i|w_1^{i-1}) = \frac{P(w_1^i)}{P(w_1^{i-1})}
\end{equation}

Un autre procédé introduisant une grammaire non contextuelle
probabiliste dans un ML consiste à la convertir en un réseau
stochastique. Dans le système TINA, les règles $X \rightarrow A B C$
et $X \rightarrow B C D$ sont par exemple transformées en un graphe
(Fig.~\ref{figTINA}) dont les arcs sont valués par des probabilités
apprises sur des exemples et dépendant de la partie gauche $X$ de la
règle utilisée ainsi que du mot qui vient d'être rencontré. Cette
méthode présente l'avantage de fournir des probabilités explicites
d'un mot, étant donné une séquence de mots \cite{Seneff92, Seneff95}.
\FigurePS{!htbp}{6.5cm}{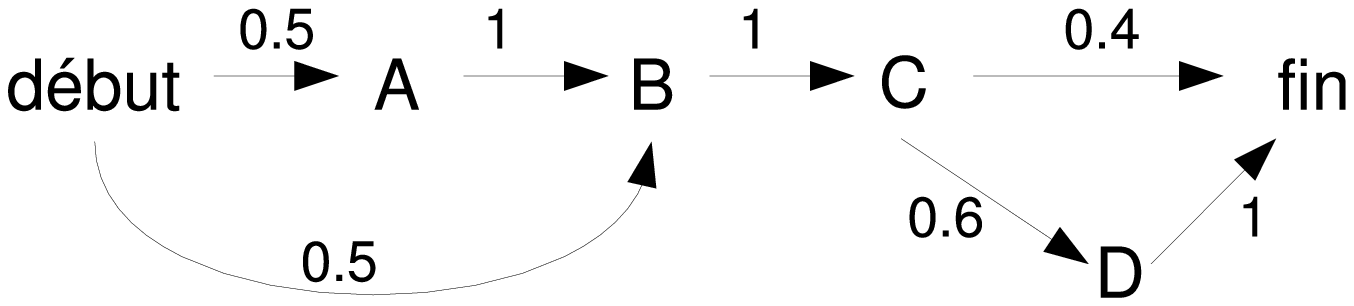}{Réseau stochastique obtenu
à partir de 2 règles ayant la même partie gauche}{figTINA}

Une autre technique, employée aussi bien par les grammaires non
contextuelles que par les lexicalisées, repose sur les probabilités
associées aux arbres de dérivation \cite{Linares04, Roark01,
Charniak01}. Si $D_{w_1^i}$ représente l'ensemble des dérivations
associées à $w_1^i$, on peut estimer $P(w_1^{i})$ par :
\begin{equation}
P(w_1^i) = \sum_{d \in D_{w_1^i}} P(d)
\end{equation}
où $P(d)$ est obtenue en faisant le produit des probabilités de toutes
les règles utilisées dans la dérivation $d$. Il en résulte que :
\begin{equation}
\label{eqDervationsProba}
P(w_i|w_1^{i-1}) = \frac {P(w_1^i)} {P(w_1^{i-1})} = \frac {\sum_{d \in
    D_{w_1^i}} P(d)} {\sum_{d \in D_{w_1^{i-1}}} P(d)}
\end{equation}
En pratique, l'ensemble $D_{w_1^i}$ peut être très grand, ce qui
conduit à utiliser une pile conservant uniquement les dérivations
les plus probables.

Les modèles de langages structurés \cite{Chelba00}, à base d'une grammaire
lexicalisée, utilisent un mode de calcul légèrement différent de
l'équation~(\ref{eqDervationsProba}) :
\begin{equation}
P(w_i|w_1^{i-1}) = \frac {\sum_{d \in D_{w_1^{i-1}}} P(w_i, h_{-1,d},
  h_{0,d}) P(d)} {\sum_{d \in D_{w_1^{i-1}}} P(d)}
\end{equation}
où $h_{-1,d}$ et $h_{0,d}$ représentent les têtes des deux derniers
constituants de la dérivation $d$. Ce calcul a l'avantage de prendre
explicitement en compte des dépendances à longue distance. Dans
l'exemple de la figure~\ref{figMLstruct}, le mot \og \textit{after}
\fg\ est ainsi prédit à partir de \og \textit{contract} \fg\ et \og
\textit{ended} \fg, et non à partir de \og \textit{7} \fg\ et \og
\textit{cents} \fg\ comme cela serait le cas avec un ML trigramme
classique.  \FigurePS{!htbp}{12cm}{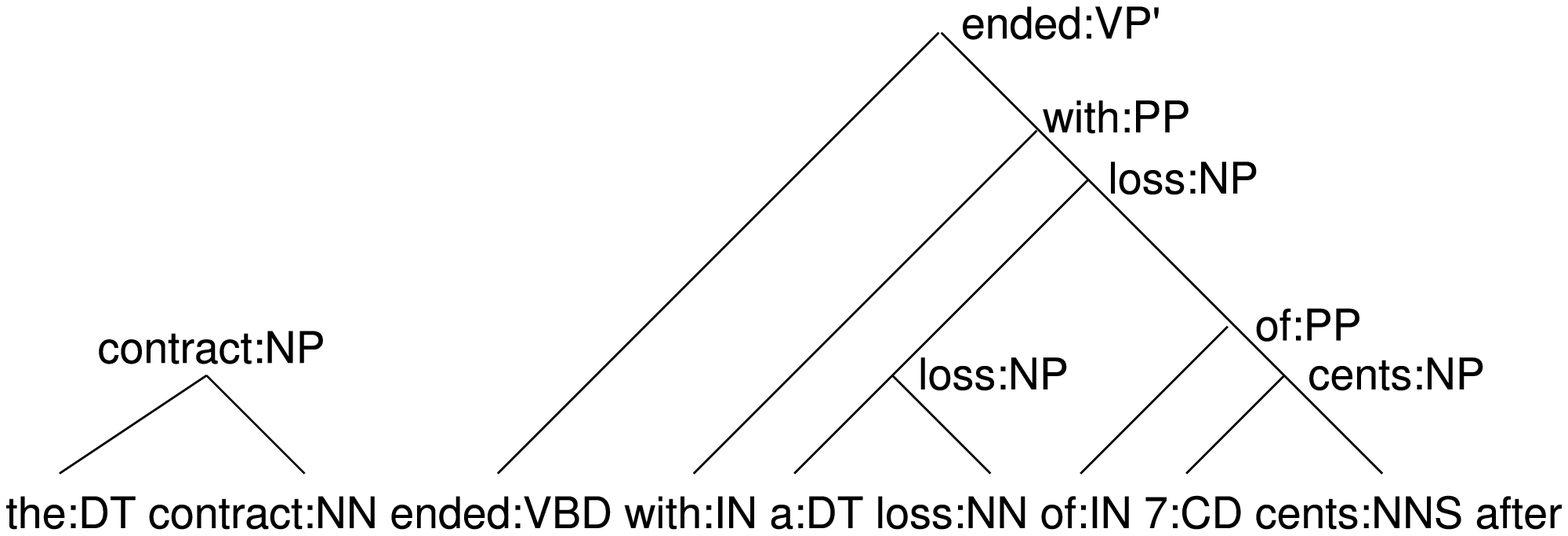}{Arbre de
dérivation partielle dont les feuilles sont des mots accompagnés de
leurs PoS et dont les n\oe uds sont annotés par la tête et le type du
constituant auquel ils sont associés
(d'après~\cite{Chelba00})}{figMLstruct}

Il existe, selon les méthodes d'analyse syntaxique employées, des
façons encore différentes d'intégrer les connaissances syntaxiques au
ML. C'est par exemple le cas pour l'analyse par grammaire de liens qui
utilisent d'autres méthodes probabilistes comme les modèles
exponentiels (\textit{cf.} section~\ref{subSubSecLingML})
\cite{Berger98b}. La fonction de contraintes déterminée sur les liens
du groupe de souffle s'exprime ici sous la forme :
\begin{equation}
f_{u \frown v}(w_1^i) = \left\{ \begin{array}{ll}
1 & \textrm{si } w = v \textrm{ et si } \exists u \in w_1^{i-2}
\textrm{ tel que } u \textrm { et } v \textrm{ sont liés}\\
0 & \textrm{sinon }
\end{array} \right.
\end{equation}
La prédiction est réalisée à partir des liens établis avec le contexte
gauche, mais elle peut également être faite en prenant en compte à la
fois les contextes gauche et droit \cite{Lafferty92}. Les modèles
construits à partir de ce type de grammaire sont très similaires aux
ML à base de mots \textit{triggers} \cite{Tillmann97}. Les paires de
mots \textit{triggers} sont deux mots qui apparaissent souvent dans le
même contexte, tels que \og \textit{demande} \fg\ et \og
\textit{répond} \fg, et qui peuvent être assimilés à des bigrammes
longue distance. Les relations entre mots \textit{triggers}
s'apparentent à des liens établis par des grammaires de liens mais
elles présentent la particularité d'être déterminées \textit{a priori}
et non en fonction de l'analyse du groupe de souffle courant.

Enfin, les grammaires de dépendance par contraintes font quant à elles
appel à des méthodes proches de celles rencontrées pour intégrer les
PoS dans un ML. Le principe de ce type d'analyse est d'attribuer à
chaque mot un rôle en fonction de sa position dans le groupe de
souffle. Ce rôle, qui contient des informations lexicales, syntaxiques
mais aussi sémantiques, est assimilable à une super-étiquette
(Fig.~\ref{figSuperEtiq}), plus informative qu'une PoS, ce qui permet
de prendre en compte ces grammaires en utilisant des modèles
N-classes, où chaque classe est associée à une super-étiquette
\cite{Wang02b}.

\begin{figure}
\begin{center}
\begin{otherlanguage*}{english}
\begin{tabular}{l}
\hline Category: Verb\\ 
\hline Features: \{verbtype=past, voice=active,
inverted=yes, gapp=yes, mood=whquestion, \\
agr=all\}\\ 
\hline
Role=G, Label=VP, (PX $>$ MX) (\textit{Gouverné par un mot à sa
gauche}) \\
Role=Need1, Label=S, (PX $<$ MX) (\textit{Nécessite un modifieur à sa
droite}) \\
Role=Need2, Label=S, (PX $<$ MX) (\textit{Nécessite un modifieur à sa
droite}) \\
Role=Need3, Label=S, (PX = MX) (\textit{Pas de modifieur})\\
\hline
Dependent Positional Constraints: \\
MX[G] $<$ PX = MX[Need3] $<$ MX[Need1] $<$ MX[Need2] \\
\hline
\end{tabular}
\end{otherlanguage*}
\end{center}
\caption{Super-étiquette de \og \textit{did} \fg\ dans le groupe de
  souffle \og \textit{what did you learn}\fg, où $G$ représente le
  rôle de gouverneur, $Need1$, $Need2$ et $Need3$ représentent des
  contraintes grammaticales sur le mot \og \textit{did} \fg, $PX$ et
  $MX$ représentent respectivement la position de \og \textit{did}
  \fg\ et d'un de ses modifieurs (d'après \cite{Wang02})}
\label{figSuperEtiq}
\end{figure}

\paragraph{}
Les grammaires ont été principalement intégrées dans les systèmes de
RAP pour des applications ciblées, du fait de leur manque de
robustesse et de leur lenteur. Les non contextuelles probabilistes ont
ainsi permis de diminuer le taux d'erreur dans les domaines de
réservations pour le trafic aérien (de 34,6\,\% à 29,6\,\% par rapport
à un modèle bigramme) \cite{Jurafsky95} et de la restauration (de
6,9\,\% à 6,7\,\% par rapport à un modèle quadrigramme)
\cite{Seneff95}. Les grammaires lexicalisées ont quant à elles conduit
à une baisse de la perplexité par rapport aux modèles trigrammes sur
un corpus du \textit{Wall Street Journal} \cite{Chelba00, Roark01,
Charniak01}. Les progrès récents de l'analyse syntaxique, obtenus en
prenant en compte un contexte de plus en plus grand lors de
l'attribution des probabilités à chacune des règles, ont ainsi
autorisé son utilisation sur de la parole lue.  En outre, leur
combinaison avec des modèles N-grammes a fait baisser davantage encore
la perplexité, ce qui semble indiquer qu'elles apportent des
connaissances supplémentaires à ces modèles. Les modèles conçus avec
ces grammaires sont cependant trop coûteux en calculs pour être
utilisés au sein de systèmes de RAP dans des applications à grande
échelle, même si l'un de ces modèles a pu être intégré dans un ML
multipasses en rescorant des graphes de mots
\cite{Hall04}. L'intégration des grammaires de dépendance par
contraintes semble quant à elle plus aisée puisqu'elle a permis
d'obtenir des ML de complexité raisonnable, et ce, en constatant une
diminution du taux d'erreur pour transcrire des émissions d'actualité
(de 14,7\,\% à 14,3\,\% par rapport à un modèle trigramme)
\cite{Wang03} mais aussi des conversations téléphoniques (avec une
baisse relative de 6,2\,\% par rapport à des ML 4-grammes et 4-classes
interpolés) \cite{Wang04}.

\subsubsection{Sémantique}
\label{subSecSemantique}

L'introduction de connaissances sémantiques vise à favoriser les
hypothèses qui possèdent plusieurs mots proches au niveau de leur
sens, en supposant que les groupes de souffle à reconnaître ont une
certaine cohérence sémantique. On attribuera ainsi un meilleur score à
des hypothèses de décodage contenant les mots \og \textit{action}
\fg\, \og \textit{obligation}\fg\ et \og \textit{bourse} \fg\ qu'à une
hypothèse n'ayant pas de champ sémantique bien défini.

Une première possibilité pour introduire ce type d'information repose
sur l'utilisation de connaissances \textit{a priori}. Un dictionnaire,
contenant les domaines d'emplois et les définitions des sens de 36~000
lemmes en anglais, a ainsi permis d'établir des associations
sé\-man\-tiques $S(x,y)$ entre paires de mots \cite{Demetriou97}. Pour
ne pas avoir à désambiguïser les mots, les descriptions sémantiques de
tous les sens associés à un même mot sont fusionnées. $S(x,y)$ est
alors fonction du nombre de mots en commun que possèdent les
descriptions sémantiques de $x$ et $y$. Un ML a été construit en
calculant pour chaque groupe de souffle $W$ un score $Score(W)$ :
\begin{equation}
Score(W) = \frac{1}{k} \sum_{i \ne j} S(w_i,w_j)
\end{equation}
où $k$ est un facteur de normalisation dépendant de la taille de $W$.
Ce ML n'a pas été comparé à des modèles N-grammes mais des expériences
ont montré qu'il était informatif lorsque le contexte pris en compte
était important, \textit{i.e.}, que le groupe de souffle à analyser
était long.

Une autre technique étudiée pour introduire des connaissances
sémantiques est l'\emph{analyse sémantique latente}
\cite{Bellegarda98}. Le principe de cette méthode, utilisée en
recherche d'information, est de trouver les relations sémantiques qui
s'établissent entre les mots d'un document. Pour ce faire, elle
suppose que deux mots sont proches s'ils ont tendance à apparaître
dans les mêmes documents. $W$, la matrice d'occurrences de mots dans
des documents d'un corpus d'apprentissage, est calculée de façon à ce
que $w_{ij}$ représente le nombre d'occurrences pondéré du mot $w_i$
dans le document $d_j$. La pondération utilise un coefficient
dépendant de $w_i$, qui vise à traduire que deux mots apparaissant
avec les mêmes nombres d'occurrences dans $d_j$ ne portent pas
nécessairement autant d'information ; cela dépend également de leur
distribution dans la collection entière de documents, ceux
apparaissant dans beaucoup de documents étant considérés comme moins
informatifs. Afin de réduire ses dimensions, $W$ est transformée dans
un espace $\mathcal{S}$ de dimension $R \times R$ (avec $100 \le R \le
200$), par des techniques proches de l'analyse en composantes
principales :
\begin{equation}
\label{eqDecompASL}
W = U S V^T
\end{equation}
où $S$ est une matrice diagonale de dimension $R \times R$, $U$ est
une matrice dont les lignes $u_i$ sont les représentations de chaque
mot $w_i$ dans $\mathcal{S}$ et $V$ est une matrice dont les lignes
$v_j$ sont les projections de chaque document $d_j$ dans
$\mathcal{S}$. D'après la définition de $W$, la manière dont $w_i$ et
$d_j$ sont associés est déterminée par $w_{ij}$. Ceci peut être
également caractérisé selon l'équation~(\ref{eqDecompASL}) par le
produit scalaire de $u_i S^{1/2}$ et $v_j S^{1/2}$, et une distance
mesurant la proximité entre $u_i$ et $v_j$, \textit{i.e.}, entre $w_i$
et $d_j$, est :
\begin{equation}
K(u_i, v_j) = cos(u_i S^{1/2}, v_j S^{1/2})
\end{equation}

L'analyse sémantique latente permet de définir sur ce principe un
nouveau type de ML \cite{Bellegarda98, Bellegarda00}. Le calcul des
probabilités $P(w_i|h_i)$ est établi en considérant que $h_i$
représente le document courant $\tilde{d}_{i-1}$ jusqu'au mot
$w_i$. $\tilde{d}_{i-1}$ pouvant être vu comme une colonne
supplémentaire de $W$, on calcule sa représentation $\tilde{v}_{i-1}$
dans $\mathcal{S}$. $P(w_i, \tilde{d}_{i-1})$ est alors déterminée à
partir de la mesure de proximité $K(w_i, \tilde{d}_{i-1})$ entre $w_i$
et l'historique $\tilde{d}_{i-1}$, en réalisant une normalisation pour
obtenir une probabilité bien définie. Grâce à ce mode de calcul,
$P(w_i, \tilde{d}_{i-1})$ est plus grande pour les mots dont le sens
se rapproche le plus de celui des mots de l'historique et plus petite
dans le cas contraire. Pour avoir des informations sur l'ordre des
mots, le calcul des probabilités prend en outre en considération les
N-grammes. L'analyse sémantique latente est une technique intéressante
puisque ce type de ML a permis une baisse relative du taux d'erreur de
16\,\% par rapport à des ML trigrammes pour transcrire de la parole
lue en anglais \cite{Bellegarda00}.

\subsubsection{Pragmatique}
\label{subSecPragmatique}

L'introduction de la pragmatique dans la RAP recouvre un ensemble de
techniques qui visent à adapter le processus de transcription au
contexte d'utilisation. L'adaptation peut se faire de deux façons :
\begin{itemize}
\item soit en concevant des modèles adaptatifs, qui se spécialisent
automatiquement au cours du processus de RAP, en fonction de ce que le
système a déjà reconnu,
\item soit en utilisant des corpus d'adaptation, obtenus par des
systèmes de recherche d'information et plus proches du texte à
transcrire que le corpus d'apprentissage.
\end{itemize}

\paragraph{Modèles adaptatifs\\}
\label{subSubSecMLAdapt}

Il existe plusieurs méthodes d'adaptation des ML, parmi lesquelles on
retrouve les modèles à base de cache (\textit{cf.}
section~\ref{subSubSecModifHisto}) et les modèles à base de mots
\textit{triggers} (\textit{cf.} section~\ref{ssSecASyntax}).  Dans le
cas des modèles à base de cache, les probabilités $P(w_i|h_i)$ sont
réestimées en fonction des mots contenus dans le cache, en supposant
que les mots qui sont apparus récemment voient leur probabilité
d'apparition augmenter. Les modèles classiques à base de cache
n'apportant pas un gain significatif, il en existe plusieurs
variantes. Il a été par exemple envisagé de ne considérer dans le
cache que les mots rares \cite{Rosenfeld94}, ou bien encore de faire
décroître les probabilités du cache de manière exponentielle en
fonction de la distance entre le mot courant et les apparitions
précédentes de ce mot \cite{Clarkson97}.

Si ces modèles modifient uniquement les probabilités des mots présents
dans le cache, ceux à base de mots \textit{triggers} font l'hypothèse
que certains mots présents dans l'historique ont une influence sur
l'apparition des mots auxquels ils sont corrélés. Le mot \og
\textit{avion}\fg\ pourra par exemple favoriser la prédiction du mot
\og \textit{vol}\fg. Ce type de modèle peut ainsi prendre en compte
des dépendances entre mots distants dans le document à transcrire
\cite{Rosenfeld96}.

Un autre type de ML adaptatif repose sur la détection du thème traité
par le document à transcrire. Il est en effet constaté que pour des
documents sonores abordant plusieurs thèmes, notamment les émissions
d'actualité, les tournures de phrase et les termes employés diffèrent
selon le sujet traité. Dans cette approche, le corpus d'apprentissage
est divisé en plusieurs ensembles correspondant chacun à un
sous-langage ou un thème ; un ML thématique est alors construit pour
chacun de ces ensembles. Pour partitionner le corpus d'apprentissage,
il est généralement supposé que chaque document du corpus est associé
à un unique thème. Ces documents sont le plus souvent regroupés en
utilisant des méthodes automatiques non supervisées \cite{Iyer99,
Clarkson97, Martin97, Florian99, Gildea99} mais peuvent être aussi
classés grâce à des méthodes automatiques supervisées \cite{Lane05,
Khudanpur99} ou de manière manuelle en annotant chaque document
\cite{Kneser93, Brun03}. Le nombre de groupes obtenu à partir du
corpus d'apprentissage est très variable. Il peut par exemple être
inférieur à 10 si les thèmes détectés sont généraux, comme l'économie,
l'histoire ou la politique \cite{Brun03, Bigi00} ou être de plus de
5~000, permettant ainsi une distinction plus fine du sujet abordé
\cite{Seymore97}.

Il existe plusieurs procédés pour utiliser les modèles adaptatifs
basés sur la détection de thèmes. Tout d'abord, la reconnaissance du
sujet abordé par le document à transcrire peut se faire à différents
niveaux. Elle est souvent réalisée à partir du document entier pour
limiter l'influence des erreurs sur les mots reconnus, puisque ce sont
des hypothèses de transcription, susceptibles d'être partiellement
erronées, qui sont utilisées pour la détection du thème. Elle peut
néanmoins être effectuée au niveau de chaque groupe de souffle. Malgré
une difficulté plus importante, ce type de détection conduit dans
certains cas à une plus grande baisse du taux d'erreur de la
transcription \cite{Khudanpur99}. Pour limiter les erreurs de
détection du thème sur un groupe de souffle, les sujets peuvent être
hiérarchisés ; on utilise alors des modèles correspondant à des thèmes
plus ou moins précis selon la confiance attribuée au groupe de souffle
\cite{Lane05}. Un autre niveau de segmentation suppose de découper le
document à transcrire pour obtenir des ensembles consécutifs de
groupes de souffle associés à un seul thème. Les segments obtenus
étant plus longs, cette solution présente l'avantage de limiter
l'influence des erreurs de transcription, tout en détectant des thèmes
plus précis qu'en prenant en compte l'ensemble du document
\cite{Chen01}.

Les modèles thématiques diffèrent en outre selon la manière dont ils
sont combinés pour calculer les probabilités. Le ML correspondant au
thème détecté à partir d'une première hypothèse de transcription peut
être directement utilisé dans la deuxième passe du système de RAP
\cite{Lane05}. Cependant, les modèles spécifiques à un thème ayant été
appris sur des corpus de taille réduite, ils sont généralement
combinés avec le modèle général, construit à partir de l'ensemble du
corpus d'apprentissage. Cette combinaison peut être réalisée au moyen
d'une interpolation linéaire \cite{Brun03, Iyer99, Seymore97} ou d'un
modèle exponentiel (\textit{cf.} section~\ref{subSubSecLingML})
\cite{Khudanpur99}. L'avantage du dernier est qu'il modifie, pour
chaque thème, les probabilités du modèle général pour un nombre limité
de mots, ce qui permet de ne pas accroître beaucoup la taille du
ML. Des méthodes propres à l'adaptation ont été de plus conçues pour
combiner plusieurs modèles spécifiques à un thème. Si $t_k$ est un
thème présent dans le corpus d'apprentissage, les ML dits à
\emph{mélange de modèles} calculent les probabilités de la manière
suivante :
\begin{equation} \label{eqMelModeles}
P(w_i|w_1^{i-1}) = \sum_k \lambda_k (w_1^{i-1}) P(w_i|w_1^{i-1}, t_k)
\end{equation}
Ce procédé diffère de l'interpolation linéaire dans la mesure où les
paramètres $\lambda_k$ sont déterminés dynamiquement en fonction des
mots rencontrés précédemment \cite{Kneser93, Martin97, Clarkson97}. Il
existe une autre variante des mélanges de modèles, consistant à
évaluer les paramètres $\lambda_k$ au niveau des phrases (ou groupes
de souffles) et non pas au niveau des N-grammes \cite{Iyer99}. Notons
que les coefficients $\lambda_k$ peuvent être assimilés aux
probabilités $P(t_k|w_1^{i-1})$ \cite{Florian99, Gildea99}. L'intérêt
de ce type de modèle est de pouvoir associer plusieurs thèmes au
document à transcrire.

D'une manière générale, les modèles adaptatifs ont permis une
réduction significative de la perplexité, mais cette réduction s'est
traduite par une baisse limitée du taux d'erreur de transcription. Une
propriété intéressante des modèles thématiques est qu'ils semblent
apporter des informations complémentaires à celles fournies par une
analyse syntaxique. En combinant ces deux sources d'informations, il a
été ainsi constaté que les gains de chaque source sont presque
additifs en ce qui concerne la baisse du taux d'erreur \cite{Wu99}.

Une technique alternative d'adaptation consiste à utiliser des corpus
spécialisés en fonction du texte à transcrire.

\paragraph{Utilisation de corpus d'adaptation\\}

Les ML statistiques sont particulièrement sensibles à l'adéquation du
corpus d'apprentissage vis-à-vis du type de document à
transcrire. Dans le cas d'émissions d'actualité, le poids attribué aux
transcriptions manuelles de programmes radio est ainsi beaucoup plus
important, eu égard à son volume de données, que celui de journaux
écrits. On voit ainsi l'intérêt d'avoir un corpus adapté au document à
transcrire.

Le corpus d'adaptation peut être déjà disponible, ce qui correspond à
une adaptation \emph{statique}. Par rapport au corpus général, ce
corpus se rapporte davantage au domaine des textes à transcrire mais
sa taille réduite, souvent de l'ordre de quelques milliers de mots, ne
lui permet pas d'être directement utilisable comme corpus
d'apprentissage du ML. Les paramètres du ML sont alors calculés
principalement à partir du corpus général et, dans les situations où
le volume de données est suffisant, à partir du corpus spécialisé
\cite{Gao00, Federico99}.

Le corpus d'adaptation peut être aussi régulièrement mis à jour ; on
parle alors d'adaptation \emph{dynamique}. Cette mise à jour est
effectuée soit à partir de documents qui ont été conçus durant la même
période que ce qui est à transcrire \cite{Allauzen03b, Kemp98}, soit à
partir de documents dont le contenu est similaire à ce que l'on
souhaite décoder \cite{Berger98, Chen03, Bigi04}. Dans le deuxième
cas, il est fait appel à un système de recherche d'information (SRI) ;
un ML général établit alors une première hypothèse de transcription
utilisée par le SRI pour retourner les documents les plus
proches. L'ensemble des résultats de la recherche forme un corpus
d'adaptation, permettant de rectifier le ML, et le ML modifié est
utilisé dans une deuxième passe du processus de transcription.

La recherche d'information peut se faire à partir de sources de types
différents. Une recherche parmi l'ensemble de documents du corpus
d'apprentissage permet d'avoir un corpus spécialisé, ce qui est
intéressant quand le corpus d'apprentissage se rapporte à plusieurs
thèmes \cite{Chen03}. Une autre possibilité repose sur l'utilisation
d'Internet \cite{Bigi04}. Cette vaste source d'informations évolutive
présente également, au sein de certains sites tels que les blogs ou
les serveurs de news, des caractéristiques proches de la langue parlée
(\textit{cf.} section~\ref{subSecLPetLE}) \cite{Vaufreydaz99}.

Le système de RAP prend en compte les corpus d'adaptation de deux
manières différentes :
\begin{itemize}
\item en modifiant son vocabulaire,
\item en adaptant le calcul des probabilités du ML au corpus
  d'adaptation.
\end{itemize}

L'adaptation du vocabulaire est particulièrement pertinente pour
transcrire des é\-mis\-sions d'actualité, des entités nom\-mées
ap\-paraissant au gré des évé\-ne\-ments \cite{Kemp98, Allauzen03b,
Bigi04}. La col\-lecte ré\-gulière d'in\-for\-ma\-tions sur des sites
de dépêches d'agences de presse ou de quotidiens nationaux permet de
réduire le taux de mots hors vocabulaire. Pour ne pas augmenter la
taille du vocabulaire, des mots présents initialement sont supprimés
pour laisser place à des mots apparaissant souvent dans l'actualité
récente. L'ajout d'un mot au système de RAP nécessite à la fois
d'associer une transcription phonétique à ce mot et de l'inclure dans
les distributions N-grammes du ML. La modification du vocabulaire
présente donc l'inconvénient de devoir effectuer un réapprentissage
fréquent des MA et des ML. Toutefois, une méthode a permis de
construire un système de RAP à vocabulaire ouvert, avec une diminution
constatée du taux d'erreur de 25,5\,\% à 24,9\,\% pour la
transcription d'émissions d'actualité en français \cite{Allauzen05}.

Le calcul des probabilités du ML peut être modifié selon plusieurs
procédés pour prendre en compte le corpus d'adaptation. Une première
possibilité consiste à utiliser les mélanges de modèles thématiques
(\textit{cf.}  section~\ref{subSubSecMLAdapt}). Les paramètres
$\lambda_k$ de l'équation~(\ref{eqMelModeles}) sont alors appris non
plus à partir des mots transcrits précédemment comme dans le cas des
modèles adaptatifs, mais à partir du corpus d'adaptation. Le mélange
de modèles peut se faire en outre de manière dynamique en adaptant le
calcul des probabilités $P(w_i|w_1^{i-1}, t_k)$ au corpus d'adaptation
\cite{Chen04b}. D'autres techniques consistent à spécialiser un ML
général sur le corpus d'adaptation en utilisant un critère de maximum
\textit{a posteriori} (MAP) \cite{Federico96, Berger98, Chen04b} ou un
critère de minimum d'information discriminante (MDI pour
\textit{Minimum Discrimination Information}) \cite{Federico99,
Chen04b}. L'utilisation de corpus d'adaptation permet une réduction
non négligeable du taux d'erreur de la transcription. Il a ainsi été
constaté une baisse de 17,1\,\% à 16,3\,\% dans la transcription
d'émissions d'actualité en anglais. Les adaptations au moyen de
mélanges dynamiques de modèles et du critère MDI semblent être les
procédés les plus performants \cite{Chen04b}.

\newpage

\section{Conclusion}
\label{SecConclusion}

La description du principe de fonctionnement de la transcription met
en évidence que les systèmes de RAP actuels s'appuient sur une
modélisation statistique. Ce type de conception a conduit à la
construction de MA à base de HMM et de ML N-grammes. Il a permis de
réaliser différents systèmes capables de transcrire des émissions
d'actualité mais aussi des dialogues spontanés, même si le traitement
de ce dernier type de documents est rendu problématique par la
présence non négligeable de phénomènes de la langue parlée tels que
les disfluences. Les connaissances linguistiques prises en compte se
limitent bien souvent à la conception d'un lexique de prononciations
et à l'apprentissage de ML N-grammes, ce qui laisse penser que des
améliorations de la qualité de la transcription sont possibles en
exploitant davantage d'informations sur le langage.

La présentation que nous avons faite des ML N-grammes et leurs
variantes a montré que deux limitations sont souvent mises en avant
quant à leur capacité à donner une valeur correcte aux probabilités de
séquences de mots. D'une part, les ML N-grammes sont basés sur
l'hypothèse un peu simpliste et arbitraire, mais conduisant à des
méthodes de calculs rapides, qui est d'examiner uniquement les $N-1$
mots précédents pour prédire le mot courant. Afin de remédier à ce
premier point, des essais ont été conduits de façon à intégrer des
connaissances sup\-plé\-men\-taires telles que les dépendances
syntaxiques ou les similarités sémantiques entre les mots d'un groupe
de souffle. Des méthodes d'adaptation tentent également de prendre en
compte des relations entre différents groupes de souffle. D'autre
part, malgré des techniques de lissage perfectionnées, les calculs
pour prédire des événements rares voire même absents du corpus sont
imprécis. Pour tenter de corriger ce problème, des études ont réuni
des mots possédant la même partie du discours ou le même lemme au sein
d'une même classe, de manière à réduire le nombre d'événements
possibles.

Au final, les résultats obtenus en intégrant des connaissances
linguistiques sup\-plé\-men\-taires montrent que les améliorations en
terme de taux d'erreur sur les mots reconnus sont gé\-né\-ra\-le\-ment
assez peu significatives, d'autant plus que les nouveaux ML proposés
ne sont presque jamais comparés avec des ML 5-grammes utilisant un
lissage performant \cite{Goodman01}. Parmi les raisons qui expliquent
cet état de fait, les nouvelles informations introduites par ces
méthodes sont souvent redondantes avec les connaissances déjà
apportées par les ML N-grammes de mots. En outre, les particularités
des transcriptions produites automatiquement, notamment la flexiblité
de la langue parlée, la segmentation en groupes de souffle ou encore
les erreurs de reconnaissance, viennent compliquer la conception de
méthodes extrayant automatiquement des connaissances linguistiques.
De plus, les techniques employées ont souvent le défaut d'augmenter
considérablement le temps de décodage du signal acoustique, ce qui
fait qu'elles sont utilisées principalement au niveau de la dernière
passe du processus de transcription.

Quelques méthodes apparaissent toutefois prometteuses pour corriger
certaines erreurs de transcription : les ML utilisant des grammaires
lexicalisées probabilistes, bien qu'ils soient encore trop coûteux au
niveau des calculs, l'introduction de connaissances sémantiques, les
modèles thématiques ou encore l'utilisation de corpus
d'adaptation. Les modèles N-classes, qui sont rapides lors de leur
utilisation, peuvent également réduire le taux d'erreur. La
combinaison de plusieurs types de connaissances semble de plus
souhaitable pour apporter des informations complémentaires.

Notons enfin que cette synthèse s'est limitée à l'amélioration de la
qualité de la transcription mais que le couplage TAL-RAP peut
également s'exprimer à l'issue de la reconnaissance. De nombreuses
recherches s'intéressent ainsi actuellement au repérage d'entités
nommées dans les textes produits, à la détection de thèmes ou encore à
la réalisation de résumés.

\newpage

\begin{otherlanguage*}{english}

\addcontentsline{toc}{section}{Références}
\bibliography{these}

\newcommand{\etalchar}[1]{$^{#1}$}
\begin{thebibliography}{BDMADG03}

\bibitem[AB05]{Alain05}
P.~\bgroup\sc Alain\egroup{} \andname{} O.~\bgroup\sc Boëffard\egroup{}.
\newblock \og Évaluation des modèles de langage {N}-gramme et
  {N}/{M}-multigramme\fg.
\newblock \Inname{} {\em Actes de la 12ème conférence sur le Traitement
  Automatique des Langues Naturelles (TALN)}, \volumename{}~1, Dourdan, France,
  2005.

\bibitem[ADHB{\etalchar{+}}04]{AddaDecker04}
M.~\bgroup\sc Adda-Decker\egroup{}, B.~\bgroup\sc Habert\egroup{},
  C.~\bgroup\sc Barras\egroup{}, G.~\bgroup\sc Adda\egroup{}, P.~\bgroup\sc
  {Boula de Mareüil}\egroup{} \andname{} P.~\bgroup\sc Paroubek\egroup{}.
\newblock \og Une étude des disfluences pour la transcription automatique de la
  parole spontanée et l'amélioration des modèles de langage\fg.
\newblock \Inname{} {\em Actes des 25èmes Journées d'Études sur la Parole
  (JEP)}, Fès, Maroc, 2004.

\bibitem[AG99]{Antoine99}
J.-Y. \bgroup\sc Antoine\egroup{} \andname{} D.~\bgroup\sc Genthial\egroup{}.
\newblock \og Méthodes hybrides issues du TALN et du TAL Parlé~: état des lieux
  et perspectives\fg.
\newblock \Inname{} {\em Actes de la 6ème conférence sur le Traitement
  Automatique des Langues Naturelles (TALN)}, Cargèse, France, 1999.

\bibitem[AG01]{Antoine01}
J.-Y. \bgroup\sc Antoine\egroup{} \andname{} J.~\bgroup\sc Goulian\egroup{}.
\newblock \og Word Order Variations and Spoken Man-Machine Dialogue in French:
  a Corpus Analysis on the ATIS Domain\fg.
\newblock \Inname{} {\em Proc. of Corpus Linguistics}, Lancaster, Royaume-Uni,
  2001.

\bibitem[AG03]{Allauzen03b}
A.~\bgroup\sc Allauzen\egroup{} \andname{} J.-L. \bgroup\sc Gauvain\egroup{}.
\newblock \og Adaptation automatique du modèle de langage d'un système de
  transcription de journaux parlés\fg.
\newblock {\em Traitement Automatique des Langues (TAL)}, 44(1):11--31, 2003.

\bibitem[AG05]{Allauzen05}
A.~\bgroup\sc Allauzen\egroup{} \andname{} J.-L. \bgroup\sc Gauvain\egroup{}.
\newblock \og Open Vocabulary {ASR} for Audiovisual Document Indexation\fg.
\newblock \Inname{} {\em Proc. of the IEEE International Conference on
  Acoustics, Speech, and Signal Processing (ICASSP)}, \volumename{}~1,
  Philadelphie, Pennsylvanie, États-Unis, 2005.

\bibitem[All94]{Allen94}
J.~B. \bgroup\sc Allen\egroup{}.
\newblock \og How do Humans Process and Recognize Speech?\fg.
\newblock {\em IEEE Transactions on Speech and Audio Processing},
  2(4):567--577, 1994.

\bibitem[AMP{\etalchar{+}}99]{Adda99}
G.~\bgroup\sc Adda\egroup{}, J.~\bgroup\sc Mariani\egroup{}, P.~\bgroup\sc
  Paroubek\egroup{}, M.~\bgroup\sc Rajman\egroup{} \andname{} J.~\bgroup\sc
  Lecomte\egroup{}.
\newblock \og Métrique et premiers résultats de l'évaluation {GRACE} des
  étiqueteurs morphosyntaxiques pour le français\fg.
\newblock \Inname{} {\em Actes de la 6ème conférence sur le Traitement
  Automatique des Langues Naturelles (TALN)}, Cargèse, France, 1999.

\bibitem[BB90]{Blanche-Benveniste90}
C.~\bgroup\sc Blanche-Benveniste\egroup{}.
\newblock {\em Le français parlé~: études grammaticales}.
\newblock Paris~: Éditions du CNRS, 1990.

\bibitem[BB97]{Blanche-Benveniste97}
C.~\bgroup\sc Blanche-Benveniste\egroup{}.
\newblock {\em Approches de la langue parlée en français}.
\newblock Gap - Paris~: Ophrys, 1997.

\bibitem[BCD{\etalchar{+}}04]{Benzitoun04b}
C.~\bgroup\sc Benzitoun\egroup{}, E.~\bgroup\sc Campione\egroup{},
  J.~\bgroup\sc Deulofeu\egroup{}, S.~\bgroup\sc Henry\egroup{}, F.~\bgroup\sc
  Sabio\egroup{}, S.~\bgroup\sc Teston\egroup{}, A.~\bgroup\sc Valli\egroup{}
  \andname{} J.~\bgroup\sc Véronis\egroup{}.
\newblock \og L'analyse syntaxique de l'oral~: problèmes et méthode\fg.
\newblock \Inname{} {\em Actes de la journée d'étude de l'ATALA sur
  l'annotation syntaxique de corpus}, Paris, France, 2004.

\bibitem[BDM98]{Boufaden98}
N.~\bgroup\sc Boufaden\egroup{}, S.~\bgroup\sc Delisle\egroup{} \andname{}
  B.~\bgroup\sc Moulin\egroup{}.
\newblock \og Analyse syntaxique robuste de dialogue retranscrits~: peut-on
  vraiment traiter l'oral à partir de l'écrit~?\fg.
\newblock \Inname{} {\em Actes de la 5ème conférence sur le Traitement
  Automatique des Langues Naturelles (TALN)}, Paris, France, 1998.

\bibitem[BDMADG03]{Boula03}
P.~\bgroup\sc Boula De~Mareüil\egroup{}, M.~\bgroup\sc Adda-Decker\egroup{}
  \andname{} V.~\bgroup\sc Gendner\egroup{}.
\newblock \og Liaisons in French: a Corpus-Based Study Using Morpho-Syntactic
  Information\fg.
\newblock \Inname{} {\em Proc. of the 15th International Congress of Phonetic
  Sciences}, Barcelone, Espagne, 2003.

\bibitem[BDMS00]{Bigi00}
B.~\bgroup\sc Bigi\egroup{}, R.~\bgroup\sc De~Mori\egroup{} \andname{}
  T.~\bgroup\sc Spriet\egroup{}.
\newblock \og Reconnaissance thématique à partir de textes dictés et adaptation
  dynamique de modèles de langage thématiques\fg.
\newblock \Inname{} {\em Actes des 23èmes Journées d'Études sur la Parole
  (JEP)}, Aussois, France, 2000.

\bibitem[BDPd{\etalchar{+}}92]{Brown92}
P.~F. \bgroup\sc Brown\egroup{}, V.~J. \bgroup\sc Della~Pietra\egroup{}, P.~V.
  \bgroup\sc deSouza\egroup{}, J.~C. \bgroup\sc Lai\egroup{} \andname{} R.~L.
  \bgroup\sc Mercer\egroup{}.
\newblock \og Class-Based {N}-Gram Models of Natural Language\fg.
\newblock {\em Computational Linguistics}, 18(4):467--480, 1992.

\bibitem[BDVJ03]{bengio03}
Y.~\bgroup\sc Bengio\egroup{}, R.~\bgroup\sc Ducharme\egroup{}, P.~\bgroup\sc
  Vincent\egroup{} \andname{} C.~\bgroup\sc Jauvin\egroup{}.
\newblock \og A Neural Probabilistic Language Model\fg.
\newblock {\em Journal of Machine Learning Research}, 3(2):1137--1155, 2003.

\bibitem[Bel98]{Bellegarda98}
J.~R. \bgroup\sc Bellegarda\egroup{}.
\newblock \og A Multispan Language Modeling Framework for Large Vocabulary
  Speech Recognition\fg.
\newblock {\em IEEE Transactions on Speech and Audio Processing},
  6(5):456--467, 1998.

\bibitem[Bel00]{Bellegarda00}
J.~R. \bgroup\sc Bellegarda\egroup{}.
\newblock \og Large Vocabulary Speech Recognition with Multispan Statistical
  Language Models\fg.
\newblock {\em IEEE Transactions on Speech and Audio Processing}, 8(1):76--84,
  2000.

\bibitem[Ben04]{Benzitoun04}
C.~\bgroup\sc Benzitoun\egroup{}.
\newblock \og L'annotation syntaxique de corpus oraux constitue-t-elle un
  problème spécifique~?\fg.
\newblock \Inname{} {\em Actes de la 8ème Rencontre des Étudiants Chercheurs en
  Informatique pour le Traitement Automatique des Langues (RECITAL)}, Fès,
  Maroc, 2004.

\bibitem[BFHM98]{Brill98}
E.~\bgroup\sc Brill\egroup{}, R.~\bgroup\sc Florian\egroup{}, J.~C. \bgroup\sc
  Henderson\egroup{} \andname{} L.~\bgroup\sc Mangu\egroup{}.
\newblock \og Beyond {N}-Grams: Can Linguistic Sophistication Improve Language
  Modeling?\fg.
\newblock \Inname{} {\em Proc. of the 36th Annual Meeting of the Association
  for Computational Linguistics and the 17th International Conference on
  Computational Linguistics (COLING-ACL)}, \volumename{}~1, Montréal, Canada,
  1998.

\bibitem[BGWL01]{Barras01}
C.~\bgroup\sc Barras\egroup{}, E.~\bgroup\sc Geoffrois\egroup{}, Z.~\bgroup\sc
  Wu\egroup{} \andname{} M.~\bgroup\sc Liberman\egroup{}.
\newblock \og Transcriber: Development and Use of a Tool for Assisting Speech
  Corpora Production\fg.
\newblock {\em Speech Communication}, 33(1-2):5--22, 2001.

\bibitem[BHDM04]{Bigi04}
B.~\bgroup\sc Bigi\egroup{}, Y.~\bgroup\sc Huang\egroup{} \andname{}
  R.~\bgroup\sc De~Mori\egroup{}.
\newblock \og Vocabulary and Language Model Adaptation Using Information
  Retrieval\fg.
\newblock \Inname{} {\em Proc. of the 8th International Conference on Spoken
  Language Processing (ICSLP)}, \volumename{}~2, île de Jeju, Corée du Sud,
  2004.

\bibitem[Bla99]{Blasig99}
R.~\bgroup\sc Blasig\egroup{}.
\newblock \og Combination of Words and Word Categories in Varigram
  Histories\fg.
\newblock \Inname{} {\em Proc. of the IEEE International Conference on
  Acoustics, Speech, and Signal Processing (ICASSP)}, \volumename{}~1, Phoenix,
  Arizona, États-Unis, 1999.

\bibitem[BM98]{Berger98}
A.~\bgroup\sc Berger\egroup{} \andname{} R.~\bgroup\sc Miller\egroup{}.
\newblock \og Just-in-Time Language Modelling\fg.
\newblock \Inname{} {\em Proc. of the IEEE International Conference on
  Acoustics, Speech, and Signal Processing (ICASSP)}, \volumename{}~2, Seattle,
  Washington, États-Unis, 1998.

\bibitem[BNSd99]{Bechet99}
F.~\bgroup\sc Béchet\egroup{}, A.~\bgroup\sc Nasr\egroup{}, T.~\bgroup\sc
  Spriet\egroup{} \andname{} R.~\bgroup\sc {de Mori}\egroup{}.
\newblock \og Modèles de langage à portée variable~: application au traitement
  des homophones\fg.
\newblock \Inname{} {\em Actes de la 6ème conférence sur le Traitement
  Automatique des Langues Naturelles (TALN)}, Cargèse, France, 1999.

\bibitem[BP98]{Berger98b}
A.~\bgroup\sc Berger\egroup{} \andname{} H.~\bgroup\sc Printz\egroup{}.
\newblock \og Recognition Performance of a Large-Scale Dependency-Grammar
  Language Model\fg.
\newblock \Inname{} {\em Proc. of the 5th International Conference on Spoken
  Language Processing (ICSLP)}, \volumename{}~6, Sydney, Australie, 1998.

\bibitem[BP03]{Beutler03}
R.~\bgroup\sc Beutler\egroup{} \andname{} B.~\bgroup\sc Pfister\egroup{}.
\newblock \og Integrating Statistical and Rule-Based Knowledge for Continuous
  German Speech Recognition\fg.
\newblock \Inname{} {\em Proc. of the 8th European Conference on Speech
  Communication and Technology (Eurospeech)}, Genève, Suisse, 2003.

\bibitem[BPLA95]{Bimbot95}
F.~\bgroup\sc Bimbot\egroup{}, R.~\bgroup\sc Pieraccini\egroup{}, E.~\bgroup\sc
  Levin\egroup{} \andname{} B.~\bgroup\sc Atal\egroup{}.
\newblock \og Variable-Length Sequence Modeling: Multigrams\fg.
\newblock {\em Signal Processing Letters, IEEE}, 2(6):111--113, 1995.

\bibitem[Bra00]{Brants00}
T.~\bgroup\sc Brants\egroup{}.
\newblock \og TnT - A Statistical Part-of-Speech Tagger\fg.
\newblock \Inname{} {\em Proc. of the Sixth Applied NLP}, Seattle, Washington,
  États-Unis, 2000.

\bibitem[Bru03]{Brun03}
A.~\bgroup\sc Brun\egroup{}.
\newblock \og {\em Détection de thème et adaptation des modèles de langage pour
  la reconnaissance automatique de la parole}\fg.
\newblock \thesename{}, Université Henri Poincaré - Nancy 1, France, 2003.

\bibitem[BV05]{Benzitoun05}
C.~\bgroup\sc Benzitoun\egroup{} \andname{} J.~\bgroup\sc Véronis\egroup{}.
\newblock \og Problèmes d'annotation d'un corpus oral dans le cadre de la
  campagne EASY\fg.
\newblock \Inname{} {\em Actes de la 12ème conférence sur le Traitement
  Automatique des Langues Naturelles (TALN)}, \volumename{}~2, Dourdan, France,
  2005.

\bibitem[Cam01]{Campione01b}
E.~\bgroup\sc Campione\egroup{}.
\newblock \og {\em Étiquetage prosodique semi-automatique de corpus oraux~:
  algorithmes et méthodologie}\fg.
\newblock \thesename{}, Université de Provence, Aix-en-Provence, France, 2001.

\bibitem[Can00]{Candea00}
M.~\bgroup\sc Candea\egroup{}.
\newblock \og {\em Contribution à l'étude des pauses silencieuses et des
  phénomènes dits d'«hésitation» en français oral spontané. Étude sur un corpus
  de récits en classe de français}\fg.
\newblock \thesename{}, Université Paris III, France, 2000.

\bibitem[CG98]{Chen98}
S.~F. \bgroup\sc Chen\egroup{} \andname{} J.~\bgroup\sc Goodman\egroup{}.
\newblock \og An Empirical Study of Smoothing Techniques for Language
  Modeling\fg.
\newblock \technicalreportname{}, Harvard University, Cambridge, Massachusetts,
  États-Unis, 1998.

\bibitem[CGL{\etalchar{+}}01]{Chen01}
L.~\bgroup\sc Chen\egroup{}, J.-L. \bgroup\sc Gauvain\egroup{}, L.~\bgroup\sc
  Lamel\egroup{}, G.~\bgroup\sc Adda\egroup{} \andname{} M.~\bgroup\sc
  Adda-Decker\egroup{}.
\newblock \og Using Information Retrieval Methods for Language Model
  Adaptation\fg.
\newblock \Inname{} {\em Proc. of the 7th European Conference on Speech
  Communication and Technology (Eurospeech)}, Aalborg, Danemark, 2001.

\bibitem[CGLA03]{Chen03}
L.~\bgroup\sc Chen\egroup{}, J.-L. \bgroup\sc Gauvain\egroup{}, L.~\bgroup\sc
  Lamel\egroup{} \andname{} G.~\bgroup\sc Adda\egroup{}.
\newblock \og Unsupervised Language Model Adaptation for Broadcast News\fg.
\newblock \Inname{} {\em Proc. of the IEEE International Conference on
  Acoustics, Speech, and Signal Processing (ICASSP)}, \volumename{}~1, Hong
  Kong, Chine, 2003.

\bibitem[CGLA04]{Chen04b}
L.~\bgroup\sc Chen\egroup{}, J.-L. \bgroup\sc Gauvain\egroup{}, L.~\bgroup\sc
  Lamel\egroup{} \andname{} G.~\bgroup\sc Adda\egroup{}.
\newblock \og Dynamic Language Modeling for Broadcast News\fg.
\newblock \Inname{} {\em Proc. of the 8th International Conference on Spoken
  Language Processing (ICSLP)}, île de Jeju, Corée du Sud, 2004.

\bibitem[Cha00]{Charniak00}
E.~\bgroup\sc Charniak\egroup{}.
\newblock \og A Maximum-Entropy-Inspired Parser\fg.
\newblock \Inname{} {\em Proc. of the 1st Conference of the North American
  Chapter of the Association for Computational Linguistics}, Seattle,
  Washington, États-Unis, 2000.

\bibitem[Cha01]{Charniak01}
E.~\bgroup\sc Charniak\egroup{}.
\newblock \og Immediate-Head Parsing for Language Models\fg.
\newblock \Inname{} {\em Proc. of the 39th Annual Meeting of the Association
  for Computational Linguistics (ACL)}, Toulouse, France, 2001.

\bibitem[CJ00]{Chelba00}
C.~\bgroup\sc Chelba\egroup{} \andname{} F.~\bgroup\sc Jelinek\egroup{}.
\newblock \og Structured Language Modeling\fg.
\newblock {\em Computer Speech and Language}, 14(4):283--332, 2000.

\bibitem[CR89]{Chow89b}
Y.-L. \bgroup\sc Chow\egroup{} \andname{} S.~\bgroup\sc Roukos\egroup{}.
\newblock \og Speech Understanding Using a Unification Grammar\fg.
\newblock \Inname{} {\em Proc. of the IEEE International Conference on
  Acoustics, Speech, and Signal Processing (ICASSP)}, \volumename{}~2, Glasgow,
  Royaume-Uni, 1989.

\bibitem[CR97]{Clarkson97}
P.~R. \bgroup\sc Clarkson\egroup{} \andname{} A.~J. \bgroup\sc
  Robinson\egroup{}.
\newblock \og Language Model Adaptation Using Mixtures and an Exponentially
  Decaying Cache\fg.
\newblock \Inname{} {\em Proc. of the IEEE International Conference on
  Acoustics, Speech, and Signal Processing (ICASSP)}, \volumename{}~2, Munich,
  Allemagne, 1997.

\bibitem[CR99]{Clarkson99}
P.~\bgroup\sc Clarkson\egroup{} \andname{} T.~\bgroup\sc Robinson\egroup{}.
\newblock \og Towards Improved Language Model Evaluation Measures\fg.
\newblock \Inname{} {\em Proc. of the 6th European Conference on Speech
  Communication and Technology (Eurospeech)}, \volumename{}~5, Budapest,
  Hongrie, 1999.

\bibitem[CRAR99]{Chappelier99}
J.-C. \bgroup\sc Chappelier\egroup{}, M.~\bgroup\sc Rajman\egroup{},
  R.~\bgroup\sc Aragüés\egroup{} \andname{} A.~\bgroup\sc Rozenknop\egroup{}.
\newblock \og Lattice Parsing for Speech Recognition\fg.
\newblock \Inname{} {\em Actes de la 6ème conférence sur le Traitement
  Automatique des Langues Naturelles (TALN)}, Cargèse, France, 1999.

\bibitem[CS89]{Chow89}
Y.-L. \bgroup\sc Chow\egroup{} \andname{} R.~\bgroup\sc Schwartz\egroup{}.
\newblock \og The N-Best Algorithm: An efficient Procedure for Finding Top N
  Sentence Hypotheses\fg.
\newblock \Inname{} {\em Proc. of the DARPA Speech and Natural Language
  Workshop}, Philadelphie, Pennsylvanie, États-Unis, 1989.

\bibitem[CV02]{Campione02}
E.~\bgroup\sc Campione\egroup{} \andname{} J.~\bgroup\sc Véronis\egroup{}.
\newblock \og Étude des relations entre pauses et ponctuations pour la synthèse
  de la parole à partir de texte\fg.
\newblock \Inname{} {\em Actes de la 9ème conférence sur le Traitement
  Automatique des Langues Naturelles (TALN)}, Nancy, France, 2002.

\bibitem[CV04]{Campione04}
E.~\bgroup\sc Campione\egroup{} \andname{} J.~\bgroup\sc Véronis\egroup{}.
\newblock \og Pauses et hésitations en français spontané\fg.
\newblock \Inname{} {\em Actes des 25èmes Journées d'Études sur la Parole
  (JEP)}, Fès, Maroc, 2004.

\bibitem[CVD05]{Campione05}
E.~\bgroup\sc Campione\egroup{}, J.~\bgroup\sc Véronis\egroup{} \andname{}
  J.~\bgroup\sc Deulofeu\egroup{}.
\newblock \og {\em C-ORAL-ROM, Integrated Reference Corpora for Spoken Romance
  Languages}\fg, \chaptername{} 3. The French corpus, \pagesname{} 111--133.
\newblock Amsterdam: John Benjamins, 2005.

\bibitem[DAS97]{Demetriou97}
G.~\bgroup\sc Demetriou\egroup{}, E.~\bgroup\sc Atwell\egroup{} \andname{}
  C.~\bgroup\sc Souter\egroup{}.
\newblock \og Large-Scale Lexical Semantics for Speech Recognition Support\fg.
\newblock \Inname{} {\em Proc. of the 5th European Conference on Speech,
  Communication, Technology (Eurospeech)}, Rhodes, Grèce, 1997.

\bibitem[DB95]{Deligne95}
S.~\bgroup\sc Deligne\egroup{} \andname{} F.~\bgroup\sc Bimbot\egroup{}.
\newblock \og Language Modeling by Variable Length Sequences: Theoretical
  Formulation and Evaluation of Multigrams\fg.
\newblock \Inname{} {\em Proc. of the IEEE International Conference on
  Acoustics, Speech, and Signal Processing (ICASSP)}, Detroit, Michigan,
  États-Unis, 1995.

\bibitem[DGA{\etalchar{+}}93]{Dowding93}
J.~\bgroup\sc Dowding\egroup{}, J.~M. \bgroup\sc Gawron\egroup{}, D.~\bgroup\sc
  Appelt\egroup{}, J.~\bgroup\sc Bear\egroup{}, L.~\bgroup\sc Cherny\egroup{},
  R.~\bgroup\sc Moore\egroup{} \andname{} D.~\bgroup\sc Moran\egroup{}.
\newblock \og Gemini: A Natural Language System for Spoken Language
  Understanding\fg.
\newblock \Inname{} {\em Proc. of the 31st Annual Meeting of the Association
  for Computational Linguistics (ACL)}, Columbus, Ohio, États-Unis, 1993.

\bibitem[DGP99]{Deshmukh99}
N.~\bgroup\sc Deshmukh\egroup{}, A.~\bgroup\sc Ganapathiraju\egroup{}
  \andname{} J.~\bgroup\sc Picone\egroup{}.
\newblock \og Hierarchical Search for Large Vocabulary Conversational Speech
  Recognition\fg.
\newblock {\em IEEE Signal Processing Magazine}, 16(5):84--107, 1999.

\bibitem[DS98]{Deligne98}
S.~\bgroup\sc Deligne\egroup{} \andname{} Y.~\bgroup\sc Sakisaga\egroup{}.
\newblock \og Learning a Syntagmatic and Paradigmatic Structure from Language
  Data with a Bi-Multigram Model\fg.
\newblock \Inname{} {\em Proc. of the 36th Annual Meeting of the Association
  for Computational Linguistics and the 17th International Conference on
  Computational Linguistics (COLING-ACL)}, \volumename{}~1, Montréal, Canada,
  1998.

\bibitem[EBD90]{Elbeze90}
M.~\bgroup\sc El-Bèze\egroup{} \andname{} A.-M. \bgroup\sc Derouault\egroup{}.
\newblock \og A Morphological Model for Large Vocabulary Speech Recognition\fg.
\newblock \Inname{} {\em Proc. of the IEEE International Conference on
  Acoustics, Speech, and Signal Processing (ICASSP)}, \volumename{}~1,
  Albuquerque, Nouveau Mexique, États-Unis, 1990.

\bibitem[Fed96]{Federico96}
M.~\bgroup\sc Federico\egroup{}.
\newblock \og Bayesian Estimation Methods for {N}-Gram Language Model
  Adaptation\fg.
\newblock \Inname{} {\em Proc. of the 4th International Conference on Spoken
  Language Processing (ICSLP)}, \volumename{}~1, Philadelphie, Pennsylvanie,
  États-Unis, 1996.

\bibitem[Fed99]{Federico99}
M.~\bgroup\sc Federico\egroup{}.
\newblock \og Efficient Language Model Adaptation through MDI Estimation\fg.
\newblock \Inname{} {\em Proc. of the 6th European Conference on Speech,
  Communication, Technology (Eurospeech)}, \volumename{}~4, Budapest, Hongrie,
  1999.

\bibitem[FIO96]{Farhat96}
A.~\bgroup\sc Farhat\egroup{}, J.-F. \bgroup\sc Isabelle\egroup{} \andname{}
  D.~\bgroup\sc O'Shaughnessy\egroup{}.
\newblock \og Clustering Words for Statistical Language Models Based on
  Contextual Word Similarity\fg.
\newblock \Inname{} {\em Proc. of the IEEE International Conference on
  Acoustics, Speech, and Signal Processing (ICASSP)}, \volumename{}~1, Atlanta,
  Géorgie, États-Unis, 1996.

\bibitem[FRWP03]{Franz03}
M.~\bgroup\sc Franz\egroup{}, B.~\bgroup\sc Ramabhadran\egroup{}, T.~\bgroup\sc
  Ward\egroup{} \andname{} M.~\bgroup\sc Picheny\egroup{}.
\newblock \og Information Access in Large Spoken Archives\fg.
\newblock \Inname{} {\em Proc. of the ISCA Multilingual Spoken Document
  Retrieval Workshop}, Macao/Hong Kong, Chine, 2003.

\bibitem[FY99]{Florian99}
R.~\bgroup\sc Florian\egroup{} \andname{} D.~\bgroup\sc Yarowsky\egroup{}.
\newblock \og Dynamic Nonlocal Language Modeling via Hierarchical Topic-Based
  Adaptation\fg.
\newblock \Inname{} {\em Proc. of 37th Annual Meeting of the Association for
  Computational Linguistics (ACL)}, College Park, Maryland, États-Unis, 1999.

\bibitem[GAAD{\etalchar{+}}05]{Gauvain05}
J.-L. \bgroup\sc Gauvain\egroup{}, G.~\bgroup\sc Adda\egroup{}, M.~\bgroup\sc
  Adda-Decker\egroup{}, A.~\bgroup\sc Allauzen\egroup{}, V.~\bgroup\sc
  Gendner\egroup{}, L.~\bgroup\sc Lamel\egroup{} \andname{} H.~\bgroup\sc
  Schwenk\egroup{}.
\newblock \og Where are we in Transcribing {F}rench Broadcast News?\fg.
\newblock \Inname{} {\em Proc. of the 9th European Conference on Speech
  Communication and Technology (Eurospeech)}, Lisbonne, Portugal, 2005.

\bibitem[GAD02]{Gendner02}
V.~\bgroup\sc Gendner\egroup{} \andname{} M.~\bgroup\sc Adda-Decker\egroup{}.
\newblock \og Analyse comparative de corpus oraux et écrits français~: mots,
  lemmes et classes morpho-syntaxiques\fg.
\newblock \Inname{} {\em Actes des 24èmes Journées d'Études sur la Parole
  (JEP)}, Nancy, France, 2002.

\bibitem[GAL{\etalchar{+}}04]{Gauvain04}
J.-L. \bgroup\sc Gauvain\egroup{}, G.~\bgroup\sc Adda\egroup{}, L.~\bgroup\sc
  Lamel\egroup{}, F.~\bgroup\sc Lefèvre\egroup{} \andname{} H.~\bgroup\sc
  Schwenk\egroup{}.
\newblock \og Transcription de la parole conversationnelle\fg.
\newblock \Inname{} {\em Actes des 25èmes Journées d'Études sur la Parole
  (JEP)}, Fès, Maroc, 2004.

\bibitem[Gar95]{Garside95}
R.~\bgroup\sc Garside\egroup{}.
\newblock \og {\em Spoken English on Computer: Transcription, Mark-up and
  Application}\fg, \chaptername{} Grammatical Tagging of the Spoken Part of the
  British National Corpus: A Progress Report, \pagesname{} 161--167.
\newblock Harlow: Longma, 1995.

\bibitem[Geu96]{Geutner96}
P.~\bgroup\sc Geutner\egroup{}.
\newblock \og Introducing Linguistic Constraints into Statistical Language
  Modeling\fg.
\newblock \Inname{} {\em Proc. of the 4th International Conference on Spoken
  Language Processing (ICSLP)}, Philadelphie, Pennsylvanie, États-Unis, 1996.

\bibitem[GH99]{Gildea99}
D.~\bgroup\sc Gildea\egroup{} \andname{} T.~\bgroup\sc Hofmann\egroup{}.
\newblock \og Topic-Based Language Models Using EM\fg.
\newblock \Inname{} {\em Proc. of the 6th European Conference on Speech
  Communication and Technology (Eurospeech)}, Budapest, Hongrie, 1999.

\bibitem[GLL00]{Gao00}
J.~\bgroup\sc Gao\egroup{}, M.~\bgroup\sc Li\egroup{} \andname{} K.-F.
  \bgroup\sc Lee\egroup{}.
\newblock \og {N}-Gram Distribution Based Language Model Adaptation\fg.
\newblock \Inname{} {\em Proc. of the 6th International Conference on Spoken
  Language Processing (ICSLP)}, \volumename{}~1, Pékin, Chine, 2000.

\bibitem[Goo01]{Goodman01}
J.~T. \bgroup\sc Goodman\egroup{}.
\newblock \og A Bit of Progress in Language Modeling, Extended Version\fg.
\newblock \technicalreportname{}, Microsoft Research, Redmond, Washington,
  États-Unis, 2001.

\bibitem[GSTN96]{Gallwitz96}
F.~\bgroup\sc Gallwitz\egroup{}, E.~G. \bgroup\sc Schukat-Talamazzini\egroup{}
  \andname{} H.~\bgroup\sc Niemann\egroup{}.
\newblock \og Integrating Large Context Language Models into a Real Time Word
  Recognizer\fg.
\newblock \Inname{} {\em Proc. of the 3rd Slovenian-German and the 2nd SDRV
  Workshop}, Ljubljana, Slovénie, 1996.

\bibitem[GT04]{Gardes04}
J.~\bgroup\sc Gardes~Tamine\egroup{}.
\newblock {\em Pour une grammaire de l'écrit}.
\newblock Paris~: Belin, 2004.

\bibitem[Gué05]{Guenot05}
M.-L. \bgroup\sc Guénot\egroup{}.
\newblock \og Parsing de l'oral~: traiter les disfluences\fg.
\newblock \Inname{} {\em Actes de la 12ème conférence sur le Traitement
  Automatique des Langues Naturelles (TALN)}, \volumename{}~1, Dourdan, France,
  2005.

\bibitem[GW98]{Gillett98}
J.~\bgroup\sc Gillett\egroup{} \andname{} W.~\bgroup\sc Ward\egroup{}.
\newblock \og A Language Model Combining Trigrams and Stochastic Context-Free
  Grammars\fg.
\newblock \Inname{} {\em Proc. of the 5th International Conference on Spoken
  Language Processing (ICSLP)}, \volumename{}~6, Sydney, Australie, 1998.

\bibitem[Hee99]{Heeman99}
P.~A. \bgroup\sc Heeman\egroup{}.
\newblock \og {POS} Tags and Decision Trees for Language Modeling\fg.
\newblock \Inname{} {\em Proc. of the Joint SIGDAT Conference on Empirical
  Methods in Natural Language Processing and Very Large Corpora}, College Park,
  Maryland, États-Unis, 1999.

\bibitem[Hen02a]{Henry02}
S.~\bgroup\sc Henry\egroup{}.
\newblock \og Quelles répétitions à l'oral~? Esquisse d'une typologie\fg.
\newblock \Inname{} {\em Actes des 2èmes Journées de Linguistique de Corpus},
  Lorient, France, 2002.

\bibitem[Hen02b]{Henry02a}
S.~\bgroup\sc Henry\egroup{}.
\newblock \og Étude des répétitions en français parlé spontané pour les
  technologies de la parole\fg.
\newblock \Inname{} {\em Actes de la 6ème Rencontre des Étudiants Chercheurs en
  Informatique pour le Traitement Automatique des Langues (RECITAL)}, Nancy,
  France, 2002.

\bibitem[HH95]{Harper95}
M.~P. \bgroup\sc Harper\egroup{} \andname{} R.~A. \bgroup\sc
  Helzerman\egroup{}.
\newblock \og Extensions to Constraint Dependency Parsing for Spoken Language
  Processing\fg.
\newblock {\em Computer Speech and Language}, \pagesname{} 187--234, 1995.

\bibitem[HJ04]{Hall04}
K.~\bgroup\sc Hall\egroup{} \andname{} M.~\bgroup\sc Johnson\egroup{}.
\newblock \og Attention Shifting for Parsing Speech\fg.
\newblock \Inname{} {\em Proc. of the 42nd Meeting of the Association for
  Computational Linguistics (ACL)}, Barcelone, Espagne, 2004.

\bibitem[HJJ{\etalchar{+}}99]{Harper99}
M.~P. \bgroup\sc Harper\egroup{}, M.~T. \bgroup\sc Johnson\egroup{}, L.~H.
  \bgroup\sc Jamieson\egroup{}, S.~A. \bgroup\sc Hockema\egroup{} \andname{}
  C.~M. \bgroup\sc White\egroup{}.
\newblock \og Interfacing a CDG Parser with an HMM Word Recognizer Using Word
  Graphs\fg.
\newblock \Inname{} {\em Proc. of the IEEE International Conference on
  Acoustics, Speech, and Signal Processing (ICASSP)}, \volumename{}~2, Phoenix,
  Arizona, États-Unis, 1999.

\bibitem[HJM{\etalchar{+}}94]{Harper94}
M.~P. \bgroup\sc Harper\egroup{}, L.~H. \bgroup\sc Jamieson\egroup{}, C.~D.
  \bgroup\sc Mitchell\egroup{}, G.~\bgroup\sc Ying\egroup{}, S.~\bgroup\sc
  Potisuk\egroup{}, P.~N. \bgroup\sc Srinivasan\egroup{}, R.~\bgroup\sc
  Chen\egroup{}, C.~B. \bgroup\sc Zoltowski\egroup{}, L.~L. \bgroup\sc
  McPheters\egroup{}, B.~\bgroup\sc Pellom\egroup{} \andname{} R.~A. \bgroup\sc
  Helzerman\egroup{}.
\newblock \og Integrating Language Models with Speech Recognition\fg.
\newblock \Inname{} {\em Proc. of the AAAI94 Workshop on the Integration of
  Natural Language and Speech Processing}, Seattle, Washington, États-Unis,
  1994.

\bibitem[HP03]{Henry03}
S.~\bgroup\sc Henry\egroup{} \andname{} B.~\bgroup\sc Pallaud\egroup{}.
\newblock \og Word Fragments and Repeats in Spontaneous Spoken French\fg.
\newblock \Inname{} {\em Proceedings of Disfluency in Spontaneous Speech
  Workshop (DISS)}, Göteborg, Suède, 2003.

\bibitem[HW94]{Hauenstein94}
A.~\bgroup\sc Hauenstein\egroup{} \andname{} H.~\bgroup\sc Weber\egroup{}.
\newblock \og An Investigation of Tightly-Coupled Time-Synchronous Speech
  Language Understanding Using a Unification Grammar\fg.
\newblock \Inname{} {\em Proc. of the 12th National Conference on Artificial
  Intelligence Workshop on the Integration of Natural Language and Speech
  Processing}, Seattle, Washington, États-Unis, 1994.

\bibitem[IM94]{Isotani94}
R.~\bgroup\sc Isotani\egroup{} \andname{} S.~\bgroup\sc Matsunaga\egroup{}.
\newblock \og Speech Recognition Using a Stochastic Language Model Integrating
  Local and Global Constraints\fg.
\newblock \Inname{} {\em Proc. of the ARPA SLT Workshop}, 1994.

\bibitem[IO99]{Iyer99}
R.~\bgroup\sc Iyer\egroup{} \andname{} M.~\bgroup\sc Ostendorf\egroup{}.
\newblock \og Modeling Long Distance Dependence in Language: Topic Mixtures
  \textit{versus} Dynamic Cache Models\fg.
\newblock {\em IEEE Transactions on Speech and Audio Processing}, 7(1):30--39,
  1999.

\bibitem[Jar96]{Jardino96}
M.~\bgroup\sc Jardino\egroup{}.
\newblock \og Multilingual Stochastic {N}-Gram Class Language Models\fg.
\newblock \Inname{} {\em Proc. of the IEEE International Conference on
  Acoustics, Speech, and Signal Processing (ICASSP)}, \volumename{}~1, Atlanta,
  Géorgie, États-Unis, 1996.

\bibitem[JC04]{Johnson04b}
M.~\bgroup\sc Johnson\egroup{} \andname{} E.~\bgroup\sc Charniak\egroup{}.
\newblock \og A TAG-Based Noisy Channel Model of Speech Repairs\fg.
\newblock \Inname{} {\em Proc. of the 42nd Annual Meeting of the Association
  for Computational Linguistics (ACL)}, Barcelone, Espagne, 2004.

\bibitem[Jel97]{Jelinek97}
F.~\bgroup\sc Jelinek\egroup{}.
\newblock {\em Statistical Methods for Speech Recognition}.
\newblock The MIT Press, 1997.

\bibitem[JL91]{Jelinek91}
F.~\bgroup\sc Jelinek\egroup{} \andname{} J.~D. \bgroup\sc Lafferty\egroup{}.
\newblock \og Computation of the Probability of Initial Substring Generation by
  Stochastic Context-Free Grammars\fg.
\newblock {\em Computation Linguistics}, 17(3):315--323, 1991.

\bibitem[JM00]{Jurafsky00}
D.~\bgroup\sc Jurafsky\egroup{} \andname{} J.~H. \bgroup\sc Martin\egroup{}.
\newblock {\em Speech and Natural Language Processing: An Introduction to
  Natural Language Processing, Computational Linguistics, and Speech
  Recognition}.
\newblock Prentice-Hall, 2000.

\bibitem[Jou96]{Jouvet96}
D.~\bgroup\sc Jouvet\egroup{}.
\newblock \og Robustesse et flexibilité en reconnaissance automatique de la
  parole\fg.
\newblock {\em L'écho des recherches}, 165:25--38, 1996.

\bibitem[JWS{\etalchar{+}}95]{Jurafsky95}
D.~\bgroup\sc Jurafsky\egroup{}, C.~\bgroup\sc Wooters\egroup{}, J.~\bgroup\sc
  Segal\egroup{}, A.~\bgroup\sc Stolcke\egroup{}, E.~\bgroup\sc
  Fosler\egroup{}, G.~\bgroup\sc Tajchman\egroup{} \andname{} N.~\bgroup\sc
  Morgan\egroup{}.
\newblock \og Using a Stochastic Context-Free Grammar as a Language Model for
  Speech Recognition\fg.
\newblock \Inname{} {\em Proc. of the IEEE International Conference on
  Acoustics, Speech, and Signal Processing (ICASSP)}, \volumename{}~1, Detroit,
  Michigan, États-Unis, 1995.

\bibitem[Kat87]{Katz87}
S.~M. \bgroup\sc Katz\egroup{}.
\newblock \og Estimation of Probabilities from Sparse Data for the Language
  Model Component of a Speech Recognizer\fg.
\newblock {\em IEEE Transactions on Acoustics, Speech, and Signal Processing},
  35(3):400--401, 1987.

\bibitem[KDM90]{Kuhn90}
R.~\bgroup\sc Kuhn\egroup{} \andname{} R.~\bgroup\sc De~Mori\egroup{}.
\newblock \og A Cache-Based Natural Language Model for Speech Recognition\fg.
\newblock {\em IEEE Transactions on Pattern Analysis and Machine Intelligence},
  12(6):570--583, 1990.

\bibitem[KKS89]{Kita89}
K.~\bgroup\sc Kita\egroup{}, T.~\bgroup\sc Kawabata\egroup{} \andname{}
  H.~\bgroup\sc Saito\egroup{}.
\newblock \og HMM Continuous Speech Recognition Using Predictive LR Parsing\fg.
\newblock \Inname{} {\em Proc. of the IEEE International Conference on
  Acoustics, Speech, and Signal Processing (ICASSP)}, Glasgow, Royaume-Uni,
  1989.

\bibitem[KN93]{Kneser93b}
R.~\bgroup\sc Kneser\egroup{} \andname{} H.~\bgroup\sc Ney\egroup{}.
\newblock \og Improved Clustering Techniques for Class-Based Statistical
  Language Modelling\fg.
\newblock \Inname{} {\em Proc. of the 3rd European Conference on Speech
  Communication and Technology (Eurospeech)}, \volumename{}~2, Berlin,
  Allemagne, 1993.

\bibitem[Kne96]{Kneser96}
R.~\bgroup\sc Kneser\egroup{}.
\newblock \og Statistical Language Modeling Using a Variable Context Length\fg.
\newblock \Inname{} {\em Proc. of the 4th International Conference on Spoken
  Language Processing (ICSLP)}, \volumename{}~1, Philadelphie, Pennsylvanie,
  États-Unis, 1996.

\bibitem[KNST94]{Kuhn94}
T.~\bgroup\sc Kuhn\egroup{}, H.~\bgroup\sc Niemann\egroup{} \andname{} E.~G.
  \bgroup\sc Schukat-Talamazzini\egroup{}.
\newblock \og Ergodic Hidden Markov Models and Polygrams for Language\fg.
\newblock \Inname{} {\em Proc. of the IEEE International Conference on
  Acoustics, Speech, and Signal Processing (ICASSP)}, \volumename{}~1,
  Adélaïde, Australie, 1994.

\bibitem[KR99]{Kuo99}
H.-K.~J. \bgroup\sc Kuo\egroup{} \andname{} W.~\bgroup\sc Reichl\egroup{}.
\newblock \og Phrase-Based Language Models for Speech Recognition\fg.
\newblock \Inname{} {\em Proc. of the 6th European Conference on Speech
  Comunication and Technology (Eurospeech)}, Budapest, Hongrie, 1999.

\bibitem[KS93]{Kneser93}
R.~\bgroup\sc Kneser\egroup{} \andname{} V.~\bgroup\sc Steinbiss\egroup{}.
\newblock \og On the Dynamic Adaptation of Stochastic Language Models\fg.
\newblock \Inname{} {\em Proc. of the IEEE International Conference on
  Acoustics, Speech, and Signal Processing (ICASSP)}, \volumename{}~2,
  Minneapolis, Minnesota, États-Unis, 1993.

\bibitem[KW98]{Kemp98}
T.~\bgroup\sc Kemp\egroup{} \andname{} A.~\bgroup\sc Waibel\egroup{}.
\newblock \og Reducing the OOV Rate in Broadcast News Speech Recognition\fg.
\newblock \Inname{} {\em Proc. of the 5th International Conference on Spoken
  Language Processing (ICSLP)}, Sydney, Australie, 1998.

\bibitem[KW99]{Khudanpur99}
S.~\bgroup\sc Khudanpur\egroup{} \andname{} J.~\bgroup\sc Wu\egroup{}.
\newblock \og A Maximum Entropy Language Model to Integrate {N}-Grams and Topic
  Dependencies for Conversational Speech Recognition\fg.
\newblock \Inname{} {\em Proc. of the IEEE International Conference on
  Acoustics, Speech, and Signal Processing (ICASSP)}, Phoenix, Arizona,
  États-Unis, 1999.

\bibitem[LBS04]{Linares04}
D.~\bgroup\sc Linares\egroup{}, J.-M. \bgroup\sc Benedí\egroup{} \andname{}
  J.-A. \bgroup\sc Sánchez\egroup{}.
\newblock \og A Hybrid Language Model Based on a Combination of {N}-Grams and
  Stochastic Context-Free Grammars\fg.
\newblock {\em ACM Transactions on Asian Language Information Processing
  (TALIP)}, 3(2):113--127, 2004.

\bibitem[LBSH03]{Langlois03}
D.~\bgroup\sc Langlois\egroup{}, A.~\bgroup\sc Brun\egroup{}, K.~\bgroup\sc
  Smaïli\egroup{} \andname{} J.-P. \bgroup\sc Haton\egroup{}.
\newblock \og Événements impossibles en modélisation stochastique du
  langage\fg.
\newblock {\em Traitement Automatique des Langues (TAL)}, 44(1):33--61, 2003.

\bibitem[LKMN05]{Lane05}
I.~R. \bgroup\sc Lane\egroup{}, T.~\bgroup\sc Kawahara\egroup{}, T.~\bgroup\sc
  Matsui\egroup{} \andname{} S.~\bgroup\sc Nakamura\egroup{}.
\newblock \og Dialogue Speech Recognition by Combining Hierarchical Topic
  Classification and Language Model Switching\fg.
\newblock {\em IEICE Transactions on Information and Systems},
  E88-D(3):446--454, 2005.

\bibitem[LMW97]{Leech97}
G.~\bgroup\sc Leech\egroup{}, A.~\bgroup\sc McEnery\egroup{} \andname{}
  M.~\bgroup\sc Wynne\egroup{}.
\newblock \og {\em Corpus Annotation}\fg, \chaptername{} Further Levels of
  Annotation, \pagesname{} 85--101.
\newblock London: Longman, 1997.

\bibitem[LSHS04]{Liu04}
Y.~\bgroup\sc Liu\egroup{}, A.~\bgroup\sc Stolcke\egroup{}, M.~P. \bgroup\sc
  Harper\egroup{} \andname{} E.~\bgroup\sc Shriberg\egroup{}.
\newblock \og Comparing and Combining Generative and Posterior Probability
  Models: Some Advances in Sentence Boundary Detection in Speech\fg.
\newblock \Inname{} {\em Proc. of the Conference on Empirical Methods in
  Natural Language Processing (EMNLP)}, Barcelone, Espagne, 2004.

\bibitem[LST92]{Lafferty92}
J.~\bgroup\sc Lafferty\egroup{}, D.~\bgroup\sc Sleator\egroup{} \andname{}
  D.~\bgroup\sc Temperley\egroup{}.
\newblock \og Grammatical Trigrams: A Probabilistic Link Grammar\fg.
\newblock \Inname{} {\em Proc. of the AAAI Fall Symposium on Probabilistic
  Approaches to Natural Language}, Cambridge, Massachusetts, États-Unis, 1992.

\bibitem[MAB03]{Mendes03}
A.~\bgroup\sc Mendes\egroup{}, R.~\bgroup\sc Amaro\egroup{} \andname{} M.~F.
  \bgroup\sc {Bacelar do Nascimento}\egroup{}.
\newblock \og Reusing Available Resources for Tagging a Spoken Portuguese
  Corpus\fg.
\newblock \Inname{} {\em Proc. of the Workshop on Tagging and Shallow
  Processing of Portuguese (TASHA)}, Lisbonne, Portugal, 2003.

\bibitem[MBS00]{Mangu00}
L.~\bgroup\sc Mangu\egroup{}, E.~\bgroup\sc Brill\egroup{} \andname{}
  A.~\bgroup\sc Stolcke\egroup{}.
\newblock \og Finding Consensus in Speech Recognition: Word Error Minimization
  and Other Applications of Confusion Networks\fg.
\newblock {\em Computer Speech and Language}, 14(4):373--400, 2000.

\bibitem[Mel00]{Melis00}
L.~\bgroup\sc Melis\egroup{}.
\newblock \og Le français parlé et le français écrit, une opposition à
  géométrie variable\fg.
\newblock {\em Romaneske}, 25(3):56--66, 2000.

\bibitem[MG03]{Moreno03}
A.~\bgroup\sc Moreno\egroup{} \andname{} J.~M. \bgroup\sc Guirao\egroup{}.
\newblock \og Tagging a Spontaneous Speech Corpus of Spanish\fg.
\newblock \Inname{} {\em Proc. of Recent Advances in Natural Language
  Processing (RANLP)}, Borovets, Bulgarie, 2003.

\bibitem[MI96]{Meteer96}
M.~\bgroup\sc Meteer\egroup{} \andname{} R.~\bgroup\sc Iyer\egroup{}.
\newblock \og Modeling Conversational Speech for Speech Recognition\fg.
\newblock \Inname{} {\em Proc. of the Conference on Empirical Methods in
  Natural Language Processing (EMNLP)}, Philadelphie, Pennsylvanie, États-Unis,
  1996.

\bibitem[MLN97]{Martin97}
S.~C. \bgroup\sc Martin\egroup{}, J.~\bgroup\sc Liermann\egroup{} \andname{}
  H.~\bgroup\sc Ney\egroup{}.
\newblock \og Adaptive Topic Dependent Language Modelling Using Word-Based
  Varigrams\fg.
\newblock \Inname{} {\em Proc. of the 5th European Conference on Speech
  Communication and Technology (Eurospeech)}, Rhodes, Grèce, 1997.

\bibitem[MM92]{Maltese92}
G.~\bgroup\sc Maltese\egroup{} \andname{} F.~\bgroup\sc Mancini\egroup{}.
\newblock \og An Automatic Technique to Include Grammatical and Morphological
  Information in a Trigram-Based Statistical Language Model\fg.
\newblock \Inname{} {\em Proc. of the IEEE International Conference on
  Acoustics, Speech, and Signal Processing (ICASSP)}, \volumename{}~1, San
  Francisco, Californie, États-Unis, 1992.

\bibitem[Moo99]{Moore99}
R.~C. \bgroup\sc Moore\egroup{}.
\newblock \og {\em Computational Models of Speech Pattern Processing}\fg,
  \chaptername{} Using Natural-Language Knowledge Sources in Speech
  Recognition, \pagesname{} 304--327.
\newblock Springer-Verlag, 1999.

\bibitem[MPM89]{Moore89}
R.~\bgroup\sc Moore\egroup{}, F.~\bgroup\sc Pereira\egroup{} \andname{}
  H.~\bgroup\sc Murveit\egroup{}.
\newblock \og Integrating Speech and Natural-Language Processing\fg.
\newblock \Inname{} {\em Proc. of the DARPA Speech and Natural Language
  Workshop}, Philadelphie, Pennsylvanie, États-Unis, 1989.

\bibitem[MSZ02]{Mou02}
X.~\bgroup\sc Mou\egroup{}, S.~\bgroup\sc Seneff\egroup{} \andname{}
  V.~\bgroup\sc Zue\egroup{}.
\newblock \og Integration of Supra-Lexical Linguistic Models with Speech
  Recognition Using Shallow Parsing and Finite State Transducers\fg.
\newblock \Inname{} {\em Proc. of the 7th International Conference on Spoken
  Language Processing (ICSLP)}, Denver, Colorado, États-Unis, 2002.

\bibitem[NEB{\etalchar{+}}99]{Nasr99}
A.~\bgroup\sc Nasr\egroup{}, Y.~\bgroup\sc Estève\egroup{}, F.~\bgroup\sc
  Béchet\egroup{}, T.~\bgroup\sc Spriet\egroup{} \andname{} R.~\bgroup\sc {de
  Mori}\egroup{}.
\newblock \og A Language Model Combining {N}-Grams and Stochastic Finite State
  Automata\fg.
\newblock \Inname{} {\em Proc. of the 6th European Conference on Speech
  Communication and Technology (Eurospeech)}, \volumename{}~5, Budapest,
  Hongrie, 1999.

\bibitem[NG01]{Nivre01}
J.~\bgroup\sc Nivre\egroup{} \andname{} L.~\bgroup\sc Grönqvist\egroup{}.
\newblock \og Tagging a Corpus of Spoken Swedish\fg.
\newblock {\em International Journal of Corpus Linguistics}, 6(1):47--78, 2001.

\bibitem[NW96a]{Niesler96}
T.~R. \bgroup\sc Niesler\egroup{} \andname{} P.~C. \bgroup\sc
  Woodland\egroup{}.
\newblock \og Combination of Word-Based and Category-Based Language Models\fg.
\newblock \Inname{} {\em Proc. of the 4th International Conference on Spoken
  Language Processing (ICSLP)}, \volumename{}~1, Philadelphie, Pennsylvanie,
  États-Unis, 1996.

\bibitem[NW96b]{Niesler96b}
T.~R. \bgroup\sc Niesler\egroup{} \andname{} P.~C. \bgroup\sc
  Woodland\egroup{}.
\newblock \og A Variable-Length Category-Based {N}-Gram Language Model\fg.
\newblock \Inname{} {\em Proc. of the IEEE International Conference on
  Acoustics, Speech, and Signal Processing (ICASSP)}, \volumename{}~1, Atlanta,
  Géorgie, États-Unis, 1996.

\bibitem[NWW98]{Niesler98}
T.~R. \bgroup\sc Niesler\egroup{}, E.~W.~D. \bgroup\sc Whittaker\egroup{}
  \andname{} P.~C. \bgroup\sc Woodland\egroup{}.
\newblock \og Comparison of Part-of-Speech and Automatically Derived
  Category-Based Language Models for Speech Recognition\fg.
\newblock \Inname{} {\em Proc. of the IEEE International Conference on
  Acoustics, Speech, and Signal Processing (ICASSP)}, Seattle, Washington,
  États-Unis, 1998.

\bibitem[ONA97]{Ortmanns97}
S.~\bgroup\sc Ortmanns\egroup{}, H.~\bgroup\sc Ney\egroup{} \andname{}
  X.~\bgroup\sc Aubert\egroup{}.
\newblock \og A Word Graph Algorithm for Large Vocabulary Continuous Speech
  Recognition\fg.
\newblock {\em Computer, Speech and Language}, 11(1):43--72, 1997.

\bibitem[Pal03]{Pallett03}
D.~S. \bgroup\sc Pallett\egroup{}.
\newblock \og A Look at NIST's Benchmark ASR Tests: Past, Present, and
  Future\fg.
\newblock \Inname{} {\em Proc. of the IEEE Workshop Automatic Speech
  Recognition and Understanding}, St. Thomas, îles Vierges, États-Unis, 2003.

\bibitem[PH04]{Pallaud04}
B.~\bgroup\sc Pallaud\egroup{} \andname{} S.~\bgroup\sc Henry\egroup{}.
\newblock \og Amorces de mots et répétitions~: des hésitations plus que des
  erreurs en français parlé\fg.
\newblock \Inname{} {\em Actes des 7èmes Journées internationales d'Analyse
  statistique des Données Textuelles (JADT)}, Louvain-la-Neuve, Belgique, 2004.

\bibitem[PMVGL03]{Perraud2003}
F.~\bgroup\sc Perraud\egroup{}, E.~\bgroup\sc Morin\egroup{}, C.~\bgroup\sc
  Viard-Gaudin\egroup{} \andname{} P.-M. \bgroup\sc Lallican\egroup{}.
\newblock \og Modèles {N}-grammes et {N}-classes pour la reconnaissance de
  l'écriture manuscrite en-ligne\fg.
\newblock {\em Traitement Automatique des Langues (TAL)}, 44(1):63--92, 2003.

\bibitem[Pol03]{Polguere03}
A.~\bgroup\sc Polguère\egroup{}.
\newblock {\em Lexicologie et sémantique lexicale~: notions fondamentales}.
\newblock Les Presses de l'Université de Montréal, 2003.

\bibitem[PPM04]{Panunzi04}
A.~\bgroup\sc Panunzi\egroup{}, E.~\bgroup\sc Picchi\egroup{} \andname{}
  M.~\bgroup\sc Moneglia\egroup{}.
\newblock \og Using PiTagger for Lemmatization and PoS Tagging of a Spontaneous
  Speech Corpus: C-Oral-Rom Italian\fg.
\newblock \Inname{} {\em Proc. of the 4th International Conference on Language
  Resources and Evaluation (LREC)}, \volumename{}~2, Lisbonne, Portugal, 2004.

\bibitem[PS01]{Peng01}
F.~\bgroup\sc Peng\egroup{} \andname{} D.~\bgroup\sc Schuurmans\egroup{}.
\newblock \og A Simple Closed-Class/Open-Class Factorization for Language
  Modeling\fg.
\newblock \Inname{} {\em Proc. of the 6th Natural Language Processing Pacific
  Rim Symposium (NLPRS)}, Tokyo, Japon, 2001.

\bibitem[Rab89]{Rabiner89}
L.~\bgroup\sc Rabiner\egroup{}.
\newblock \og A Tutorial on Hidden Markov Models and Selected Applications in
  Speech Recognition\fg.
\newblock {\em Proc. of the IEEE}, 77(2):257--285, 1989.

\bibitem[RBW96]{RIes96}
K.~\bgroup\sc Ries\egroup{}, F.~D. \bgroup\sc Buø\egroup{} \andname{}
  A.~\bgroup\sc Waibel\egroup{}.
\newblock \og Class Phrase Models for Language Modeling\fg.
\newblock \Inname{} {\em Proc. of the 4th International Conference on Spoken
  Language Processing (ICSLP)}, \volumename{}~1, Philadelphie, Pennsylvanie,
  États-Unis, 1996.

\bibitem[Roa01]{Roark01}
B.~\bgroup\sc Roark\egroup{}.
\newblock \og Probabilistic Top-Down Parsing and Language Modelling\fg.
\newblock {\em Computational Linguistics}, 27(2):249--276, 2001.

\bibitem[Ros94]{Rosenfeld94}
R.~\bgroup\sc Rosenfeld\egroup{}.
\newblock \og A Hybrid Approach to Adaptive Statistical Language Modeling\fg.
\newblock \Inname{} {\em Proc. of the ARPA Workshop on Human Language
  Technology}, Plainsboro, New Jersey, États-Unis, 1994.

\bibitem[Ros96]{Rosenfeld96}
R.~\bgroup\sc Rosenfeld\egroup{}.
\newblock \og A Maximum Entropy Approach to Adaptive Statistical Language
  Modeling\fg.
\newblock {\em Computer, Speech and Language}, 10:187--228, 1996.

\bibitem[Ros00a]{Rosenfeld00b}
R.~\bgroup\sc Rosenfeld\egroup{}.
\newblock \og Incorporating Linguistic Structure into Statistical Language
  Models\fg.
\newblock {\em Philosophical Transactions: Mathematical, Physical and
  Engineering Sciences}, 358:1311--1324, 2000.

\bibitem[Ros00b]{Rosenfeld00}
R.~\bgroup\sc Rosenfeld\egroup{}.
\newblock \og Two Decades of Statistical Language Modeling: Where do we Go from
  Here?\fg.
\newblock {\em Proc. of the IEEE}, 88(8):1270--1278, 2000.

\bibitem[SA90]{Schwartz90}
R.~\bgroup\sc Schwartz\egroup{} \andname{} S.~\bgroup\sc Austin\egroup{}.
\newblock \og Efficient, High-Performance Algorithms for N-Best Search\fg.
\newblock \Inname{} {\em Proc. of the DARPA Speech and Natural Language
  Workshop}, Hidden Valley, Pennsylvanie, États-Unis, 1990.

\bibitem[Sch94]{Schmid94}
H.~\bgroup\sc Schmid\egroup{}.
\newblock \og Probabilistic Part-of-Speech Tagging Using Decision Trees\fg.
\newblock \Inname{} {\em Proc. of the International Conference on New Methods
  in Language Processing}, Manchester, Royaume-Uni, 1994.

\bibitem[Sch95]{Schmid95}
H.~\bgroup\sc Schmid\egroup{}.
\newblock \og Improvements in Part-of-Speech Tagging with an Application to
  German\fg.
\newblock \Inname{} {\em Proc. of the ACL SIGDAT Workshop}, Dublin, Irlande,
  1995.

\bibitem[SCL92]{Su92}
K.-Y. \bgroup\sc Su\egroup{}, T.-H. \bgroup\sc Chiang\egroup{} \andname{} Y.-C.
  \bgroup\sc Lin\egroup{}.
\newblock \og A Unified Framework to Incorporate Speech and Language
  Information in Spoken Language Processing\fg.
\newblock \Inname{} {\em Proc. of the IEEE International Conference on
  Acoustics, Speech, and Signal Processing (ICASSP)}, \volumename{}~1, San
  Francisco, Californie, États-Unis, 1992.

\bibitem[Sen92]{Seneff92}
S.~\bgroup\sc Seneff\egroup{}.
\newblock \og TINA: a Natural Language System for Spoken Language
  Applications\fg.
\newblock {\em Computational Linguistics}, 18(1):61--86, 1992.

\bibitem[SG04]{Schwenk04}
H.~\bgroup\sc Schwenk\egroup{} \andname{} J.-L. \bgroup\sc Gauvain\egroup{}.
\newblock \og Neural Network Language Models for Conversational Speech
  Recognition\fg.
\newblock \Inname{} {\em Proc. of the 8th International Conference on Spoken
  Language Processing (ICSLP)}, île de Jeju, Corée du Sud, 2004.

\bibitem[Shr94]{Shriberg94}
E.~\bgroup\sc Shriberg\egroup{}.
\newblock \og {\em Preliminaries to a Theory of Speech Disfluencies}\fg.
\newblock \thesename{}, University of California, Berkeley, Californie,
  États-Unis, 1994.

\bibitem[Shr01]{Shriberg01}
E.~\bgroup\sc Shriberg\egroup{}.
\newblock \og To ``Errrr'' is Human: Ecology and Acoustics of Speech
  Disfluencies\fg.
\newblock {\em Journal of the International Phonetic Association},
  31(1):153--169, 2001.

\bibitem[SMZ95]{Seneff95}
S.~\bgroup\sc Seneff\egroup{}, M.~\bgroup\sc McCandless\egroup{} \andname{}
  V.~\bgroup\sc Zue\egroup{}.
\newblock \og Integrating Natural Language into the Word Graph Search for
  Simultaneous Speech Recognition and Understanding\fg.
\newblock \Inname{} {\em Proc. of the 4th European Conference on Speech
  Communication and Technology (Eurospeech)}, Madrid, Espagne, 1995.

\bibitem[SO96]{Siu96}
M.-H. \bgroup\sc Siu\egroup{} \andname{} M.~\bgroup\sc Ostendorf\egroup{}.
\newblock \og Modeling Disfluencies in Conversational Speech\fg.
\newblock \Inname{} {\em Proc. of the 4th International Conference on Spoken
  Language Processing (ICSLP)}, \volumename{}~1, Philadelphie, Pennsylvanie,
  États-Unis, 1996.

\bibitem[SO00]{Siu00}
M.-H. \bgroup\sc Siu\egroup{} \andname{} M.~\bgroup\sc Ostendorf\egroup{}.
\newblock \og Variable {N}-Grams and Extensions for Conversational Speech
  Language Modeling\fg.
\newblock {\em IEEE Transactions on Speech and Audio Processing}, 8(1):63--75,
  2000.

\bibitem[SR97]{Seymore97}
K.~\bgroup\sc Seymore\egroup{} \andname{} R.~\bgroup\sc Rosenfeld\egroup{}.
\newblock \og Using Story Topics for Language Model Adaptation\fg.
\newblock \Inname{} {\em Proc. of the 5th European Conference on Speech
  Communication and Technology (Eurospeech)}, Rhodes, Grèce, 1997.

\bibitem[SR99]{Samuelsson99}
C.~\bgroup\sc Samuelsson\egroup{} \andname{} W.~\bgroup\sc Reichl\egroup{}.
\newblock \og Class-Based Language Model for Large-Vocabulary Speech
  Recognition Extracted from Part-of-Speech Statistics\fg.
\newblock \Inname{} {\em Proc. of the IEEE International Conference on
  Acoustics, Speech, and Signal Processing (ICASSP)}, \volumename{}~1, Phoenix,
  Arizona, États-Unis, 1999.

\bibitem[SS94]{Stolcke94}
A.~\bgroup\sc Stolcke\egroup{} \andname{} J.~\bgroup\sc Segal\egroup{}.
\newblock \og Precise {N}-Gram Probabilities from Stochastic Context-Free
  Grammars\fg.
\newblock \Inname{} {\em Proc. of the 32nd Annual Meeting of the Association
  for Computational Linguistics (ACL)}, Las Cruces, Nouveau Mexique,
  États-Unis, 1994.

\bibitem[SS96]{Stolcke96b}
A.~\bgroup\sc Stolcke\egroup{} \andname{} E.~\bgroup\sc Shriberg\egroup{}.
\newblock \og Statistical Language Modeling for Speech Disfluencies\fg.
\newblock \Inname{} {\em Proc. of the IEEE International Conference on
  Acoustics, Speech, and Signal Processing (ICASSP)}, \volumename{}~1, Atlanta,
  Géorgie, États-Unis, 1996.

\bibitem[STHKN95]{Schukat95}
E.~G. \bgroup\sc Schukat-Talamazzini\egroup{}, R.~\bgroup\sc Hendrych\egroup{},
  R.~\bgroup\sc Kompe\egroup{} \andname{} H.~\bgroup\sc Niemann\egroup{}.
\newblock \og Permugram Language Models\fg.
\newblock \Inname{} {\em Proc. of the 4th European Conference on Speech
  Communication and Technology (Eurospeech)}, \volumename{}~3, Madrid, Espagne,
  1995.

\bibitem[Sto95]{Stolcke95}
A.~\bgroup\sc Stolcke\egroup{}.
\newblock \og An Efficient Probabilistic Context-Free Parsing Algorithm that
  Computes Prefix Probabilities\fg.
\newblock {\em Computational Linguistics}, 21(2):165--202, 1995.

\bibitem[Sto97]{Stolcke97}
A.~\bgroup\sc Stolcke\egroup{}.
\newblock \og Linguistic Knowledge and Empirical Methods in Speech Recognition
  - Natural Language Processing\fg.
\newblock {\em AI Magazine}, 18(4), 1997.

\bibitem[Str03]{Strassel03}
S.~\bgroup\sc Strassel\egroup{}.
\newblock \og {\em Simple Metadata Annotation Specification. Version 5.0}\fg.
\newblock Linguistic Data Consortium, 2003.

\bibitem[SW94]{Suhm94}
B.~\bgroup\sc Suhm\egroup{} \andname{} A.~\bgroup\sc Waibel\egroup{}.
\newblock \og Towards Better Language Models for Spontaneous Speech\fg.
\newblock \Inname{} {\em Proc. of the 3rd International Conference on Spoken
  Language Processing (ICSLP)}, \volumename{}~2, Yokohama, Japon, 1994.

\bibitem[SWH03]{Seneff03}
S.~\bgroup\sc Seneff\egroup{}, C.~\bgroup\sc Wang\egroup{} \andname{} T.~J.
  \bgroup\sc Hazen\egroup{}.
\newblock \og Automatic Induction of {N}-Gram Language Models from a Natural
  Language Grammar\fg.
\newblock \Inname{} {\em Proc. of the 8th European Conference on Speech
  Communication and Technology (Eurospeech)}, Genève, Suisse, 2003.

\bibitem[TK95]{Tamoto95}
M.~\bgroup\sc Tamoto\egroup{} \andname{} T.~\bgroup\sc Kawabata\egroup{}.
\newblock \og Clustering Word Category Based on Binomial Posteriori
  Cooccurrence Distribution\fg.
\newblock \Inname{} {\em Proc. of the IEEE International Conference on
  Acoustics, Speech, and Signal Processing (ICASSP)}, Detroit, Michigan,
  États-Unis, 1995.

\bibitem[TN97]{Tillmann97}
C.~\bgroup\sc Tillmann\egroup{} \andname{} H.~\bgroup\sc Ney\egroup{}.
\newblock \og Word Triggers and the EM Algorithm\fg.
\newblock \Inname{} {\em Proc. of the Workshop Computational Natural Language
  Learning (CoNLL 97)}, Madrid, Espagne, 1997.

\bibitem[UNY{\etalchar{+}}02]{Uchimoto02}
K.~\bgroup\sc Uchimoto\egroup{}, C.~\bgroup\sc Nobata\egroup{}, A.~\bgroup\sc
  Yamada\egroup{}, S.~\bgroup\sc Sekine\egroup{} \andname{} H.~\bgroup\sc
  Isahara\egroup{}.
\newblock \og Morphological Analysis of the Spontaneous Speech Corpus\fg.
\newblock \Inname{} {\em Proc. of the 19th International Conference on
  Computational Linguistics (COLING)}, \volumename{}~2, Taipei, Taiwan, 2002.

\bibitem[VAR99]{Vaufreydaz99}
D.~\bgroup\sc Vaufreydaz\egroup{}, M.~\bgroup\sc Akbar\egroup{} \andname{}
  J.~\bgroup\sc Rouillard\egroup{}.
\newblock \og Internet Documents: a Rich Source for Spoken Language
  Modeling\fg.
\newblock \Inname{} {\em Proc. of the IEEE Workshop Automatic Speech
  Recognition and Understanding (ASRU)}, Keystone, Colorado, États-Unis, 1999.

\bibitem[VEZD00]{VanEynde00}
F.~\bgroup\sc Van~Eynde\egroup{}, J.~\bgroup\sc Zavrel\egroup{} \andname{}
  W.~\bgroup\sc Daelemans\egroup{}.
\newblock \og Part of Speech Tagging and Lemmatisation for the Spoken Dutch
  Corpus\fg.
\newblock \Inname{} {\em Proc. of the Conference on Language Resources and
  Evaluation (LREC)}, Athènes, Grèce, 2000.

\bibitem[VKDS04]{Vergyri04}
D.~\bgroup\sc Vergyri\egroup{}, K.~\bgroup\sc Kirchhoff\egroup{}, K.~\bgroup\sc
  Duh\egroup{} \andname{} A.~\bgroup\sc Stolcke\egroup{}.
\newblock \og Morphology-Based Language Modeling for Arabic Speech
  Recognition\fg.
\newblock \Inname{} {\em Proc. of the 8th International Conference on Spoken
  Language Processing (ICSLP)}, île de Jeju, Corée du Sud, 2004.

\bibitem[VV99]{Valli99}
A.~\bgroup\sc Valli\egroup{} \andname{} J.~\bgroup\sc Véronis\egroup{}.
\newblock \og Étiquetage grammatical de corpus oraux~: problèmes et
  perspectives\fg.
\newblock {\em Revue française de linguistique appliquée}, 4(2):113--133, 1999.

\bibitem[Vér04]{Veronis04}
J.~\bgroup\sc Véronis\egroup{}.
\newblock \og Le traitement automatique des corpus oraux\fg.
\newblock {\em TAL}, 45(2):7--14, 2004.

\bibitem[WH02]{Wang02b}
W.~\bgroup\sc Wang\egroup{} \andname{} M.~P. \bgroup\sc Harper\egroup{}.
\newblock \og The SuperARV Language Model: Investigating the Effectiveness of
  Tightly Integrating Multiple Knowledge Sources\fg.
\newblock \Inname{} {\em Proc. of the Empirical Methods in Natural Language
  Processing Conference (EMNLP)}, Philadelphie, Pennsylvanie, États-Unis, 2002.

\bibitem[WHS03]{Wang03}
W.~\bgroup\sc Wang\egroup{}, M.~P. \bgroup\sc Harper\egroup{} \andname{}
  A.~\bgroup\sc Stolcke\egroup{}.
\newblock \og The Robustness of an Almost-Parsing Language Model Given Errorful
  Training Data\fg.
\newblock \Inname{} {\em Proc. of the IEEE International Conference on
  Acoustics, Speech, and Signal Processing (ICASSP)}, \volumename{}~1, Hong
  Kong, Chine, 2003.

\bibitem[WK99]{Wu99}
J.~\bgroup\sc Wu\egroup{} \andname{} S.~\bgroup\sc Khudanpur\egroup{}.
\newblock \og Combining Nonlocal, Syntactic and {N}-Gram Dependencies in
  Language Modeling\fg.
\newblock \Inname{} {\em Proc. of the 5th European Conference on Speech
  Communication and Technology (Eurospeech)}, Budapest, Hongrie, 1999.

\bibitem[WLH02]{Wang02}
W.~\bgroup\sc Wang\egroup{}, Y.~\bgroup\sc Liu\egroup{} \andname{} M.~P.
  \bgroup\sc Harper\egroup{}.
\newblock \og Rescoring Effectiveness of Language Models Using Different Levels
  of Knowledge and their Integration\fg.
\newblock \Inname{} {\em Proc. of the IEEE International Conference on
  Acoustics, Speech, and Signal Processing (ICASSP)}, \volumename{}~1, Orlando,
  Florida, États-Unis, 2002.

\bibitem[WMH00]{Wang00}
Y.-Y. \bgroup\sc Wang\egroup{}, M.~\bgroup\sc Mahajan\egroup{} \andname{}
  X.~\bgroup\sc Huang\egroup{}.
\newblock \og A Unified Context-Free Grammar and {N}-Gram Model for Spoken
  Language Processing\fg.
\newblock \Inname{} {\em Proc. of the IEEE International Conference on
  Acoustics, Speech, and Signal Processing (ICASSP)}, \volumename{}~3,
  Istanbul, Turquie, 2000.

\bibitem[WSH04]{Wang04}
W.~\bgroup\sc Wang\egroup{}, A.~\bgroup\sc Stolcke\egroup{} \andname{} M.~P.
  \bgroup\sc Harper\egroup{}.
\newblock \og The Use of a Linguistically Motivated Language Model in
  Conversational Speech Recognition\fg.
\newblock \Inname{} {\em Proc. of the IEEE International Conference on
  Acoustics, Speech, and Signal Processing (ICASSP)}, \volumename{}~1,
  Montréal, Canada, 2004.

\bibitem[WW01]{Whittaker01}
E.~W.~D. \bgroup\sc Whittaker\egroup{} \andname{} P.~C. \bgroup\sc
  Woodland\egroup{}.
\newblock \og Efficient Class-Based Language Modelling for Very Large
  Vocabularies\fg.
\newblock \Inname{} {\em Proc. of the IEEE International Conference on
  Acoustics, Speech, and Signal Processing (ICASSP)}, \volumename{}~1, Salt
  Lake City, Utah, États-Unis, 2001.

\bibitem[YS99]{Yamamoto99}
H.~\bgroup\sc Yamamoto\egroup{} \andname{} Y.~\bgroup\sc Sagisaka\egroup{}.
\newblock \og Multi-Class Composite {N}-Gram Based on Connection Direction\fg.
\newblock \Inname{} {\em Proc. of the IEEE International Conference on
  Acoustics, Speech, and Signal Processing (ICASSP)}, \volumename{}~1, Phoenix,
  Arizona, États-Unis, 1999.

\bibitem[ZW98]{Zechner98}
K.~\bgroup\sc Zechner\egroup{} \andname{} A.~\bgroup\sc Waibel\egroup{}.
\newblock \og Using Chunk Based Partial Parsing of Spontaneous Speech in
  Unrestricted Domains for Reducing Word Error Rate in Speech Recognition\fg.
\newblock \Inname{} {\em Proc. of the 36th Annual Meeting of the Association
  for Computational Linguistics and the 17th International Conference on
  Computational Linguistics (COLING-ACL)}, Montréal, Canada, 1998.

\bibitem[ÉD04]{delic04}
Équipe \bgroup\sc {DELIC}\egroup{}.
\newblock \og Présentation du corpus de référence du français parlé\fg.
\newblock {\em Recherches sur le français parlé}, 18, 2004.

\end{thebibliography}

\end{otherlanguage*}

\end{document}